%% file: main.tex
\pdfoutput=1
\documentclass[12pt,a4paper]{article}
\usepackage[nice]{nicefrac}
\usepackage{amsmath}
\usepackage{xfrac}
\usepackage{booktabs}
\usepackage{ifthen} % for conditional statements
\newboolean{pdflatex}
\setboolean{pdflatex}{true} % False for eps figures 

\newboolean{articletitles}
\setboolean{articletitles}{true} % False removes titles in references

\newboolean{uprightparticles}
\setboolean{uprightparticles}{false} %True for upright particle symbols

\def\lg{\lambda_\gamma}
\def\lb{\lambda_b}
\def\lp{\lambda_p}
\def\lL{\lambda_\Lz}
\def\jg{J_\gamma}
\def\jb{J_b}

\def\jL{J_\Lz}
\def\LbLg{\decay{\Lb}{\Lz\g}}
\def\LpK{\decay{\Lz}{\proton\Km}}
\def\LbpKg{\decay{\Lb}{\proton\Km\g}}
\def\LbpKll{\decay{\Lb}{\proton\Km\ellell}}
\def\LbpKJpsi{\decay{\Lb}{\jpsi\proton\Km}}
\def\pK{\proton\Km}
\def\sWeights{\mbox{\em sWeights}\xspace}

\def\paperauthors{LHCb collaboration} % Leave as is for PAPER, CONF and FIGURE
\def\paperasciititle{Amplitude analysis of the Lb 2 pKgamma decay} % Set ASCII title here !! MAKE sure it's only ASCII characters !! 
\def\papertitle{Amplitude analysis of the \LbpKg decay} % Latex formatted title
\def\paperkeywords{{High Energy Physics}, {LHCb}} % Comma separated list
\def\papercopyright{\the\year\ CERN for the benefit of the LHCb collaboration}
\def\paperlicence{CC BY 4.0 licence}
\def\paperlicenceurl{https://creativecommons.org/licenses/by/4.0/}

\input{preamble}
\usepackage{pdflscape}
\usepackage{multicol}
\usepackage{multirow}
\begin{document}

%%%%%%%%%%%%%%%%%%%%%%%%%
%%%%% Title     %%%%%%%%%
%%%%%%%%%%%%%%%%%%%%%%%%%
\renewcommand{\thefootnote}{\fnsymbol{footnote}}
\setcounter{footnote}{1}

% %%%%%%% CHOOSE TITLE PAGE--------
%\onecolumn
\input{title-LHCb-PAPER}

%\twocolumn
% %%%%%%%%%%%%% ---------

\renewcommand{\thefootnote}{\arabic{footnote}}
\setcounter{footnote}{0}

%%%%%%%%%%%%%%%%%%%%%%%%%%%%%%%%
%%%%%  Table of Content   %%%%%%
%%%%%%%%%%%%%%%%%%%%%%%%%%%%%%%%
%%%% Uncomment if desired
%\tableofcontents
\cleardoublepage

%%%%%%%%%%%%%%%%%%%%%%%%%
%%%%% Main text %%%%%%%%%
%%%%%%%%%%%%%%%%%%%%%%%%%

\pagestyle{plain} % restore page numbers for the main text
\setcounter{page}{1}
\pagenumbering{arabic}

%% Uncomment during review phase. 
%% Comment before a final submission.
%\linenumbers

\input{body}
\cleardoublepage

\input{acknowledgements}
\input{appendix}
\cleardoublepage

\addcontentsline{toc}{section}{References}
%\setboolean{inbibliography}{true}
\bibliographystyle{LHCb}
\bibliography{main,standard,LHCb-PAPER,LHCb-CONF,LHCb-DP,LHCb-TDR}

%\newpage
\input{Authorship_LHCb-PAPER-2023-036}

\end{document}

%% file: preamble.tex
\usepackage[top=1in, bottom=1.25in, left=1in, right=1in]{geometry}

\usepackage{microtype}
\usepackage{lineno}  % for line numbering during review
\usepackage{xspace} % To avoid problems with missing or double spaces after
\usepackage{caption} %these three command get the figure and table captions automatically small

\usepackage{graphicx}  % to include figures (can also use other packages)
\usepackage{color}
\usepackage{colortbl}
\graphicspath{{./figs/}} % Make Latex search fig subdir for figures
\usepackage{amsmath} % Adds a large collection of math symbols
\usepackage{amssymb}
\usepackage{amsfonts}
\usepackage{upgreek} % Adds in support for greek letters in roman typeset

\newcommand*\patchAmsMathEnvironmentForLineno[1]{%
\expandafter\let\csname old#1\expandafter\endcsname\csname #1\endcsname
\expandafter\let\csname oldend#1\expandafter\endcsname\csname
end#1\endcsname
 \renewenvironment{#1}%
   {\linenomath\csname old#1\endcsname}%
   {\csname oldend#1\endcsname\endlinenomath}%
}
\newcommand*\patchBothAmsMathEnvironmentsForLineno[1]{%
  \patchAmsMathEnvironmentForLineno{#1}%
  \patchAmsMathEnvironmentForLineno{#1*}%
}
\AtBeginDocument{%
\patchBothAmsMathEnvironmentsForLineno{equation}%
\patchBothAmsMathEnvironmentsForLineno{align}%
\patchBothAmsMathEnvironmentsForLineno{flalign}%
\patchBothAmsMathEnvironmentsForLineno{alignat}%
\patchBothAmsMathEnvironmentsForLineno{gather}%
\patchBothAmsMathEnvironmentsForLineno{multline}%
\patchBothAmsMathEnvironmentsForLineno{eqnarray}%
}

\usepackage{hyperxmp}

\usepackage[pdftex,
            pdfauthor={\paperauthors},
            pdftitle={\paperasciititle},
            pdfkeywords={\paperkeywords},
            pdfcopyright={Copyright (C) \papercopyright},
            pdflicenseurl={\paperlicenceurl}]{hyperref}
\usepackage[colorinlistoftodos,textsize=scriptsize]{todonotes}

\usepackage[bottom,flushmargin,hang,multiple]{footmisc}

\usepackage[all]{hypcap} % Internal hyperlinks to floats.

\input{lhcb-symbols-def} % Add in the predefined LHCb symbols

\usepackage{cite} % Allows for ranges in citations
\usepackage{mciteplus}

%% file: lhcb-symbols-def.tex
\usepackage{xspace} 
\usepackage{upgreek}

\def\lhcb   {\mbox{LHCb}\xspace}

\def\spd    {SPD\xspace}

\def\ecal   {ECAL\xspace}

\def\MagUp {\mbox{\em Mag\kern -0.05em Up}\xspace}

\ifthenelse{\boolean{uprightparticles}}%
{
 
 \def\Pgamma      {\ensuremath{\upgamma}\xspace}

 \def\Ppi         {\ensuremath{\uppi}\xspace}

 \def\Ppsi        {\ensuremath{\uppsi}\xspace}

 \def\PDelta      {\ensuremath{\Delta}\xspace}                 
 \def\PXi         {\ensuremath{\Xi}\xspace}                 
 \def\PLambda     {\ensuremath{\Lambda}\xspace}                 
 \def\PSigma      {\ensuremath{\Sigma}\xspace}                 
 \def\POmega      {\ensuremath{\Omega}\xspace}                 
 \def\PUpsilon    {\ensuremath{\Upsilon}\xspace}
 \let\oldPi\Pi
 \def\PPi         {\ensuremath{\oldPi}\xspace}

 \def\PB      {\ensuremath{\mathrm{B}}\xspace}                 
                  
 \def\PD      {\ensuremath{\mathrm{D}}\xspace}

 \def\PJ      {\ensuremath{\mathrm{J}}\xspace}                 
 \def\PK      {\ensuremath{\mathrm{K}}\xspace}

 \def\Pb      {\ensuremath{\mathrm{b}}\xspace}                 
 \def\Pc      {\ensuremath{\mathrm{c}}\xspace}

 \def\Pi      {\ensuremath{\mathrm{i}}\xspace}

 \def\Pp      {\ensuremath{\mathrm{p}}\xspace}

 \def\Ps      {\ensuremath{\mathrm{s}}\xspace}

 \def\thebaroffset{0.0em}
}
{
 
 \def\Pgamma      {\ensuremath{\gamma}\xspace}

 \def\Ppi         {\ensuremath{\pi}\xspace}

 \def\Ppsi        {\ensuremath{\psi}\xspace}                 
                  
 \mathchardef\PDelta="7101
 \mathchardef\PXi="7104
 \mathchardef\PLambda="7103
 \mathchardef\PSigma="7106
 \mathchardef\POmega="710A
 \mathchardef\PUpsilon="7107
 \mathchardef\PPi="7105
                  
 \def\PB      {\ensuremath{B}\xspace}                 
                  
 \def\PD      {\ensuremath{D}\xspace}

 \def\PJ      {\ensuremath{J}\xspace}                 
 \def\PK      {\ensuremath{K}\xspace}

 \def\Pb      {\ensuremath{b}\xspace}                 
 \def\Pc      {\ensuremath{c}\xspace}

 \def\Pi      {\ensuremath{i}\xspace}

 \def\Pp      {\ensuremath{p}\xspace}

 \def\Ps      {\ensuremath{s}\xspace}

 \def\thebaroffset{0.18em}
}
\newcommand{\offsetoverline}[2][\thebaroffset]{\kern #1\overline{\kern -#1 #2}}%

\makeatletter
\ifcase \@ptsize \relax% 10pt
  \newcommand{\miniscule}{\@setfontsize\miniscule{4}{5}}% \tiny: 5/6
\or% 11pt
  \newcommand{\miniscule}{\@setfontsize\miniscule{5}{6}}% \tiny: 6/7
\or% 12pt
  \newcommand{\miniscule}{\@setfontsize\miniscule{5}{6}}% \tiny: 6/7
\fi
\makeatother

\DeclareRobustCommand{\optbar}[1]{\shortstack{{\miniscule (\rule[.5ex]{1.25em}{.18mm})}
  \\ [-.7ex] $#1$}}

\def\ellell     {\ensuremath{\ell^+ \ell^-}\xspace}

\def\g      {{\ensuremath{\Pgamma}}\xspace}

\def\squark    {{\ensuremath{\Ps}}\xspace}

\def\cquark    {{\ensuremath{\Pc}}\xspace}

\def\bquark    {{\ensuremath{\Pb}}\xspace}

\def\pion   {{\ensuremath{\Ppi}}\xspace}
\def\piz    {{\ensuremath{\pion^0}}\xspace}
\def\pip    {{\ensuremath{\pion^+}}\xspace}
\def\pim    {{\ensuremath{\pion^-}}\xspace}

\def\kaon    {{\ensuremath{\PK}}\xspace}
\def\KorKbar {\kern \thebaroffset\optbar{\kern -\thebaroffset \PK}{}\xspace}

\def\Kp      {{\ensuremath{\kaon^+}}\xspace}
\def\Km      {{\ensuremath{\kaon^-}}\xspace}

\def\Kstarm  {{\ensuremath{\kaon^{*-}}}\xspace}

\def\D       {{\ensuremath{\PD}}\xspace}

\def\DorDbar {\kern \thebaroffset\optbar{\kern -\thebaroffset \PD}\xspace}
\def\Dz      {{\ensuremath{\D^0}}\xspace}

\def\Dp      {{\ensuremath{\D^+}}\xspace}
\def\Dm      {{\ensuremath{\D^-}}\xspace}

\def\DpDm    {\ensuremath{\Dp {\kern -0.16em \Dm}}\xspace}

\def\B       {{\ensuremath{\PB}}\xspace}

\def\BorBbar {\kern \thebaroffset\optbar{\kern -\thebaroffset \PB}\xspace}
\def\Bz      {{\ensuremath{\B^0}}\xspace}

\def\Bd      {{\ensuremath{\B^0}}\xspace}

\def\BdorBdbar {\kern \thebaroffset\optbar{\kern -\thebaroffset \Bd}\xspace}

\def\Bs      {{\ensuremath{\B^0_\squark}}\xspace}

\def\BsorBsbar {\kern \thebaroffset\optbar{\kern -\thebaroffset \Bs}\xspace}

\def\jpsi     {{\ensuremath{{\PJ\mskip -3mu/\mskip -2mu\Ppsi}}}\xspace}

\def\Y#1S{\ensuremath{\PUpsilon{(#1S)}}\xspace}

\def\proton      {{\ensuremath{\Pp}}\xspace}
\def\antiproton  {{\ensuremath{\overline \proton}}\xspace}

\def\Lz          {{\ensuremath{\PLambda}}\xspace}
\def\Lbar        {{\ensuremath{\offsetoverline{\PLambda}}}\xspace}
\def\LorLbar     {\kern \thebaroffset\optbar{\kern -\thebaroffset \PLambda}\xspace}

\def\Xires       {{\ensuremath{\PXi}}\xspace}

\def\Lc          {{\ensuremath{\Lz^+_\cquark}}\xspace}

\def\Lb           {{\ensuremath{\Lz^0_\bquark}}\xspace}
\def\Lbbar        {{\ensuremath{\Lbar{}^0_\bquark}}\xspace}

\def\Xibz         {{\ensuremath{\Xires^0_\bquark}}\xspace}

\newcommand{\decay}[2]{\ensuremath{#1\!\to #2}\xspace} 

\def\to                 {\ensuremath{\rightarrow}\xspace}

\def\qsq       {{\ensuremath{q^2}}\xspace}

\def\CP                {{\ensuremath{C\!P}}\xspace}

\def\bsll     {\decay{\bquark}{\squark \ell^+ \ell^-}}

\def\AT#1     {\ensuremath{A_{\mathrm{T}}^{#1}}\xspace}           % 2
\def\btosgam  {\decay{\bquark}{\squark \g}}

\def\C#1      {\ensuremath{\mathcal{C}_{#1}}\xspace}                       % 9
\def\Cp#1     {\ensuremath{\mathcal{C}_{#1}^{'}}\xspace}                    % 7
\def\Ceff#1   {\ensuremath{\mathcal{C}_{#1}^{\mathrm{(eff)}}}\xspace}        % 9  
\def\Cpeff#1  {\ensuremath{\mathcal{C}_{#1}^{'\mathrm{(eff)}}}\xspace}       % 7
\def\Ope#1    {\ensuremath{\mathcal{O}_{#1}}\xspace}                       % 2
\def\Opep#1   {\ensuremath{\mathcal{O}_{#1}^{'}}\xspace}                    % 7

\newcommand{\aunit}[1]{\ensuremath{\text{\,#1}}}       
\newcommand{\tev}{\aunit{Te\kern -0.1em V}\xspace}
\newcommand{\gev}{\aunit{Ge\kern -0.1em V}\xspace}
\newcommand{\mev}{\aunit{Me\kern -0.1em V}\xspace}
\newcommand{\kev}{\aunit{ke\kern -0.1em V}\xspace}
\newcommand{\ev}{\aunit{e\kern -0.1em V}\xspace}
 
\newcommand{\mevc}{\ensuremath{\aunit{Me\kern -0.1em V\!/}c}\xspace}
\newcommand{\gevc}{\ensuremath{\aunit{Ge\kern -0.1em V\!/}c}\xspace}
\newcommand{\mevcc}{\ensuremath{\aunit{Me\kern -0.1em V\!/}c^2}\xspace}
\newcommand{\gevcc}{\ensuremath{\aunit{Ge\kern -0.1em V\!/}c^2}\xspace}
\newcommand{\gevgevcccc}{\ensuremath{\gev^2\!/c^4}\xspace} % for q^2

\def\mm   {\aunit{mm}\xspace}

\def\fb   {\ensuremath{\aunit{fb}}\xspace}
\def\invfb   {\ensuremath{\fb^{-1}}\xspace}

\newcommand{\chisq}{\ensuremath{\chi^2}\xspace}

\def\deriv {\ensuremath{\mathrm{d}}}

\def\gsim{{~\raise.15em\hbox{$>$}\kern-.85em
          \lower.35em\hbox{$\sim$}~}\xspace}
\def\lsim{{~\raise.15em\hbox{$<$}\kern-.85em
          \lower.35em\hbox{$\sim$}~}\xspace}

\def\sPlot{\mbox{\em sPlot}\xspace}

\def\pt         {\ensuremath{p_{\mathrm{T}}}\xspace}

\def\et         {\ensuremath{E_{\mathrm{T}}}\xspace}

\def\rad{\aunit{rad}\xspace}

\def\pythia     {\mbox{\textsc{Pythia}}\xspace}

\def\tell1  {TELL1\xspace}
\def\ukl1   {UKL1\xspace}

\newcommand{\ie}{\mbox{\itshape i.e.}\xspace}

\newcommand{\lhcborcid}[1]{\href{https://orcid.org/#1}{\hspace*{0.1em}\raisebox{-0.45ex}{\includegraphics[width=1em]{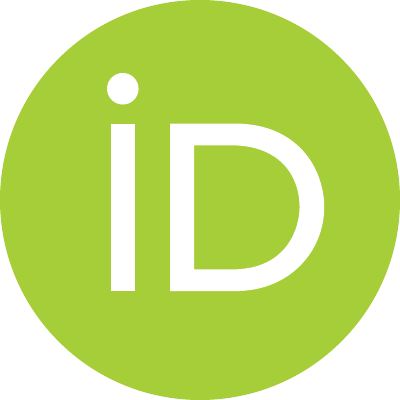}}}}

%% file: title-LHCb-PAPER.tex
% ===============================================================================
% Purpose: LHCb-PAPER journal paper title page template
% Author: 
% Created on: 2010-09-25
% ===============================================================================

%%%%%%%%%%%%%%%%%%%%%%%%%
%%%%%  TITLE PAGE  %%%%%%
%%%%%%%%%%%%%%%%%%%%%%%%%
\begin{titlepage}
\pagenumbering{roman}

% Header ---------------------------------------------------
\vspace*{-1.5cm}
\centerline{\large EUROPEAN ORGANIZATION FOR NUCLEAR RESEARCH (CERN)}
\vspace*{1.5cm}
\noindent
\begin{tabular*}{\linewidth}{lc@{\extracolsep{\fill}}r@{\extracolsep{0pt}}}
\ifthenelse{\boolean{pdflatex}}% Logo format choice
{\vspace*{-1.5cm}\mbox{\!\!\!\includegraphics[width=.14\textwidth]{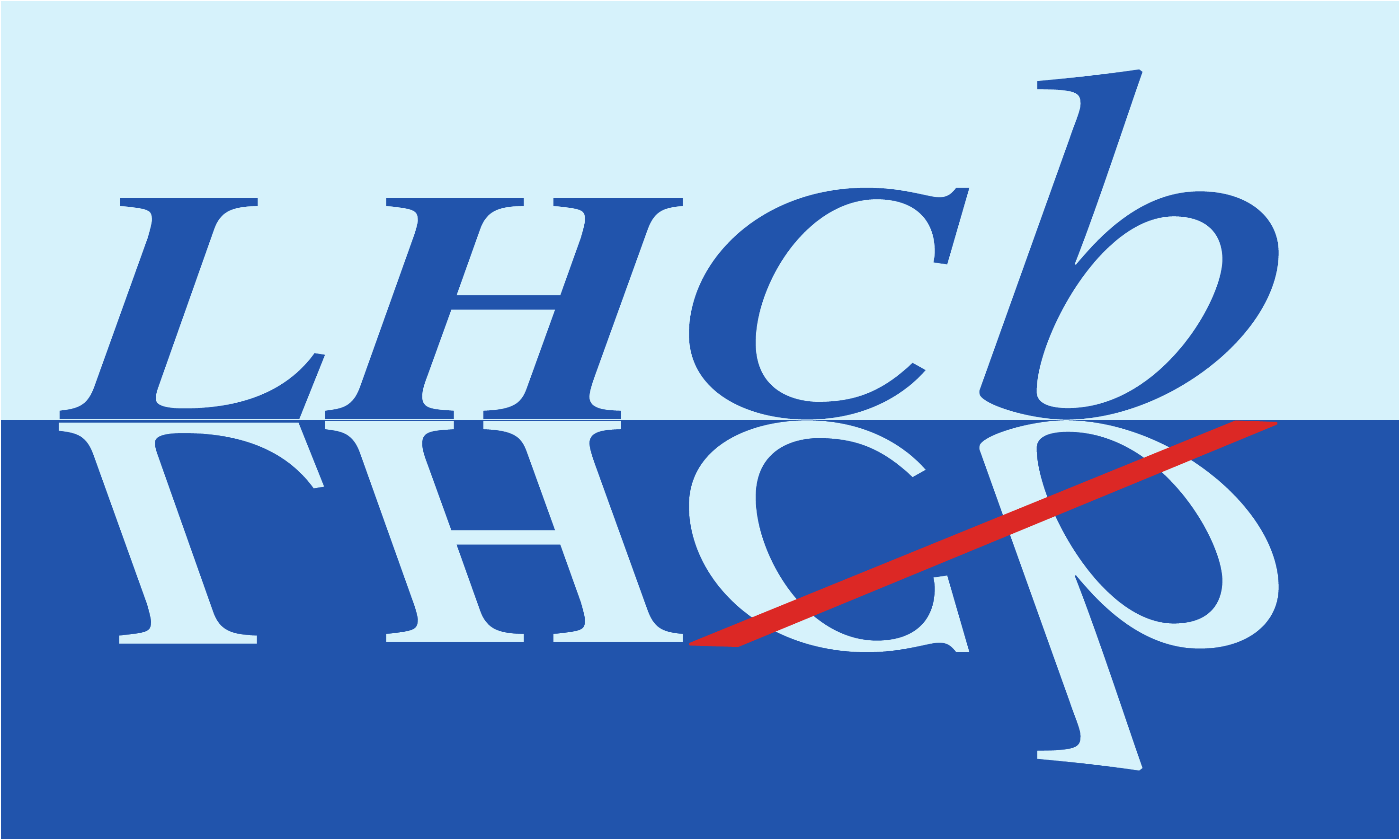}} & &}%
{\vspace*{-1.2cm}\mbox{\!\!\!\includegraphics[width=.12\textwidth]{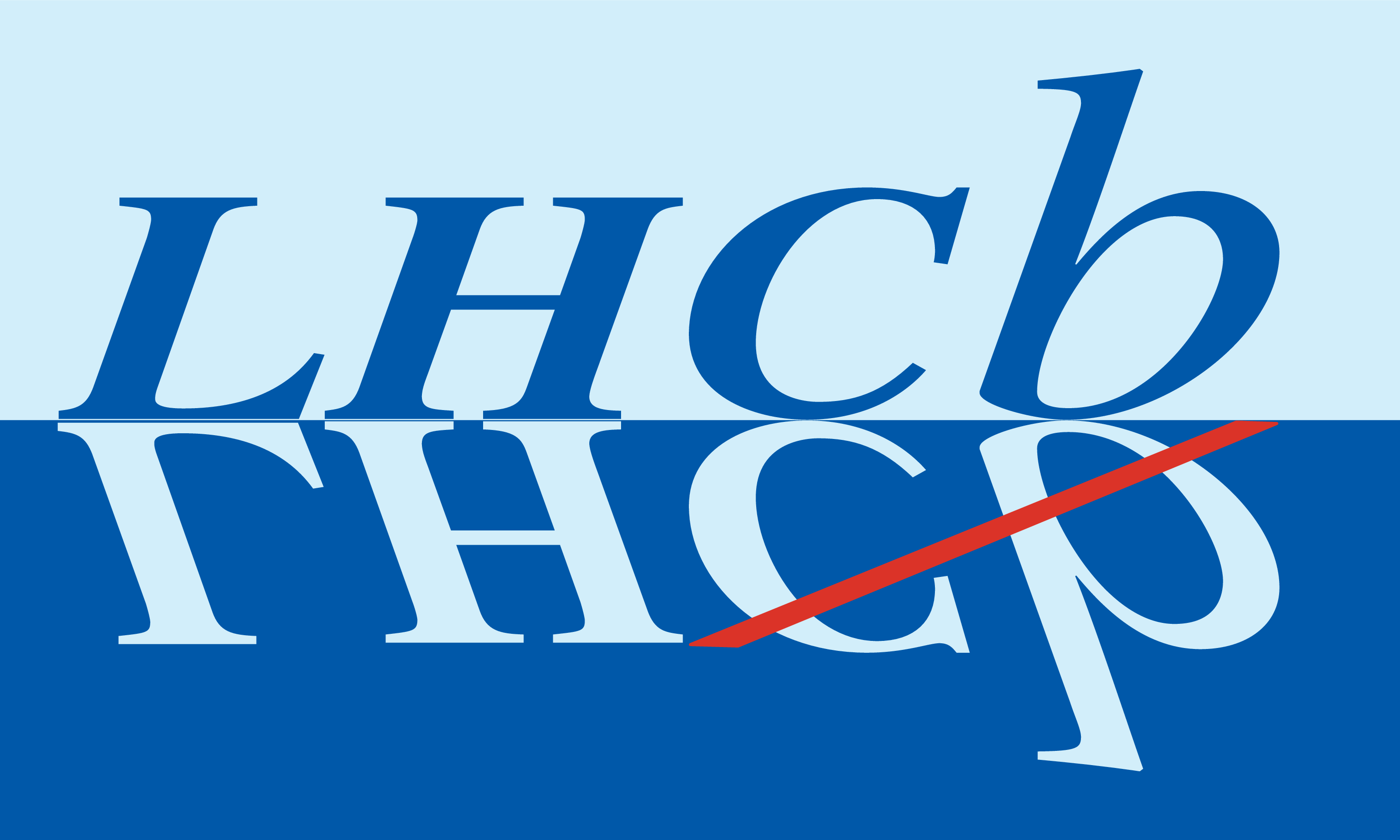}} & &}%
\\
 & & CERN-EP-2023-253 \\  % ID 
 & & LHCb-PAPER-2023-036 \\  % ID 
 & & 18 June 2024 \\%6 March 2024 \\ % Date - Can also hardwire e.g.: 23 March 2010
 & & \\
% not in paper \hline
\end{tabular*}

\vspace*{4.0cm}

% Title --------------------------------------------------
{\normalfont\bfseries\boldmath\huge
\begin{center}
% DO NOT EDIT HERE. Instead edit macro in main.tex to keep metadata correct
  \papertitle 
\end{center}
}

\vspace*{2.0cm}

% Authors -------------------------------------------------
\begin{center}
%In the footnote, replace 'paper' by 'Letter' in case of submission to PRL or PLB 
% Edit macro in main.tex to keep metadata correct
\paperauthors\footnote{Authors are listed at the end of this paper.}
\end{center}

\vspace{\fill}

% Abstract -----------------------------------------------
\begin{abstract}
  \noindent
  \noindent
The resonant structure of the radiative decay \LbpKg in the region of proton-kaon invariant-mass up to 2.5\gevcc is studied using proton-proton collision data recorded at centre-of-mass energies of 7, 8, and 13 \tev collected with the LHCb detector, corresponding to a total integrated luminosity of 9\invfb.
Results are given in terms of fit and interference fractions between the different components contributing to this final state.
Only $\Lz$ resonances decaying to \pK are found to be relevant, where the largest contributions stem from the $\Lz(1520)$, $\Lz(1600)$, $\Lz(1800)$, and $\Lz(1890)$ states.
\end{abstract}

\vspace*{2.0cm}

\begin{center}
  Published in JHEP 06 (2024) 098
\end{center}

\vspace{\fill}

{\footnotesize 
% Edit macro in main.tex to keep metadata correct
\centerline{\copyright~\papercopyright. \href{\paperlicenceurl}{\paperlicence}.}}
\vspace*{2mm}

\end{titlepage}

%%%%%%%%%%%%%%%%%%%%%%%%%%%%%%%%
%%%%%  EOD OF TITLE PAGE  %%%%%%
%%%%%%%%%%%%%%%%%%%%%%%%%%%%%%%%

%  empty page follows the title page ----
\newpage
\setcounter{page}{2}
\mbox{~}
%\newpage
%
%% Author List ----------------------------
%%  You need to get a new author list!
%\input{LHCb_authorlist.tex}
%
%The author list for journal publications is provided by the Membership Committee shortly after 'approval to go to paper' has been given.
%%It will be made available on the page
%%\verb!http://www.physik.uzh.ch/~strauman/forMemCo/LHCb-PAPER-XXXX-XXX/! .
%It will be sent to you by email shortly after a paper number has beens assigned.
%The author list should be included already at first circulation, 
%to allow new members of the collaboration to verify whether they have been included correctly.
%Occasionally a misspelled name is corrected or associated institutions become full members.
%In that case, a new author list will be sent to you.
%In case line numbering doesn't work well after including the authorlist, try moving the \verb!\bigskip! after the last author to a separate line.
%
%
%The authorship for Conference Reports should be ``The LHCb
%  collaboration'', with a footnote giving the name(s) of the contact
%  author(s), but without the full list of collaboration names.

%% file: body.tex
\section{Introduction}
\label{sec:Introduction}
Rare decays of \bquark hadrons  involving flavour-changing neutral currents, such as \bsll and \btosgam transitions, are forbidden at tree level in the Standard Model and further suppressed at loop-level through the GIM mechanism.
As a consequence, these decays are very sensitive to potential new particles that can enter virtually through loop-level processes or allow tree-level diagrams, affecting properties of these decays such as branching fractions and angular distributions.
The measurements of these processes can probe higher energy scales than those accessible via direct searches.

Thanks to the abundant production of \bquark baryons at the LHC, precision measurements of rare \bquark-baryon decays have become possible for the first time.
For example, the LHCb collaboration has performed tests of lepton universality using \LbpKll decays\footnote{The inclusion of charge-conjugate processes is implied throughout the text.} in the dilepton invariant-mass squared range \mbox{$0.1 < \qsq < 6.0  \gevgevcccc$}  and the \pK invariant-mass range \mbox{$m_{\proton\Km} < 2.6 \gevcc$}~\cite{LHCb-PAPER-2019-040}.
Moreover, the LHCb collaboration has searched for \CP violation in $\Lb\to\pK\mu^+\mu^-$ decays~\cite{LHCb-PAPER-2016-059} and measured the branching fraction of the $\Lb\to\Lz(1520)\mu^+\mu^-$ decay~\cite{LHCb-PAPER-2022-050}.
Direct interpretations of these results regarding models for physics beyond the Standard Model are difficult given the lack of detailed knowledge of the resonant structure of the \pK spectrum in different regions of the dilepton invariant-mass spectrum.

An observation of the \LbpKg decay was first reported unofficially in a thesis using Run 1 data (taken during 2011 and 2012) without giving a significance~\cite{RivesMolina:2230991}.
This paper presents an amplitude analysis of the \LbpKg decay which constitutes the first official observation of this mode.
This analysis measures \LbpKg decay properties for the first time and characterises the \pK spectrum at the photon pole of the recoiling system.
Theoretical knowledge of the \pK spectrum from \Lb decays, in particular the modelling of form factors, is limited to quark-model calculations~\cite{Mott:2011cx,Mott:2015zma}.
Predictions obtained from lattice QCD~\cite{Meinel:2020owd,Meinel:2021mdj}, HQET~\cite{Bordone:2021bop} and dispersive bounds analyses~\cite{Amhis:2022vcd} are only available for the decay via the $\Lz(1520)$ state.
The different $\Lz$ resonances in the \pK spectrum have been studied using fixed target experiments with incident kaons~\cite{Sarantsev:2019xxm,Matveev:2019igl}.
An amplitude analysis of \LbpKJpsi decays, which led to the discovery of states compatible with pentaquarks~\cite{LHCb-PAPER-2015-029}, studied the \pK spectrum from \Lb decays at the \jpsi resonance in the dimuon invariant-mass spectrum.
Additionally, if the amplitudes of the \LbpKg decay are known precisely, this measurement could constitute useful input to a future measurement of the photon polarisation, involving polarised \Lb baryons, for example from $Z$ boson decays~\cite{Hiller:2007ur}.

The \LbpKg decay provides an opportunity to complement the knowledge of the $\pK$ spectrum, including unique access to heavier states with masses larger than about 2~\gevcc that cannot be accessed with \LbpKJpsi decays due to the kinematic restrictions.
Measurements of resonance properties are vital inputs to the theoretical description of low-energy QCD as discussed in Ref.~\cite{Lutz:2015ejy}.
Employing data collected by the \lhcb detector in \proton\proton collisions during the years 2011--2012 (Run 1) and 2015--2018 (Run 2), corresponding to an integrated luminosity of about 9\invfb, this paper presents the first amplitude analysis of \LbpKg decays.

\section{Detector and selection}
The LHCb detector is a single-arm forward spectrometer covering the pseudorapidity range $2 < \eta < 5$, designed for the study of particles containing \bquark or \cquark quarks.
The detector includes a high-precision tracking system consisting of a silicon-strip vertex detector surrounding the proton-proton interaction region, a large-area silicon-strip detector located upstream of a dipole magnet with a bending power of about 4~Tm, and three stations of silicon-strip detectors and straw drift tubes placed downstream of the magnet.
The tracking system provides a measurement of the momentum of charged particles with a relative uncertainty that varies from 0.5\% at low momentum to 1.0\% at 200\gevc.
The minimum distance of a track to a primary proton-proton collision vertex (PV), the impact parameter (IP), is measured with a resolution of $(15 + 29/\pt)$~$\mu$m, where \pt is the component of the momentum transverse to the beam, in \gevc.
Different types of charged hadrons are distinguished using information from two ring-imaging Cherenkov detectors.
Photons, electrons and hadrons are identified by a calorimeter system consisting of scintillating-pad (\spd) and preshower detectors, an electromagnetic (\ecal) and a hadronic calorimeter.
In addition, a muon system allows the identification of muons.

Samples of simulated events are used to optimise selection requirements and estimate the efficiencies of the signal and backgrounds.
The simulated proton-proton collisions are generated using \pythia~\cite{Sjostrand:2007gs} with a specific LHCb configuration~\cite{LHCb-PROC-2010-056}.
Decays of hadronic particles are described by EvtGen~\cite{Lange:2001uf}, in which final-state radiation is generated using PHOTOS~\cite{davidson2015photos}.
The interaction of the generated particles with the
detector, and its response, are implemented in the Geant4 toolkit~\cite{Agostinelli:2002hh} as described in Ref.~\cite{Allison:2006ve}.
The \LbpKg decay is generated uniformly in phase space without assumptions on the decay dynamics.

The online event selection is performed by a trigger~\cite{LHCb-DP-2012-004,LHCb-DP-2019-001}, which consists of a hardware stage, based on information from the
calorimeter and muon systems, followed by a software stage, which applies a full event reconstruction.
The trigger exploits the presence of a high energy photon reconstructed from clusters in the \ecal.
In order to reduce background and improve the mass resolution for the \Lb and two-body invariant masses, clusters are required to have a transverse energy of $\et>2.5$--$2.96~\gev$ in Run 1 and $\et>2.11$--$2.7~\gev$ in Run 2, respectively, at the hardware trigger level.
%Depending on the specific trigger configuration, the thresholds can be as high as 2.96\gev in Run 1 and 2.7\gev in Run 2.
Moreover, the hardware trigger selects only events with fewer than 600 (450) hits in the \spd for Run 1 (2) to facilitate the reconstruction in the software trigger.
In the software trigger, the candidate must contain two high-\pt hadrons that are significantly displaced from the interaction point, as well as a high-\et photon.
During Run 2, a multivariate classifier based on topological criteria complements the cut-based software trigger selection~\cite{LHCb-PROC-2015-018}.
The Run 1 software trigger requires the di-hadron invariant mass, assuming both hadrons are kaons, to be below 2~\gevcc.
This severely affects the shape of the efficiency as a function of the proton-kaon invariant mass, resulting in the need for separate treatment of Run 1 and Run 2.
In addition to this cut in the Run 1 trigger, the large threshold for the photon energy results in low efficiency at high proton-kaon invariant mass.
As a consequence, the proton-kaon invariant mass range up to 2.5\gevcc is considered.

The reconstructed \Lb candidate is required to have good-quality track and vertex fits.
Two tracks, compatible with the kaon and proton hypotheses, are required to have an impact parameter larger than $0.1~\mm$, a transverse momentum larger than $1~\gevc$ as well as momentum larger than $5~\gevc$.
The photon must have $\et>3~\gev$.
The \Lb decay vertex isolation is used to reject partially reconstructed backgrounds.
Specifically, an upper limit is applied on the $\chi^2$ increase in the $\Lb$ decay vertex fit when adding the most compatible additional track, referred to in the following as $\Delta\chi^2_\text{vtx}(\Lb)$.
The \Lb momentum is further required to point back to the associated primary vertex.

Background candidates resulting from combinations of unrelated protons, kaons, and photons can be suppressed using kinematic variables.
A Boosted Decision Tree classifier (BDT)~\cite{Breiman} is trained on simulated events as signal proxy and on data candidates with \mbox{$m(pK\gamma) > m_\Lb+300\mevcc$} as background proxy, to suppress combinatorial background by exploiting mainly kinematic variables.
The input variables to the classifier are the momentum, pseudorapidity $\eta$, flight distance (FD), $\Delta\chi^2_{\rm vtx}$ of the \Lb baryon, IP and \pt of the hadrons, and IP, momentum, and \pt of the proton-kaon combination.
Additionally, the difference in the vertex-fit \chisq of the PV associated with the \Lb baryon reconstructed with and without the \Lb candidate is used.
A further input to the BDT in Run 2 is the isolation variable
\begin{equation}
    I_{\pt} = \frac{\pt(\Lb) - \sum \pt}{\pt(\Lb) + \sum \pt}
\end{equation}
for which the sum is taken over tracks that are not part of the signal candidate but are associated to the same PV and fall within a cone of half-angle $\Delta R < 1.7\rad$.
The half-angle of a track is defined as $(\Delta R)^2 = (\Delta\theta)^2 + (\Delta\phi)^2$, where $\Delta\theta$ and $\Delta\phi$ are the differences in the polar and azimuthal angles of each track with respect to the \Lb candidate direction.
The optimal BDT working point is determined by maximising the ratio $S/\sqrt{S+B}$, where $S$ is the number of expected signal candidates estimated from simulation samples and $B$ is the number of background candidates in the signal region estimated based on the background-dominated regions on either side of the \Lb peak.

Requirements on the particle identification variables decrease backgrounds stemming from misidentification.
Nevertheless, a large amount of misidentified \mbox{$\Bs\to\phi(\to \Kp\Km)\gamma$} decays passes all particle identification selections and pollutes the sample.
These are suppressed by vetoing candidates with a $\Kp\Km$ invariant mass, calculated by interpreting the proton candidate as a kaon, between $1.01$ and $1.04$\gevcc.
Remaining contributions from misidentified \Bs\to\Km\Kp$\gamma$ and \Bd\to\Km\pip$\gamma$ decays are estimated to contribute less than 0.5\% of the signal yield and are therefore not included in the baseline model.
Background stemming from photon misidentification, such as $\Lb\to p\Km\piz$ or \mbox{$\Lb\to p\Km\eta$}, is difficult to quantify due to their unknown resonant structures.
Estimates using simulation samples assuming a uniform distribution in the respective phase space indicate a contamination of 1--2\% relative to the signal decay.
Limiting the analysis to a proton-kaon invariant mass of 2.5\gevcc removes at least the contributions from potential proton-photon and kaon-photon resonances of these backgrounds which would be the most distorting.
Potential contamination from $\Xibz\to \pK\gamma$ decays are investigated and found to be negligible.
The data are checked for remaining misidentified backgrounds by applying higher thresholds to the proton and kaon particle identification selection requirements, and by comparing different two-body invariant mass distributions under various alternative mass hypotheses.
This reveals misidentified $\Dz\to \Kp\Km$ and $\Dz\to \Kp\pim$ decays combined with an unrelated photon, which populate the low mass side band of the signal \Lb mass peak.
A veto on these decays has a strong impact on the shape of the signal acceptance in the Dalitz plane.
For this reason, the candidates are retained and treated as part of the combinatorial background.
The effect of this treatment is considered as a systematic uncertainty.
Partially reconstructed decays, such as \Lb\to \proton\Kstarm{}(\to\Km\piz)$\gamma$, where the pion is not reconstructed, are also a source of background, which is included in the fit to the three-body invariant-mass distribution described in the following.

\section{Invariant mass fit}
\label{sec:massfit}
\begin{figure}
	\centering
	\includegraphics[width=.45\textwidth]{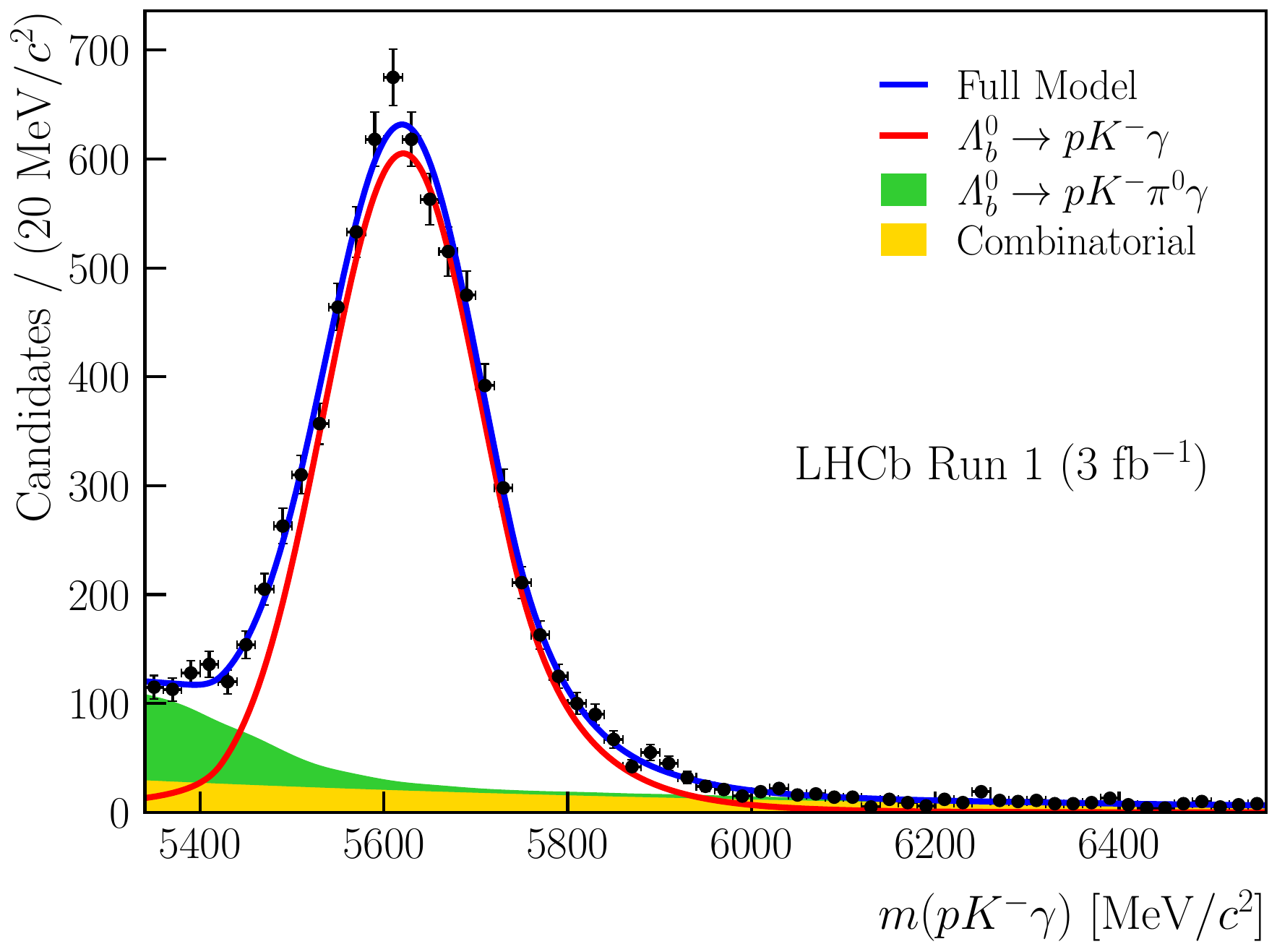}\hspace{.05\textwidth}%
	\includegraphics[width=.45\textwidth]{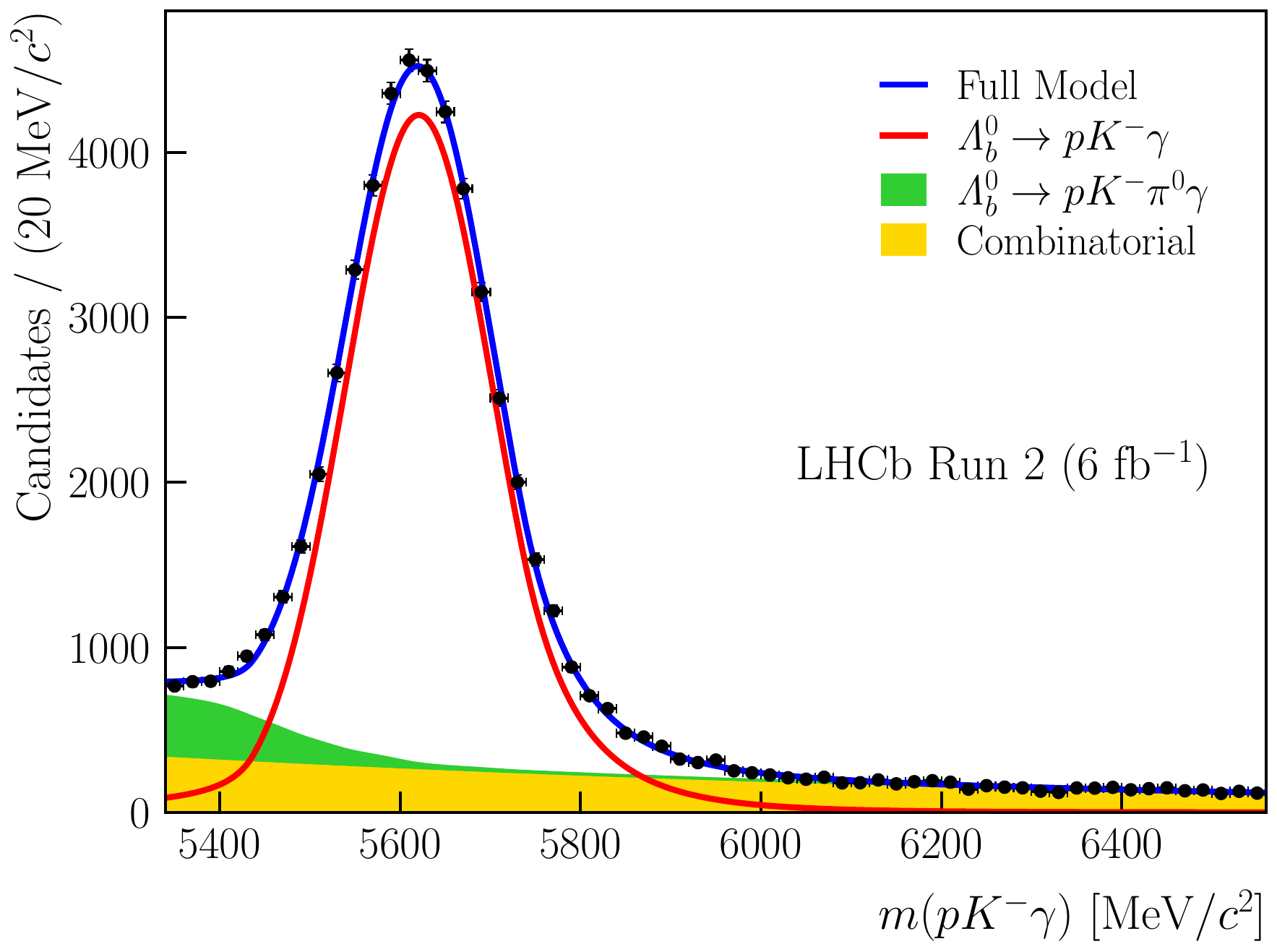}
	\caption{Distribution of the three-body invariant mass of the candidates in the (left) Run 1 and (right) Run 2 data sets.
    The results of the fits are overlaid.}
	\label{fig:massfit}
\end{figure}

The three-body invariant mass distribution of the candidates fulfilling all selection criteria is shown in Fig.~\ref{fig:massfit}.
An unbinned maximum-likelihood fit to these candidates is performed.
Following the \sPlot technique~\cite{Pivk:2004ty}, a signal weight ({\em sWeight}) is assigned to each candidate to statistically disentangle the signal and background components in the subsequent amplitude analysis.
The invariant mass fit is performed separately for Run 1 and Run 2, due to differences in the trigger configurations that affect the Dalitz plane distributions and hence require a separate treatment in the amplitude fit.

The signal is modelled by a double-sided Crystal-Ball~\cite{Oreglia:1980cs} function comprising a Gaussian core with asymmetric tails.
The tail parameters are determined using \LbpKg simulation samples.
The remaining background due to random combinations of particles is modelled using a decreasing exponential function where the slope and yield are allowed to vary freely in the fit to data.
The shape of the background from partially reconstructed decays is taken from simulation samples of \Lb\to \proton\Kstarm{}(\to\Km\piz)$\gamma$ decays generated uniformly in phase space, reconstructed as signal candidates, and modelled using a kernel density estimator~\cite{Rosenblatt:KDE} with Gaussian kernels.

Figure~\ref{fig:massfit} also shows the result of the invariant-mass fits to the Run 1 and Run 2 data sets.
The signal yields are determined to be $6855\pm93$ and $45558\pm247$, in Run 1 and Run 2, respectively.

\begin{figure}
    \centering
    \includegraphics[width=.45\textwidth]{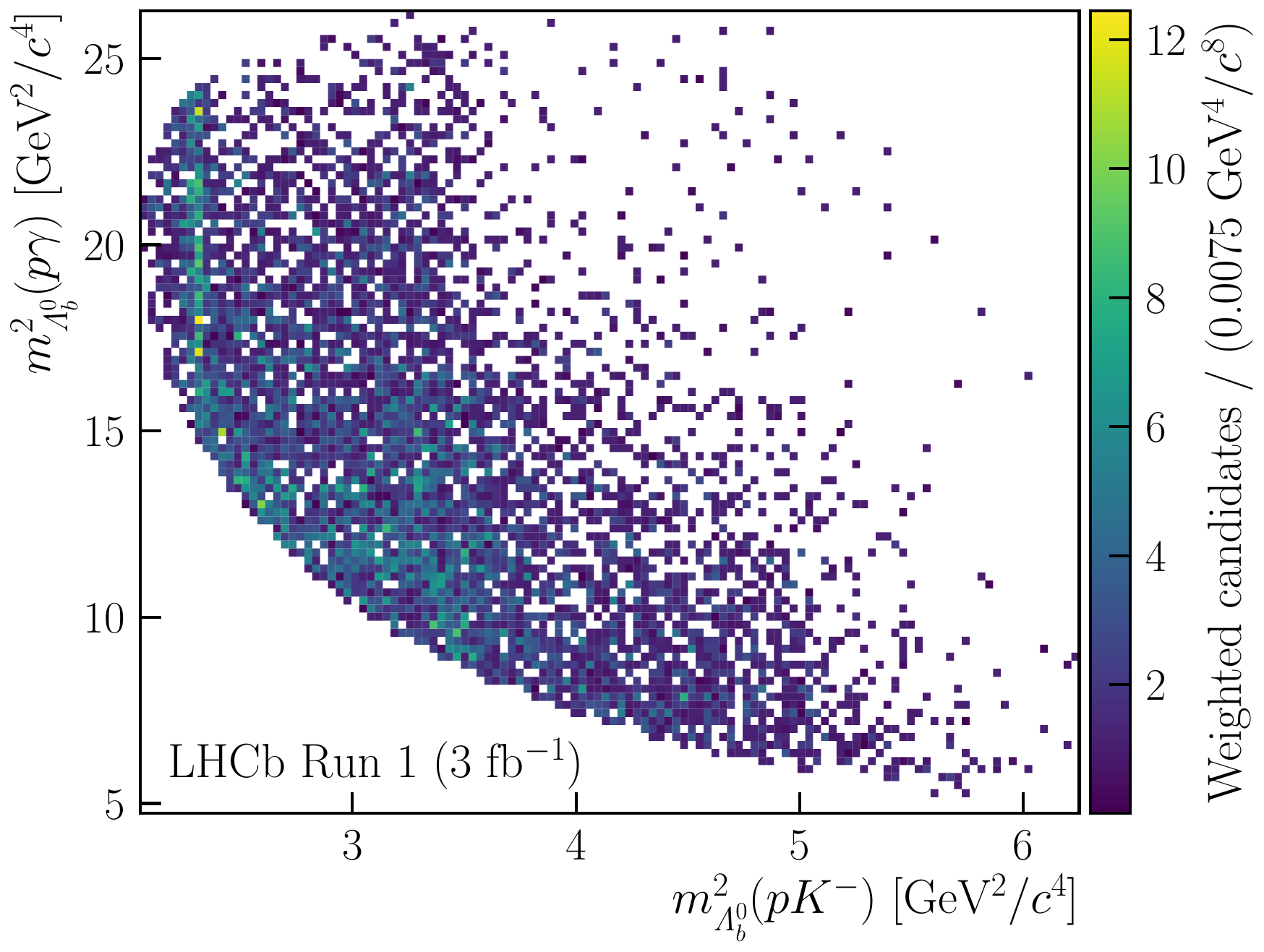}\hspace{.05\textwidth}%
    \includegraphics[width=.45\textwidth]{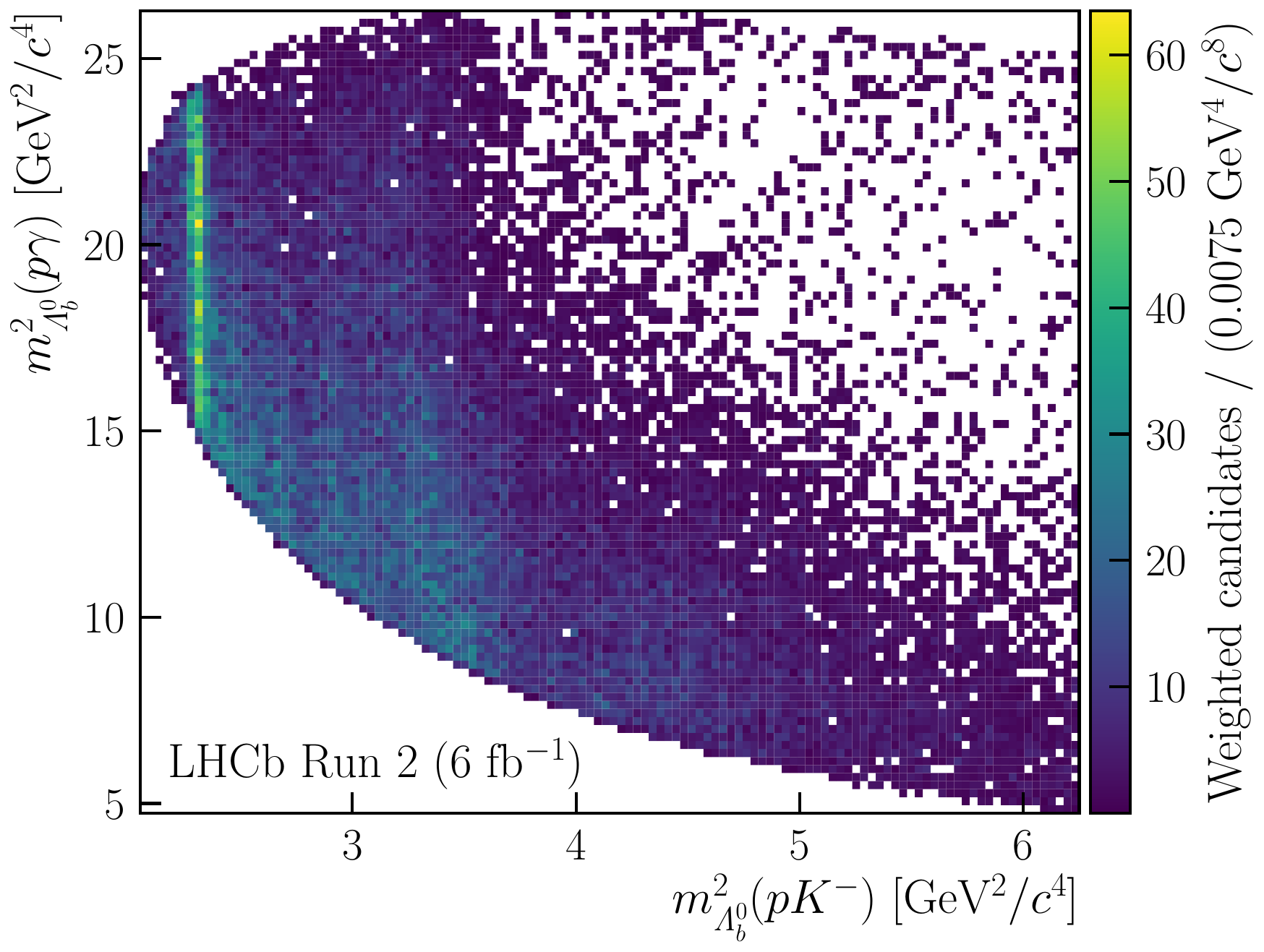}
    \caption{Distribution of the \LbpKg candidates in the Dalitz plane, defined by $m^2_\Lb(\pK)$ and $m^2_\Lb(p\gamma)$, after background-subtraction using the \sPlot method for (left) Run 1 and (right) Run 2.}
    \label{fig:data_dalitz}
\end{figure}

The observed width of the \Lb mass peak is large compared to the width reconstructed using, for example, \LbpKJpsi decays~\cite{LHCb-PAPER-2015-029}.
This is a consequence of the large uncertainty in the photon momentum reconstruction, which is based on the \ecal cluster providing only limited directional information.
Repeating the vertex fit while fixing the invariant mass of the \Lb candidate to the known \Lb mass value~\cite{PDG2022} reduces the uncertainty in the photon momentum for correctly identified \LbpKg candidates given the excellent precision of the reconstructed proton and kaon momenta~\cite{Hulsbergen:2005pu}.
The background-subtracted data in the Dalitz plane are shown in Fig.~\ref{fig:data_dalitz}.
The two-body invariant masses displayed here, and used in the amplitude fit later, are calculated using the \Lb mass constraint as indicated by the \Lb subscript.

As a cross-check for the combination of data within a single run period, the fit to the three-body invariant mass is performed on the full data set and the data set of each year individually.
No significant discrepancies between the fit results are observed.
In order to validate the \sPlot technique, fits to the three-body invariant mass are also performed in bins of the proton-kaon and the proton-photon invariant masses; these fits yield compatible results.

\section{Amplitude model}
\label{sec:model}
The structures in the data shown in Fig.~\ref{fig:data_dalitz} are described using an amplitude model following the prescription of Ref.~\cite{Albrecht:2020azd}.
The intermediate $\Lz$ resonances decaying to $p\Km$ are modelled assuming Breit--Wigner lineshapes, while their spin-dependent angular distributions are described by the helicity formalism.

The three-body decay of a particle with non-zero spin results in five independent phase-space dimensions.
Given that the \Lb baryons observed by \lhcb are produced unpolarised~\cite{LHCb-PAPER-2020-005}, the dimensionality of the phase space relevant to this analysis is reduced from five to two~\cite{Albrecht:2020azd}.
In the following, the phase-space position is denoted $\mathcal{D}$.
This position can be expressed in terms of the Dalitz variables~\cite{Dalitz:1953cp} as shown in Fig.~\ref{fig:data_dalitz} for the background-subtracted data.
Equivalently, the phase-space position can be given by the proton-kaon invariant mass, $m(p\Km)$ and the cosine of the proton helicity angle, $\cos\theta_p$, as is used in Ref.~\cite{Aubert:2005sk}.
The helicity angle of the proton, $\theta_p$, is the polar angle of the proton momentum in the proton-kaon rest frame where the $z$ axis coincides with the \Lz resonance polarisation axis.
This angle can be calculated using two steps.
First, the proton and resonance momentum are boosted into the \Lb rest frame where the coordinate system is defined such that the resonance momentum direction coincides with the $z$ axis.
Second, the proton momentum is boosted into the proton-kaon rest frame.
The magnitude of the $z$ component of the obtained proton momentum, $\vec{p}$, defines the cosine of the proton helicity angle
\begin{equation}
    \cos\theta_p = \frac{p_z}{\mid \vec{p} \mid} \ .
\end{equation}

The amplitude of the decay chain $\Lb\to\Lz(\to p\Km)\g$ with resonance spin $\jL$ and particle helicities denoted by $\lambda_i$ is
\begin{align}\label{eq:helamp}
	\mathcal{A}^\Lz_{\lg,\lL,\lp} = d^{\jL}_{\lL\lp}(\theta_p) H_{\lL,\lg}^{\Lz} h_{\lp}^{\Lz} X_{\jL}\left(m(p\Km)\right) \ .
\end{align}
The function $X_{\jL}\left(m(p\Km)\right)$ represents the resonance dynamics.
The Wigner $d$-matrix elements, $d^{\jL}_{\lL\lp}(\theta_p)$~\cite{Wigner:1931}, describe the rotation of spin states from the $\Lz$ helicity frame into the proton helicity frame.
The helicity-coupling amplitudes $H$ and $h$ contain the information about the dynamics of the decays \LbLg and \LpK, respectively.
Given that the kaon has spin-0, its helicity is also zero and is omitted in the index of the \LpK helicity amplitude, $h$.

Helicity conservation, defined as $\lb=\lL-\lg$, must be fulfilled.
As a result, the resonance helicities can only take the values $\lL=\pm\tfrac{1}{2}$ for $\jL=\pm\tfrac{1}{2}$ and \mbox{$\lL=\pm\tfrac{1}{2},\pm\tfrac{3}{2}$} for $\jL\geq\tfrac{3}{2}$.
Moreover, the resonance and photon helicities must have the same sign.
Subsequently, there are two (four) helicity couplings for each resonance with \mbox{spin-$\tfrac{1}{2}$} ($\geq\tfrac{3}{2}$).
Standard parametrisations of resonance dynamics depend on the orbital angular momentum between the children in a decay requiring a transformation of Eq.~\eqref{eq:helamp} from the helicity to the canonical basis
\begin{align}\label{eq:ls}
\begin{split}
\mathcal{A}^\Lz_{\lg,\lL,\lp} =
d^{\jL}_{\lL\lp}(\theta_p) &\sum_{l=|\jL-s|}^{|\jL+s|}\sum_{s=|J_p-J_K|}^{|J_p+J_K|}
C^\Lz_{ls} h^{\Lz}_{ls} \\
\times&\sum_{L=|\jb-S|}^{|\jb+S|}\sum_{S=|\jL-\jg|}^{|\jL+\jg|}
C^\Lz_{LS} H^{\Lz}_{LS} X_{Ll}^\Lz\left(m(p\Km\right)) \ ,
\end{split}
\end{align}
where the angular dependence remains unchanged.
This transformation couples the spins of the child particles in a decay to a total spin which is then coupled with their orbital angular momentum.
The factors $C^\Lz_{LS}$ and $C^\Lz_{ls}$ are the products of the Clebsch-Gordan coefficients required in the spin-spin and spin-orbital-angular-momentum coupling for the resonance-photon and proton-kaon systems, respectively.
In the resonance-photon system, the total spin, $S$, and orbital angular momentum, $L$, can take different values such that the product of the Clebsch-Gordan coefficients is
\begin{equation}
    C^\Lz_{LS} = \sqrt{\frac{2L+1}{2\jb+1}} \langle \jL,\lL;\jg,-\lg|S,(\lL-\lg) \rangle \cdot \langle L,0;S,(\lL-\lg)|\jb,\lb\rangle \ .
\end{equation}
The total spin of the $p\Km$ system is $s=\tfrac{1}{2}$, as the kaon carries no spin.
The orbital angular momentum between the proton and the kaon, $l$, is fixed for a given spin-parity combination due to angular momentum and parity conservation in the strong decay $\LpK$.
The corresponding Clebsch-Gordan coefficients in the proton-kaon system are
\begin{equation}
    C^\Lz_{ls} = 1 \cdot \langle l,0;J_p,\lambda_p|\jL,\lp\rangle \ .
\end{equation}
Hence, the summation over the spin and orbital angular momentum of the $p\Km$ system can be dropped and only one coupling $h^{\Lz}_{ls}$ remains and is absorbed into the $H_{LS}^{\Lz}$ couplings:
\begin{equation}
    A_{LS}^\Lz = H^{\Lz}_{LS}h^{\Lz}_{ls} \ .
\end{equation}

A standard parametrisation of resonance dynamics as employed in previous amplitude analyses (for example in Refs. \cite{LHCb-PAPER-2015-029,LHCb-PAPER-2016-061}) is used:
\begin{align}\label{eq:lineshape}
X_{Ll}^\Lz(m) = \underbrace{\left(\frac{|\vec{q}|}{q_0}\right)^LB_L(|\vec{q}|,q_0)}_{\LbLg}\underbrace{\left(\frac{|\vec{p}|}{p_0}\right)^lB_l(|\vec{p}|,p_0)}_{\LpK}\text{BW}(m) \ ,
\end{align}
where $\vec{q}$ ($\vec{p}$) is the momentum of the resonance (proton) in the \Lb ($\Lz$) rest frame and $B_l$ and $B_L$ are Blatt--Weisskopf form factors~\cite{Blatt:1952ije}.
Accordingly, the magnitudes of the momenta at the nominal resonance mass are $q_0$ and $p_0$.
The resonance is modelled using a Breit--Wigner (BW) distribution~\cite{PhysRev.49.519}
\begin{align}\label{eq:BW}
\text{BW}(m) = \frac{1}{m_0^2-m^2-im_0\Gamma(m)} \ ,\quad \Gamma(m) = \Gamma_0\left(\frac{p}{p_0}\right)^{2l+1}\frac{m_0}{m}\left[B_l(p,p_0)\right]^2 \ ,
\end{align}
with resonance mass $m_0$ and width $\Gamma_0$.
For the $\Lz(1405)$ resonance, with a pole-mass below the \pK threshold, a similar approach as the amplitude analyses of \mbox{\LbpKJpsi}~\cite{LHCb-PAPER-2015-029} and $\Lc\to\proton\Km\pip$~\cite{LHCb-PAPER-2022-002} is employed, \ie using a two-component width equivalent to the Flatt\'e parametrisation \cite{Flatte:438869}.
The barrier factors, $(|\vec{q}|/q_0)^L$ and $(|\vec{p}|/p_0)^l$, suppress high orbital angular momenta compared to low ones, which will be exploited to simplify the model later on.
The Blatt--Weisskopf form factors are equal to unity at the resonance pole and shape the resonance peak depending on the orbital angular momentum.
This analysis uses the same parametrisation of the Blatt--Weisskopf functions as Ref.~\cite{LHCb-PAPER-2015-029}.
Following the choice made in Ref.~\cite{LHCb-PAPER-2016-061}, the radius of the \Lb baryon is taken to be $5~(\!\gev/(c \hbar))^{-1}$ and the radius of the \Lz resonances is taken to be $1.5~(\!\gev/(c\hbar))^{-1}$.

The final decay rate is the sum over all appearing $\Lz$ resonances and their possible helicities, $\lL$, as well as the initial and final state helicities, $\lb,\lg,\lp$
\begin{align}\label{eq:decrate}
\begin{split}
	\frac{\deriv\Gamma}{\deriv\mathcal{D}} = \frac{1}{2}\sum_{\lb,\lg,\lp}&\left|\sum_{\Lz}\sum_{\lL}d^{\jL}_{\lL\lp}(\theta_p) C^\Lz_{ls}
    \sum_{L=|\jb-S|}^{|\jb+S|}\sum_{S=|\jL-\jg|}^{|\jL+\jg|}C^\Lz_{LS} A^{\Lz}_{LS} X_{Ll}^\Lz\left(m(p\Km)\right)\right|^2 \ .
\end{split}
\end{align}
The decay is assumed to be \CP-conserving such that the amplitudes of the decay \mbox{$\Lb\to p\Km\g$} and \mbox{$\Lbbar\to\antiproton\Kp\g$} have the same helicity couplings.
As a consequence of isospin suppression, investigated experimentally in Ref.~\cite{LHCb-PAPER-2019-039} and theoretically in Ref.~\cite{Dery:2020lbc}, the \LbpKg decay is dominated by the $\Lz$ states and therefore $\PSigma$ resonances, which have the same quark content but different isospin, are not considered in this analysis.
Additionally, resonances in the proton-photon and kaon-photon invariant masses are not included as they almost exclusively populate the region at $m(pK)>2.5$\gevcc.

Besides resonances, additional nonresonant components may be necessary to achieve a satisfactory description of the data.
Such nonresonant contributions are modelled similarly to the resonances, where the Breit--Wigner peak is replaced by an exponential function or a constant
\begin{align}\label{eq:lineshape_nr}
\begin{split}
	X_{Ll}^\text{NR,exp}(m) &= \underbrace{\left(\frac{|\vec{q}|}{q_0}\right)^L\left(\frac{|\vec{p}|}{p_0}\right)^l}_\text{barrier factors}\exp\left(-\alpha (m^2-m_\text{NR}^2)\right) \, , \\
	X_{Ll}^\text{NR,const}(m) &= \underbrace{\left(\frac{|\vec{q}|}{q_0}\right)^L\left(\frac{|\vec{p}|}{p_0}\right)^l}_\text{barrier factors} \, .
\end{split}
\end{align}
The mass parameter used in the computation of $p_0$ and $q_0$ of the nonresonant component is set to the centre of the possible proton-kaon invariant-mass range: $m_\text{NR}=3.5\gevcc$.
The parameter $\alpha$ is determined by the fit.
To incorporate this into the decay rate, the sum over all resonances in Eq.~\eqref{eq:decrate} needs to include the nonresonant component.
The corresponding coherent sum over the helicity states resembles the sum of a resonant contribution where only the lineshape in Eq.~\eqref{eq:lineshape} is replaced by the one in Eq.~\eqref{eq:lineshape_nr}.

Finally, the transformation into the $LS$ basis must conserve the number of degrees of freedom (two (four) helicity couplings for each $\Lz$ with spin $\tfrac{1}{2}$ ($\geq\tfrac{3}{2}$)).
However, given that the angular momentum coupling is a purely mathematical transformation and lacks physics knowledge such as $\lg\not=0$, there are four (six) $LS$ combinations for spin $\tfrac{1}{2}$ ($\geq\tfrac{3}{2}$) resonances.
Translating $H_{\pm1/2,0}=0$ into a combination of $LS$ couplings is non trivial.
An approximation omitting all dynamical terms in Eq.~\eqref{eq:ls} is obtained by expressing the two couplings with highest $S$ in terms of the other two or four:
\begin{align}\label{eq:photon_constraint}
\begin{split}
\begin{pmatrix}
A_{L_\text{max}, S_\text{max}}^{\Lz}\\
A_{L_\text{max}-1, S_\text{max}}^{\Lz}
\end{pmatrix}
= -
\begin{pmatrix}
C^+_{L_\text{max},S_\text{max}} & C^+_{L_\text{max}-1,S_\text{max}}\\
C^-_{L_\text{max},S_\text{max}} & C^-_{L_\text{max}-1,S_\text{max}}
\end{pmatrix}^{-1}
\sum_{L,S<S_\text{max}} A_{L,S}^{\Lz}
\begin{pmatrix}
C_{LS}^{+}\\
C_{LS}^{-}
\end{pmatrix} \ .
\end{split}
\end{align}
The constants $C_{LS}^\pm$ are the Clebsch-Gordan coefficients $C_{LS}^{\Lz}$ with resonance helicity $\lL=\pm\tfrac{1}{2}$ and photon helicity $\lg=0$.
In the case of a spin-$\tfrac{3}{2}$ resonance for example, there are six $LS$ combinations: $(0,\tfrac{1}{2}),(1,\tfrac{1}{2}),(1,\tfrac{3}{2}),(2,\tfrac{3}{2}),(2,\tfrac{5}{2}),(3,\tfrac{5}{2})$.
The transformation in Eq.~\eqref{eq:photon_constraint} replaces the latter two and ensures that the amplitude vanishes exactly at the nominal mass of the resonance.

Two interesting quantities that can be extracted from the model are the fit fraction, the relative contribution of a single resonance to the determined full amplitude computed by
\begin{equation}
	\text{FF}(n) = \frac{\int_{\mathcal{D}}\left(\frac{\deriv\Gamma(n)}{\deriv\mathcal{D}}\right)\deriv\mathcal{D}}{\int_{\mathcal{D}}\left(\frac{\deriv\Gamma}{\deriv\mathcal{D}}\right)\deriv\mathcal{D}} \ ,
\end{equation}
and the interference fit fraction
\begin{equation}
	\text{IFF}(n,m) = \frac{\int_{\mathcal{D}}\left(\frac{\deriv\Gamma(n,m)}{\deriv\mathcal{D}}\right)\deriv\mathcal{D}}{\int_{\mathcal{D}}\left(\frac{\deriv\Gamma}{\deriv\mathcal{D}}\right)\deriv\mathcal{D}} - \text{FF}(n) - \text{FF}(m) \ .
\end{equation}
Here, $\deriv\Gamma(n)/\deriv\mathcal{D}$ is the decay rate for a single state $n$, \ie where the sum in Eq.~\eqref{eq:decrate} only contains the state $n$.
Similarly, $\deriv\Gamma(n,m)/\deriv\mathcal{D}$ is the decay rate of two states $n,m$, \ie where the sum in Eq.~\eqref{eq:decrate} only contains the states $n$ and $m$.
In contrast, $\deriv\Gamma/\deriv\mathcal{D}$ is the decay rate containing all states of a given model.

\section{Amplitude fit}
A simultaneous, unbinned, maximum-likelihood fit of the amplitude model to the Run 1 and Run 2 data sets determines the $LS$ couplings $A_{LS}$.
The negative logarithm of the likelihood function (NLL) is defined as~\cite{Xie:2009rka}
\begin{align}
\text{NLL} \equiv -\log(\mathcal{L}) = &-\sum_\text{Run 1}\log\left(f_1\left(\mathcal{D}\right)\right)w_s -\sum_\text{Run 2}\log\left(f_2\left(\mathcal{D}\right)\right)w_s \ .
\end{align}
The weights $w_s$ are the \sPlot weights presented in Sec.~\ref{sec:massfit} normalised to the effective sample size~\cite{Langenbruch:2019nwe}.
The probability distribution functions $f_i$ correspond to the normalised rate in Eq.~\eqref{eq:decrate}, multiplied by the efficiency map $\varepsilon_i(\mathcal{D})$ of Run 1 or Run 2:
\begin{equation}
    f_i(\mathcal{D}) = \frac{\varepsilon_i(\mathcal{D})}{I_i}\frac{\deriv\Gamma}{\deriv\mathcal{D}} \ ,
\end{equation}
where the normalisation factor is calculated as
\begin{equation}
    I_i = \int_{\mathcal{D}}\varepsilon_i(\mathcal{D})\frac{\deriv\Gamma}{\deriv\mathcal{D}}\deriv\mathcal{D} \ .
\end{equation}
The efficiency maps, obtained from simulation samples, are implemented as interpolated histograms.
The fit is performed using the \textsc{TensorFlowAnalysis} package~\cite{morris_adam_2018_1415413}.

Table~\ref{tab:lambdas} lists all $\Lz$ resonances whose existence ranges from very likely to certain according to Ref.~\cite{PDG2022}.
Such states are rated three or four stars and are derived from analyses of data sets that include precision differential cross sections and polarisation observables, and are confirmed by independent analyses.
The allowed values of the orbital angular momenta between the proton and the kaon, $l$, and the resonance and the photon, $L$, are given explicitly in the rightmost columns.

\begin{table}
	\centering
	\caption{List of well-established $\Lz$ resonances and their properties as given in Ref.~\cite{PDG2022}.
    $J$ and $P$ are spin and parity of the resonance.
    The mass $m_0$ and width $\Gamma_0$ correspond to the Breit--Wigner parameters and are given in \mevcc and \mev respectively.
    The possible mass and width ranges, $\Delta m_0$ and $\Delta\Gamma_0$, are also given.
    If a measurement of mass and width is available, the uncertainties are given instead of a range.
    The columns $\sigma_{m_0}$ and $\sigma_{\Gamma_0}$ contain the $\sigma$ values used to estimate the systematic uncertainty related to the resonance parameters.
    The rightmost columns contain the allowed values of $l$ and $L$.}
	\label{tab:lambdas}
    \input{tabs/resonances}
\end{table}

The fit parameters are the couplings $A^\Lz_{LS}$, resulting in 45 independent complex variables when including all listed $\Lz$ resonances.
A baseline fit comprising these contributions determines $\Lz(1800)$ as the largest component.
To fix the overall phase and magnitude of the full amplitude, its coupling with lowest $L$ is therefore set to $|A^{1800}_{0,1/2}|=1$ and $\arg(A^{1800}_{0,1/2})=0$.

Due to the complexity of the amplitude model, the NLL function has many local minima.
Depending on the exact combination of initial values, the fit may converge to different minima.
When determining the \textit{best model}, the fit is repeated ten times starting from randomised initial values.
Only the result with the lowest NLL out of these ten is compared to the other models.
This procedure reduces the risk of choosing the wrong best model based on convergence to a local minimum.
While the different minima correspond to different values of the couplings, the values for the fit fractions and interference fit fractions are similar for different minima.
As a result of this instability with respect to the couplings, this analysis treats them as nuisance parameters while the derived fit fractions and interference fit fractions are the observables.

The quality of the fit is determined using a binned $\chi^2$ test comparing the two-dimensional weighted data histogram in ($m_{\Lb}(\pK),\cos\theta_p$) with the fit result.
The latter is obtained by generating a large sample of 6$\times 10^6$ data points --- more than 100 times the combined signal yield --- from the fitted pdf.
Because some kinematic regions are only sparsely populated, the histogram is defined using a non-uniform binning with at least 100 observed signal events in each bin.
Due to the differences in the Run 1 and Run 2 acceptance shapes, this binning is calculated separately for the two subsets.

\begin{figure}
    \centering
    \includegraphics[width=.45\textwidth]{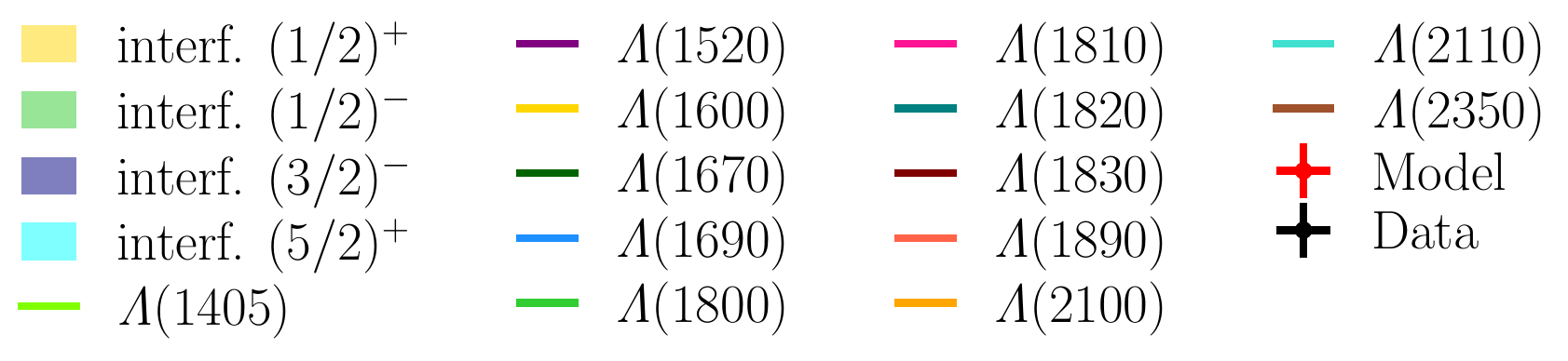} \\
    \includegraphics[width=.45\textwidth]{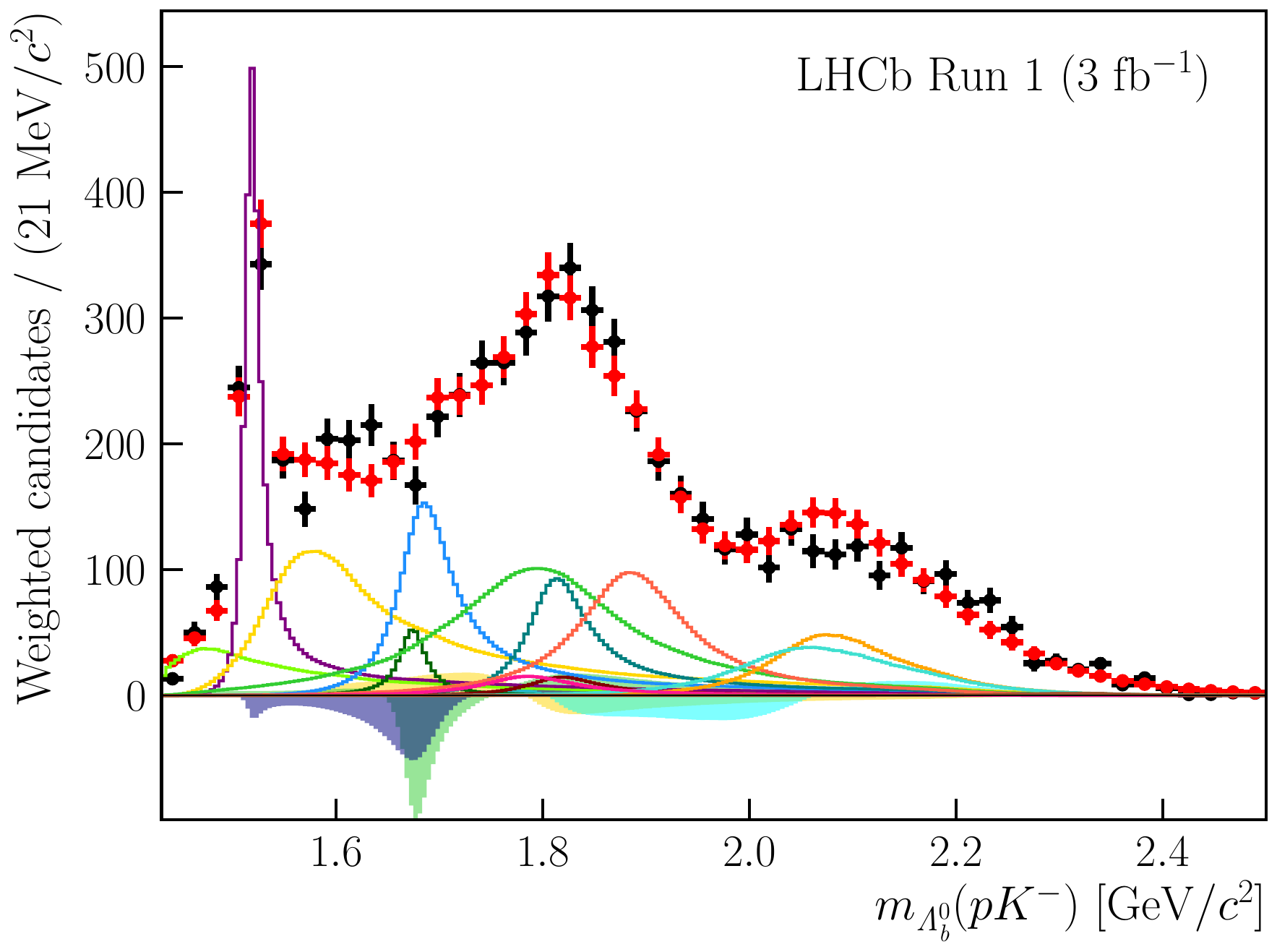}\hspace{.05\textwidth}%
    \includegraphics[width=.45\textwidth]{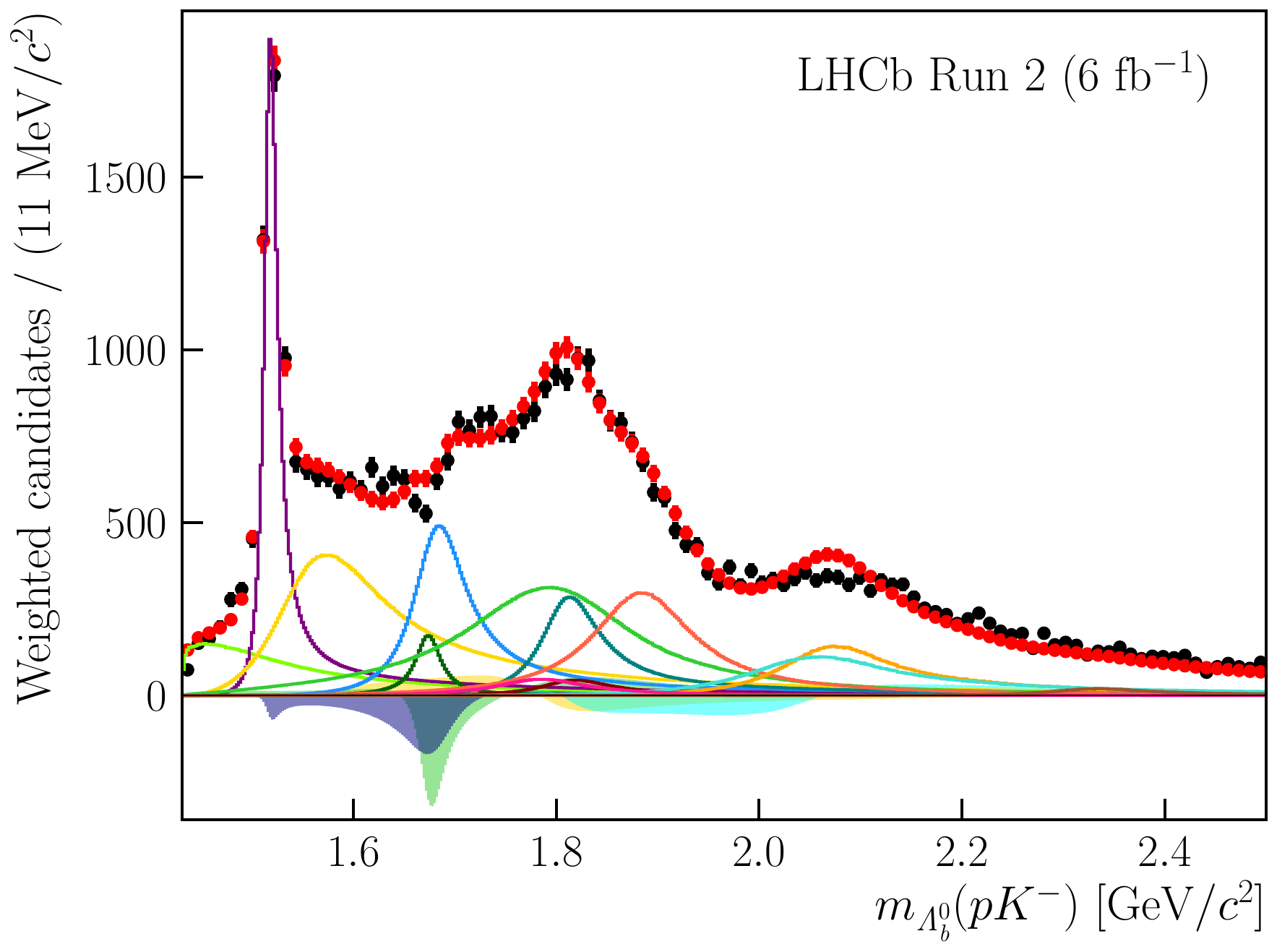}
    \caption{%
        Background-subtracted distribution of the proton-kaon invariant-mass (black dots) for the (left) Run 1 and (right) Run 2 data samples.
        Also shown is a sample generated according to the result of a simultaneous fit of the \textit{reduced model} to the data (red dots) and its components (lines) as well as the contributions due to interference between states with the same quantum numbers $J^P$ (shaded areas).}
    \label{fig:reduced:mpk}
\end{figure}

The initial model contains all well-known $\Lz$ resonances (see Table~\ref{tab:lambdas}) and no other components.
This gives an good description of the major structures in the data.
This model is referred to as the \textit{reduced model}.
The distribution of the proton-kaon invariant-mass in Run 1 and Run 2 is shown in Fig.~\ref{fig:reduced:mpk}.
The projection of the \textit{reduced model} including all its components is overlaid.
While the \textit{reduced model} overall describes the data spectrum well, the model is not satisfactory in the region $m_{\Lb}(\pK)>2\gevcc$.
As the heavy $\Lz$ states are poorly known, the mass and width of different combinations of heavy states are floated with Gaussian constraints of $100\mevcc$ around the values obtained from Ref.~\cite{PDG2022} in order to improve the fit quality.
Allowing the mass and width of the $\Lz(2100)$ and $\Lz(2110)$ states to vary, while keeping those of the $\Lz(2350)$ state fixed, yields the biggest improvement.

Another option to improve the fit quality is the addition of nonresonant contributions.
The nonresonant components can affect the entire region of the phase space, and are especially important in regions where resonances with the matching spin-parity may interfere.
Nonresonant components with spins up to $\tfrac{5}{2}$ and both parities, using an exponential or constant lineshape (see Eq.~\ref{eq:lineshape_nr}), are tested.
Both lineshape functions tested yield very similar results for a given set of quantum numbers and the constant one is taken as the default lineshape.
The model including a nonresonant component with quantum numbers $J^{P} = \tfrac{3}{2}^-$ results in the best fit quality for either lineshape.
The fit quality of this model is better than the fit quality of the \textit{reduced model} with floating resonance masses and widths.

\begin{figure}
    \centering
	\includegraphics[width=.45\textwidth]{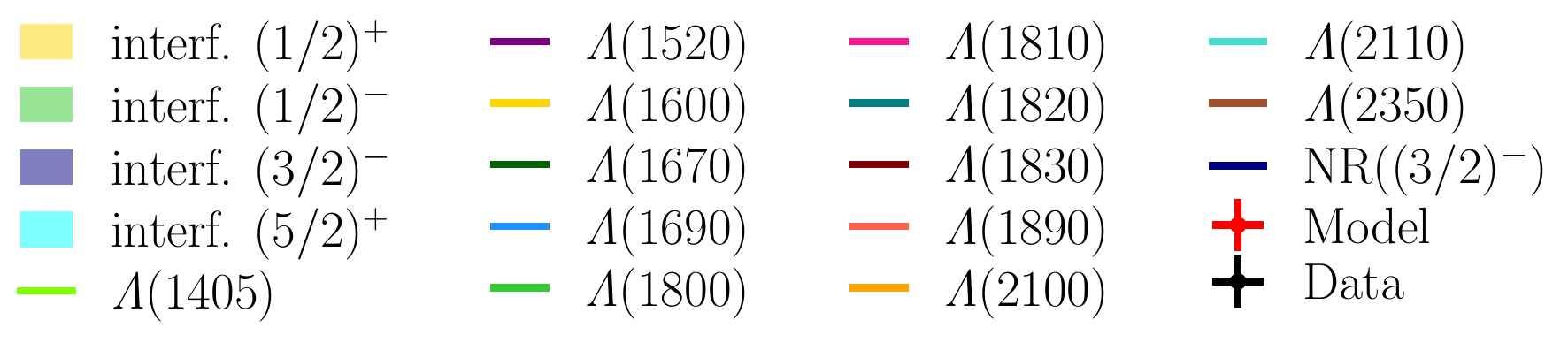} \\
	\includegraphics[width=.45\textwidth]{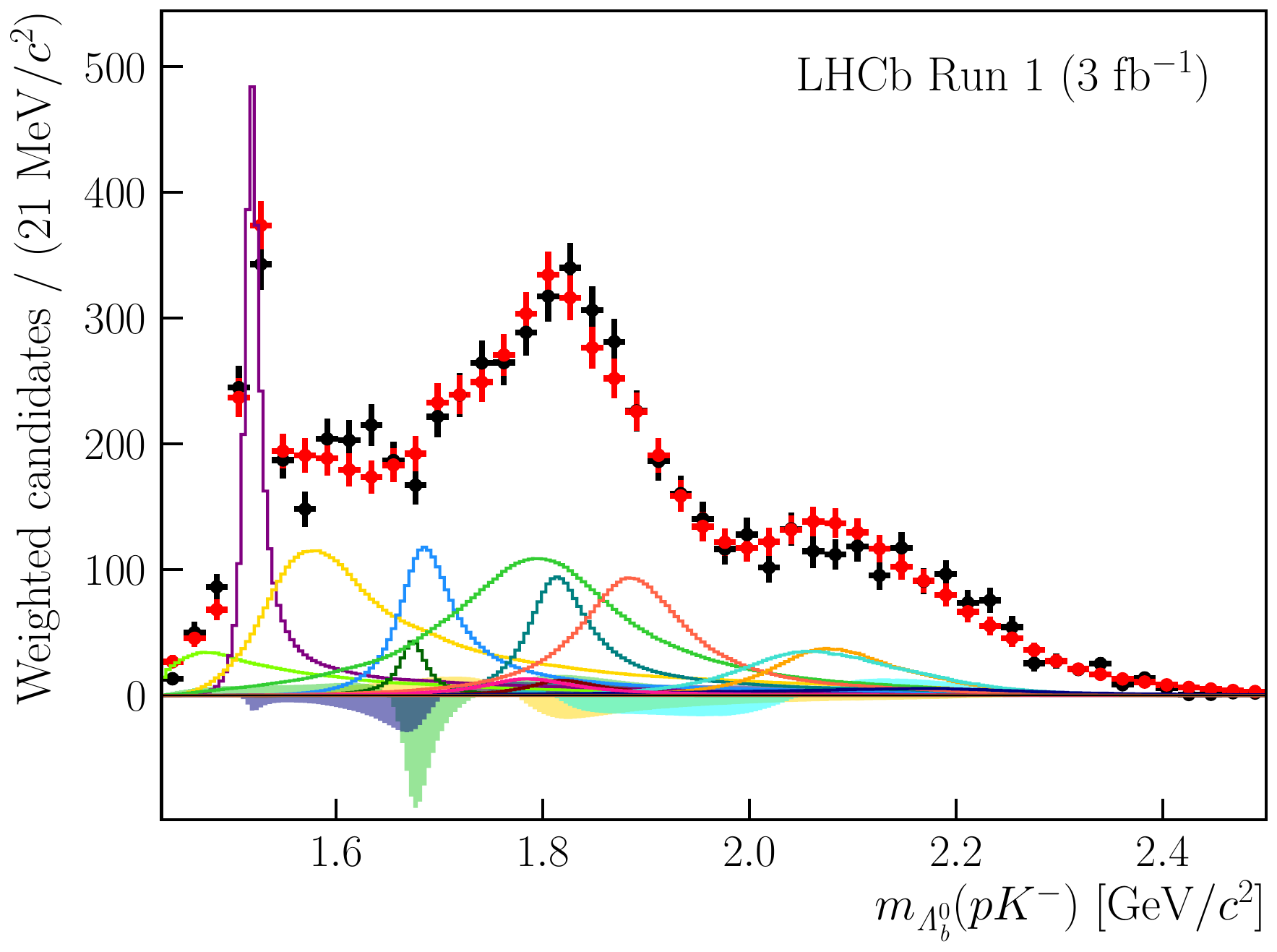}\hspace{.05\textwidth}%
	\includegraphics[width=.45\textwidth]{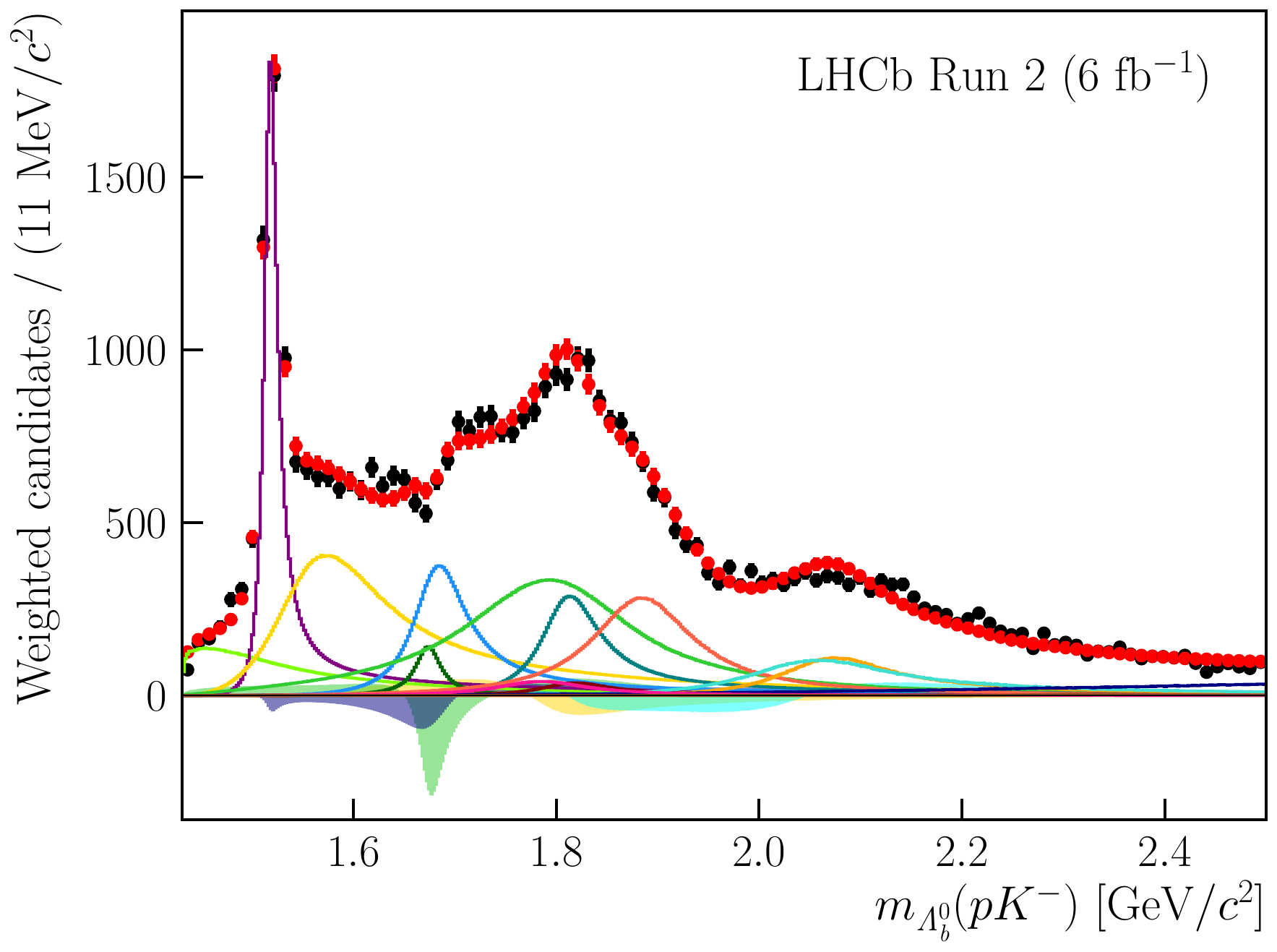}
	\includegraphics[width=.45\textwidth]{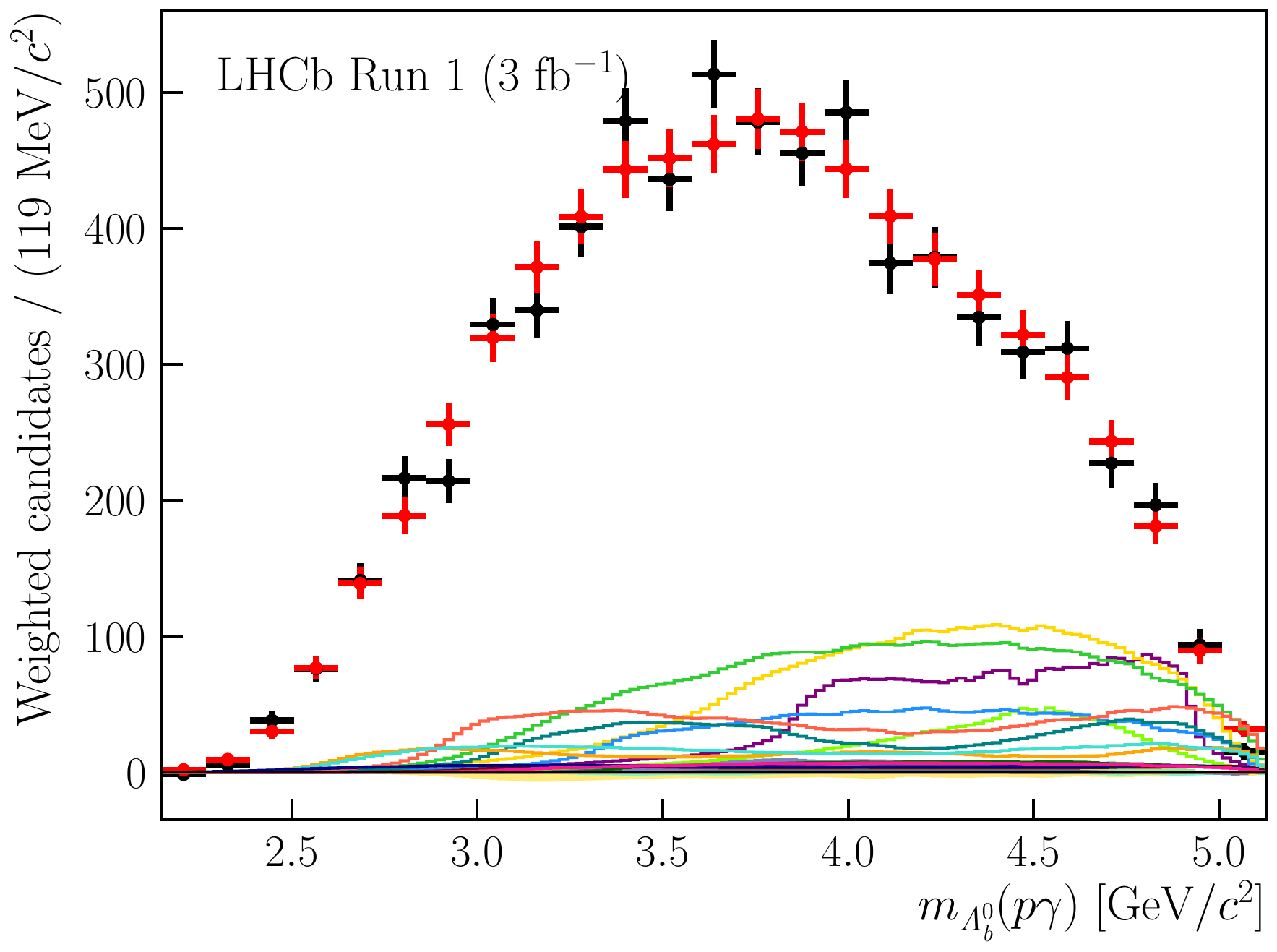}\hspace{.05\textwidth}%
	\includegraphics[width=.45\textwidth]{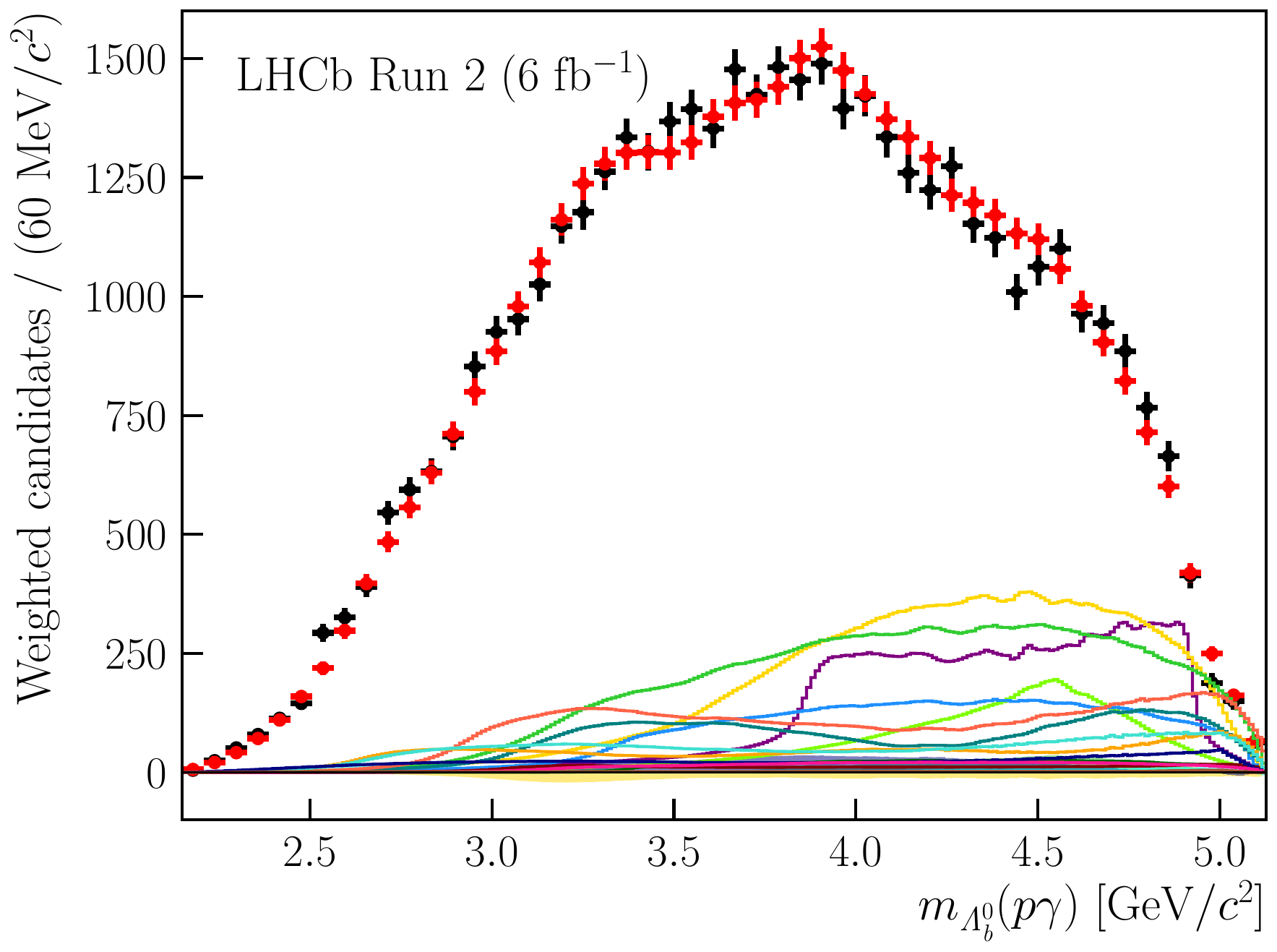}
    \caption{%
        Background-subtracted distribution of the (top) proton-kaon and (bottom) proton-photon invariant-mass (black dots) for the (left) Run 1 and (right) Run 2 data samples.
        Also shown is a sample generated according to the result of a simultaneous fit of the default model to the data (red dots) and its components (lines) as well as the contributions due to interference between states with the same quantum numbers $J^P$ (shaded areas).}
    \label{fig:best-fits}
\end{figure}

\begin{figure}
    \centering
	\includegraphics[width=.45\textwidth]{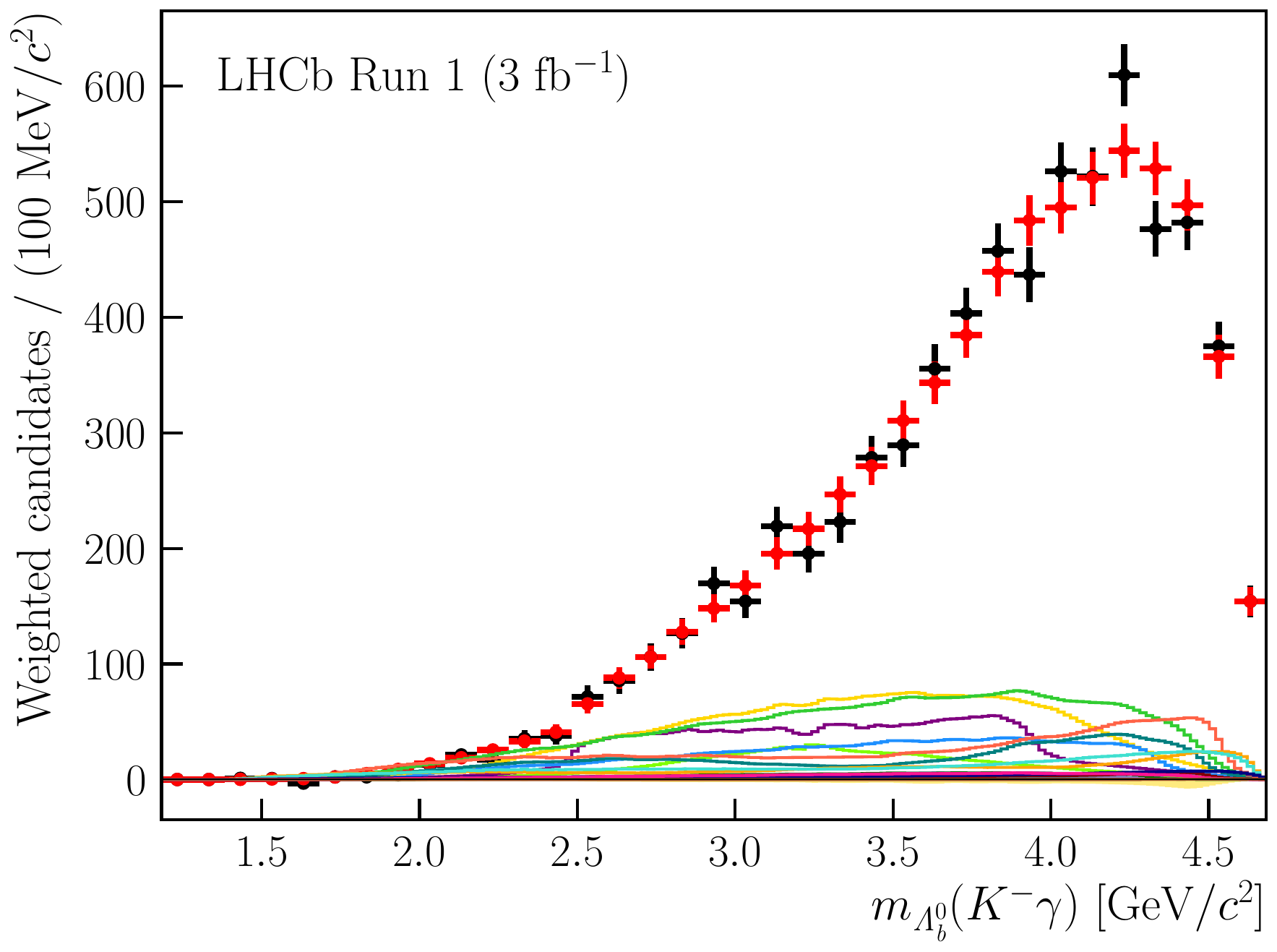}\hspace{.05\textwidth}%
	\includegraphics[width=.45\textwidth]{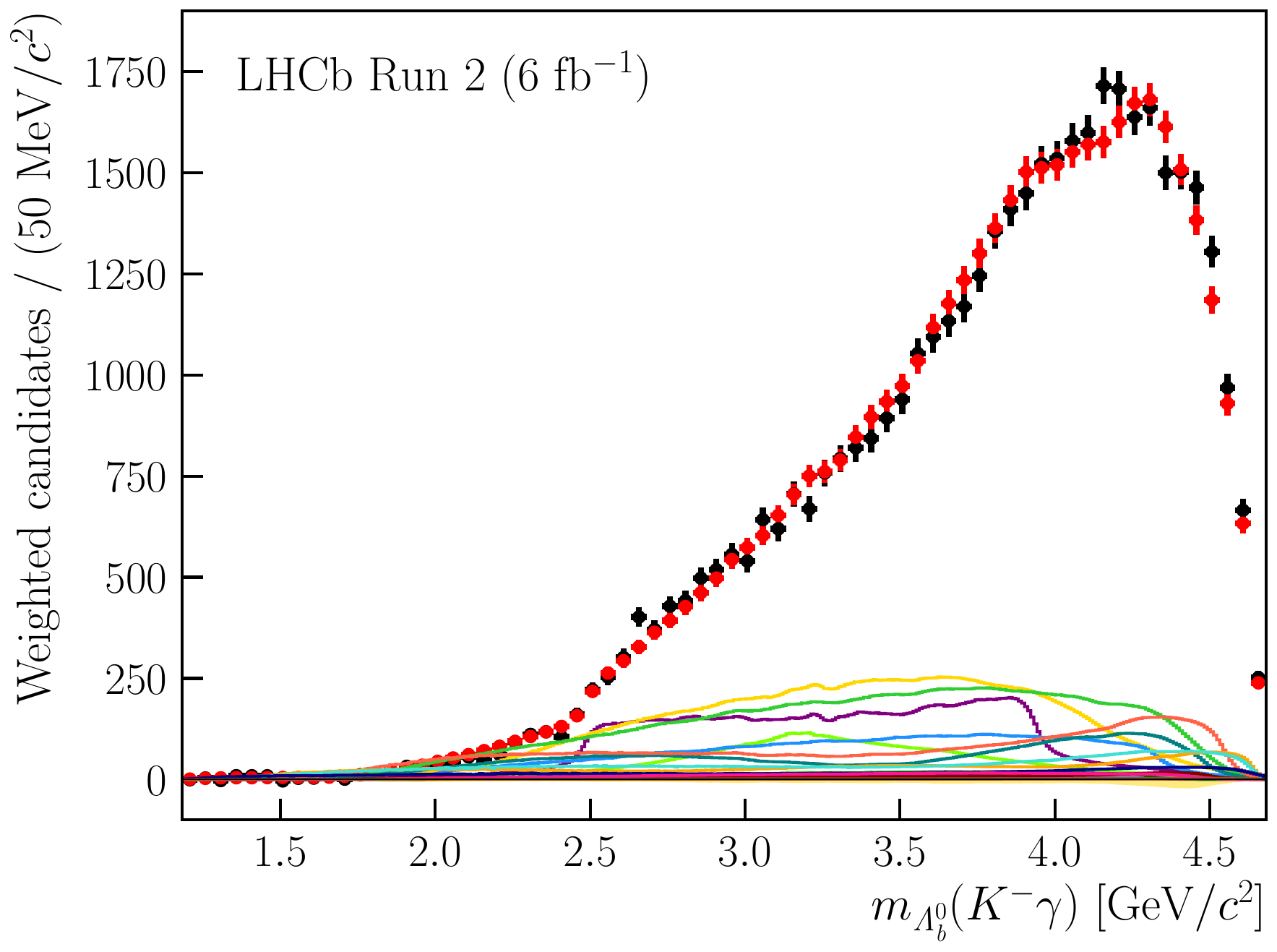}
	\includegraphics[width=.45\textwidth]{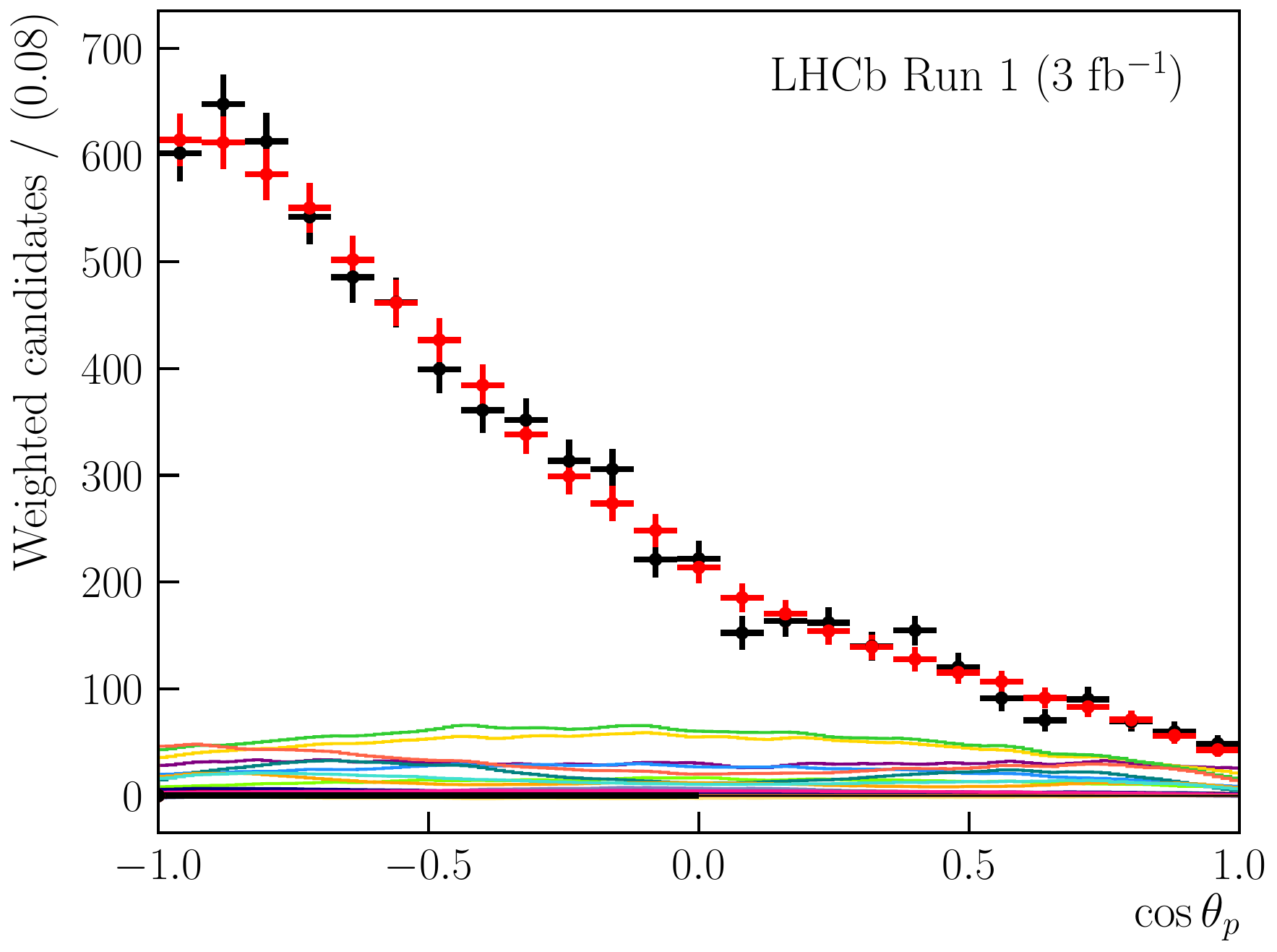}\hspace{.05\textwidth}%
	\includegraphics[width=.45\textwidth]{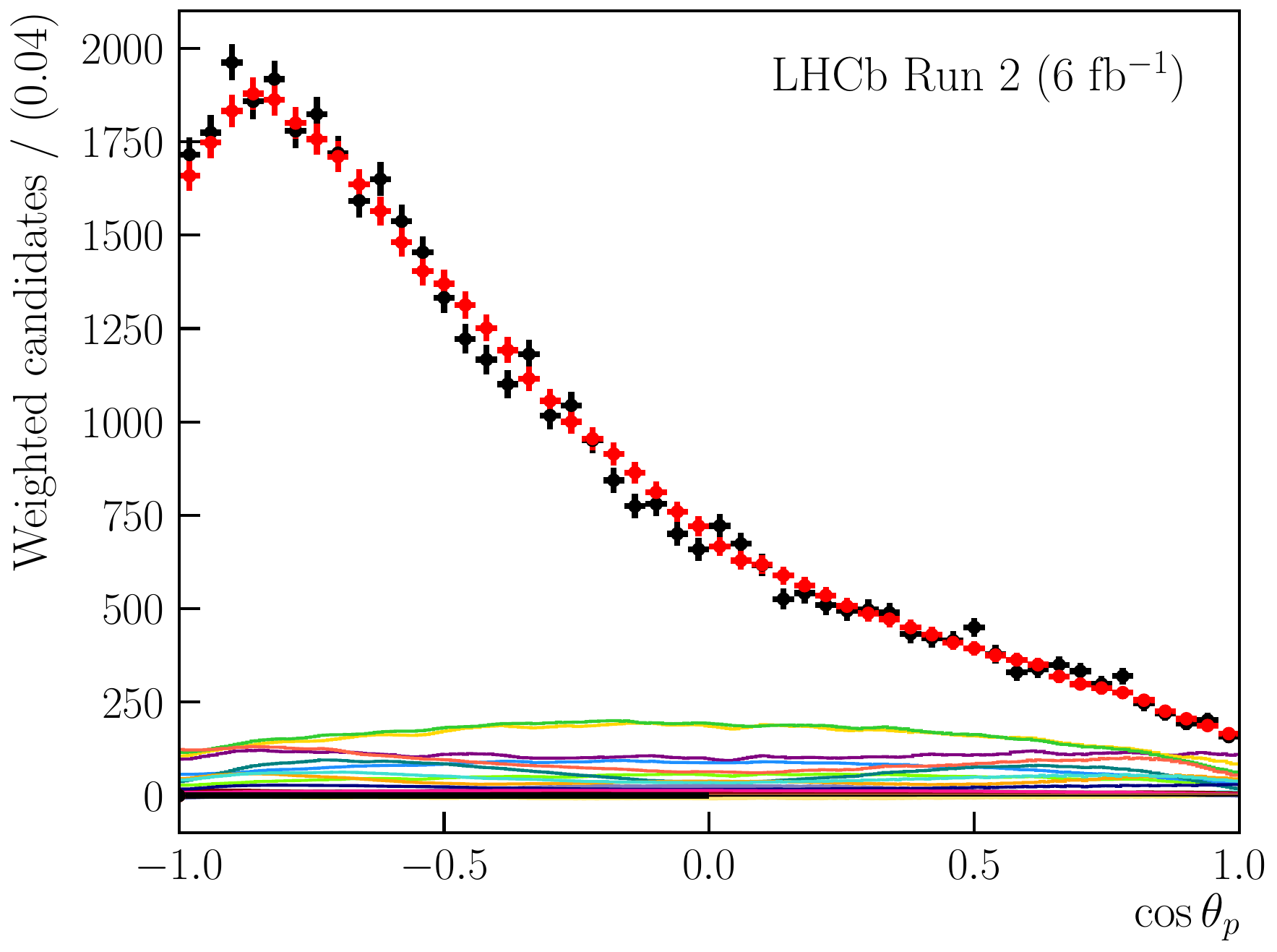}
       \caption{%
        Background-subtracted distribution of (top) the kaon-photon invariant-mass and (bottom) the proton helicity angle (black dots) for the (left) Run 1 and (right) Run 2 data samples.
        Also shown is a sample generated according to the result of a simultaneous fit of the default model to the data (red dots) and its components (lines) as well as the contributions due to interference between states with the same quantum numbers $J^P$ (shaded areas).
        See Fig.~\ref{fig:best-fits} for the legend.}
    \label{fig:best-fits:additional}
\end{figure}

As a result, the \textit{best model} used to determine the default result consists of the \textit{reduced model} containing all $\Lz$ states with mass and width fixed to the values given in Table~\ref{tab:lambdas} and a nonresonant component with quantum numbers $J^P=\tfrac{3}{2}^-$.
Figures~\ref{fig:best-fits} and \ref{fig:best-fits:additional} contain projections of the data and the model with its components onto all two-body invariant masses as well as the proton helicity angle for Run 1 and Run 2.
Appendix~\ref{app:logscale:best} shows the projections onto the proton-kaon invariant mass using a logarithmic vertical axis.
The same set of plots is provided in Appendix~\ref{app:plots:altmodels} for the fit with floating resonances representing the second best model.

The statistical uncertainties on the fit fractions and interference fit fractions are determined by bootstrapping the data 250 times.
This means that the data set is resampled and a new set of \sWeights is calculated from a fit to the three-body invariant mass of each bootstrap sample.
Running the amplitude fit on each sample with its respective \sWeights results in a distribution for each observable.
The value for the statistical confidence interval given later is obtained by finding the shortest 68\% interval around the maximum of this distribution.

\section{Systematic uncertainties}
Systematic uncertainties arise from four major categories: the choice of amplitude model, the acceptance model, the invariant-mass fit model, and potential remaining backgrounds.
The individual uncertainties are listed in Table~\ref{tab:syst} and outlined in the following.

\subsection{Amplitude model}
% fixing mass and width
In the default fit, the masses and widths of the resonances are set to their world averages and fixed in the fit.
To assess the impact of this choice, alternative masses and widths are sampled from Gaussian distributions. The widths of the Gaussians are given in Table~\ref{tab:lambdas} as $\sigma_{m_0}$ and $\sigma_{\Gamma_0}$ and are chosen based on the ranges $\Delta m_0$ and $\Delta \Gamma_0$.
Pseudoexperiments, generated using these alternative mass and width values, are fitted with the default model.
The shortest 68\% interval around the maximum of the distribution of the difference between the generated and fitted values is taken as a systematic uncertainty.

% fixing resonance diameter
Similar to the treatment  of the masses and widths, also the \Lb and \Lz radii used in the Blatt--Weisskopf functions are fixed to $d_\Lb=5~(\gev/(c\hbar))^{-1}$ and \mbox{$d_\Lz=1.5~(\gev/(c\hbar))^{-1}$} in the default fit.
The impact of this choice is assessed by generating samples with \mbox{$d_\Lb=3,5,7~(\!\gev/(c\hbar))^{-1}$} and $d_\Lz=0.5,1.5,2.5,3.5$ and $4.5~(\!\gev/(c\hbar))^{-1}$.
These samples are fitted with the default model.
The bias and standard deviation of the differences between the generated and fitted values for each combination of $d_\Lb$ and $d_\Lz$ is taken as systematic uncertainty.

% This table lives here because it should
% - appear as soon as possible after first mention
% - while not appear before figures that are mentioned earlier
\begin{table}
    \centering
    \caption{Systematic uncertainties on the fit fractions (top part of the table) and interference fit fractions (bottom part of the table).
    The values are given in \%.
    The subscripts ``BW'', ``radius'', ``amp.'', and ``res.'' refer to the systematic uncertainty due to fixing the resonance mass and width, fixing the radius of the hadrons, the choice of amplitude model, and the neglected resolution in the amplitude fit, respectively.
    The subscripts ``finite'', ``acc.'', and ``kin.'' refer to the systematic uncertainties due to the finite simulation sample used to determine the acceptance model, the choice of acceptance model, and the kinematic reweighting, respectively.
    The subscripts ``$pK$'', ``$p\gamma$'', and ``comb.'' refer to the systematic uncertainty due to calculating the \sWeights in bins of the proton-kaon invariant mass, the proton-gamma invariant mass, and the choice of model for the combinatorial background in the three-body invariant mass fit respectively.}
    \label{systematics}
\bgroup
\def\arraystretch{1.3} % this stretches the rows a bit so they don't look as cramped
\input{tabs/systematics.tex}
    \egroup
    \label{tab:syst}
\end{table}

% model choice
Besides the default model, several other models result in a good description of the data.
The systematic effects due to choosing certain components and shapes over others are quantified by generating samples using an alternative model and fitting the default model to the generated pseudosample.
The five alternative models are:
\begin{itemize}
    \item [-] removing the nonresonant component and instead floating mass and width of the $\Lz(2100)$ and $\Lz(2110)$ states using Gaussian constraints (this is the second best model);
    \item  [-] using an exponential function instead of a constant for the lineshape of the nonresonant component;
    \item  [-] employing a sub-threshold Breit--Wigner for the lineshape of the $\Lz(1405)$ state instead of the Flatt\'e shape;
    \item  [-] adding a second nonresonant component with constant lineshape and $J^P=\tfrac{5}{2}^+$;
    \item  [-] adding a second nonresonant component with constant lineshape and $J^P=\tfrac{1}{2}^+$.
\end{itemize}
The systematic uncertainty due to the model choices is calculated based on the mean and spread of the results obtained using the five alternative models.

% resolution
Because the resolution is much smaller than the width of the resonances in all regions of the Dalitz plane, the amplitude model does not include resolution effects in the two-body invariant masses, $m_{\Lb}(pK)$ and $m_{\Lb}(p\gamma)$.
The systematic impact of this choice is tested by generating samples with the default model and smearing the masses according to the resolution determined on simulation samples.
Both the unsmeared and smeared samples are fit with the default model which does not account for the resolution.
The shortest 68\% interval around the maximum of the distribution of the difference between the two results is taken as a systematic uncertainty.

\subsection{Acceptance model}
The acceptance map is created using a simulation sample generated uniformly in phase space.
The finite size of the sample, the choices regarding particle identification and kinematic modelling, as well as the number of bins in the acceptance histogram are varied individually to assess their impact on the observables.
The fit to data is repeated with each alternative acceptance map and the difference between the default and alternative result calculated.
The spread of the differences for the acceptance-related systematic effects, namely the finite sample size, the acceptance model, and the kinematic weights are taken as individual systematic uncertainties.
The uncertainty associated with the particle identification weights is found to be negligible.

\subsection{Mass fit model}
In order to quantify systematic effects due to choices in the fit to the \Lb invariant mass, the analysis is performed for each of these alternatives:
\begin{itemize}
    \item [-] modelling the combinatorial background using a polynomial instead of an exponential function;
    \item [-] modelling the partially reconstructed background using an Argus function~\cite{Albrecht:1990am} instead of a kernel density estimator obtained from simulation samples;
    \item [-] letting the signal tail parameters vary in the fit to data using a Gaussian constraint instead of fixing them;
    \item [-] calculating the \sWeights in bins of $m_\Lb(p\Km)$ and $m_\Lb(p\gamma)$ to account for possible correlations between the Dalitz variables and the three-body invariant mass.
\end{itemize}
Only changing the shape of the combinatorial background and calculating the \sWeights in bins of the two-body invariant masses results in a difference with respect to the default result; this difference is added as a systematic uncertainty.

\subsection{Additional background contamination}
After the selection, a small number of candidates from misidentified \mbox{\Dz\to $\Kp\Km$} and \mbox{\Dz\to $\Kp\pim$} decays combined with a random photon remain in the data sample.
In the three-body invariant mass, they are predominantly located below the \Lb mass peak.
The full analysis chain is repeated vetoing both \Dz decays in order to determine the systematic effect of this choice and no difference is observed.

The contamination due to misidentified $\Bs\to \Kp\Km\gamma$ and $\Bz\to \Kp\pim\gamma$ decays is estimated to not exceed 0.5\% of the signal yield.
The resulting structures are wide and spread across large parts of the phase space.
Nevertheless, the two backgrounds are included in the mass fit constraining their yield to 0.5\% of the signal yield.
The amplitude fit is repeated using the obtained alternative set of \sWeights.
No difference to the default result is observed.

\subsection{Systematic uncertainty combinations}
Table~\ref{systematics} contains all individual systematic uncertainties considered for the final result.
Sources of systematic uncertainty found to have no impact on the default values are neglected: the limited simulation sample size and the particle identification weights used to determine the acceptance model, the shape of the signal and combinatorial background in the fit to the three-body invariant mass, and the consideration of additional misidentified \Bs and \Bz backgrounds in the mass fit as well as vetoing misidentified backgrounds from \Dz decays.
All acceptance-related systematic uncertainties are assumed to be Gaussian and centred around the default value.
All mass fit systematic uncertainties are also assumed to be Gaussian and centred around the default value, however only allowing values on either the positive or negative side as the nature of these systematic effects is a one-sided bias instead of a double-sided uncertainty.
The systematic uncertainty related to the amplitude model is considered to be Gaussian but biased with respect to the default value.
Similarly to the statistical uncertainty, the systematic uncertainty due to fixing the resonance parameters and neglecting the resolution are neither centred around the default value nor symmetric.

As a consequence of asymmetric non-Gaussian behaviour of these distributions, the uncertainties can not be combined by taking the square root of the sum of the individual uncertainties squared.
Instead, they are combined by numerically convolving the distributions and taking the shortest 68\% interval around the maximum of the resulting distribution as the combined uncertainty interval.
The uncertainty due to the poorly known $\Lz$ resonance parameters dominates the combination.
In order to differentiate between the uncertainty due to this external input and the systematic uncertainty related to the analysis choices, the combination is performed once for all sources of systematic uncertainty and once for all but $\sigma^\Lz_\text{BW}$ and $\sigma_\text{radius}$.
The two external uncertainties $\sigma^\Lz_\text{BW}$ and $\sigma^\Lz_\text{radius}$ are combined to give $\sigma_\text{syst}^\text{external}$.

\section{Results and conclusion}
\begin{figure}
    \centering
    \includegraphics[width=.8\textwidth]{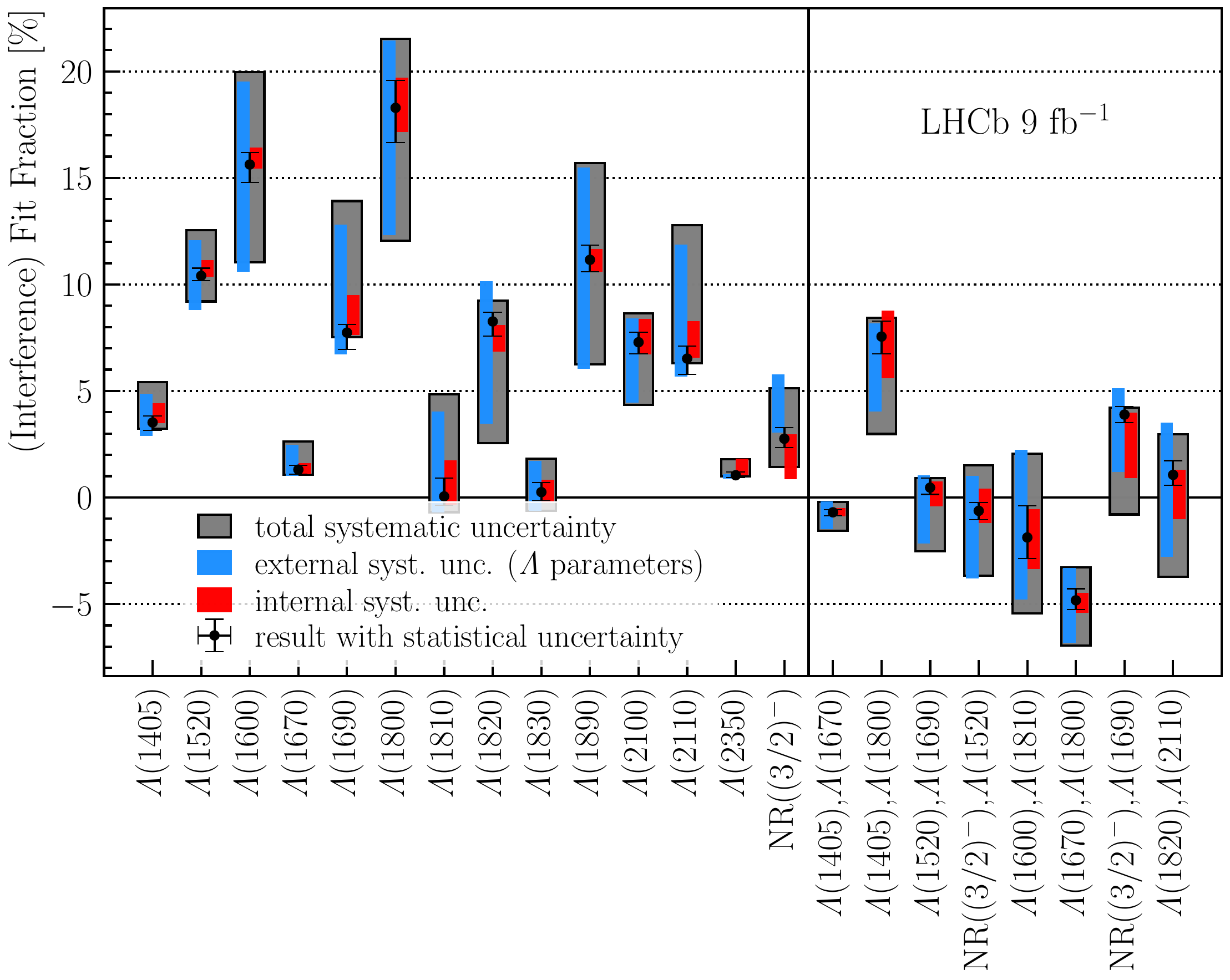}
    \caption{Final results for the fit fractions and interference fit fractions.
    The vertical line separates the fit from the interference fit fractions.
    The error bars represent the different sources of uncertainty.}
    \label{fig:finalresult}
\end{figure}

The results of this analysis, including statistical and systematic uncertainties, are presented in Fig.~\ref{fig:finalresult} and  Table~\ref{tab:finalresult}.
The statistical correlations between the observables are given in Appendix~\ref{app:correlations}.
The data and model projections on the invariant masses and proton helicity angle are shown in Figs.~\ref{fig:best-fits} and \ref{fig:best-fits:additional}.
The largest resonant contributions to the \LbpKg decay are found to arise from the $\Lz(1800)$, $\Lz(1600)$, $\Lz(1890)$ and $\Lz(1520)$ states, in decreasing order.
The largest interference term involves the $\Lz(1405)$ and $\Lz(1800)$ baryons.

The uncertainties for most observables are dominated by external inputs, specifically the masses and widths of the $\Lz$ states.
A future measurement including improved knowledge of the different $\Lz$ baryons and more data will result in a significant reduction of the uncertainties.

The analysis of \LbpKg decays provides information about the composition of the $\pK$ spectrum with unique access to the heavier $\Lz$ states.
A comparison between the composition of the spectrum in \LbpKg and \LbpKJpsi decays, see Ref.~\cite{LHCb-PAPER-2015-029}, is complicated due to the different extent of the phase space and the prominent pentaquark contributions in the latter.
Three notable differences are explained in the following.
First, the contribution of the sub-threshold resonance $\Lz(1405)$ is much smaller in the radiative mode.
Second, the $\Lz(1810)$ state appears small in decays to a photon but large in the \jpsi case; the neighbouring $\Lz(1820)$ state behaves in the opposite way.
This observation reveals a potential ambiguity between the two resonances also echoed in the systematic uncertainties on their fit fractions presented in this paper.
Third, the heavy resonances $\Lz(1890)$, $\Lz(2100)$, $\Lz(2110)$, and $\Lz(2350)$ are much larger in the radiative case, which is in part due to the phase space enhancement.

In conclusion, an amplitude analysis of the decay \LbpKg is presented for the first time, based on the helicity formalism.
A sample of around 50~000  signal candidates is selected from proton-proton collisions recorded by the LHCb experiment at centre-of-mass energies of 7, 8 and 13 TeV.
The default fit model comprises all known \Lz resonances as well as a nonresonant contribution with quantum numbers $J^P = \tfrac{3}{2}^-$.
The presented amplitude model provides a detailed description of the \LbpKg decay with possible applications ranging from searches for beyond the Standard Model physics in \LbpKll decays to QCD studies and a possible measurement of the photon polarisation in \LbpKg decays using polarised \Lb baryons from $Z$ decays at future $e^+e^-$ colliders.

\begin{table}[h!]
    \centering
    \caption{Fit fractions (top) and interference fit fractions (bottom) determined using the amplitude model.
    The values are given in \%.
    The uncertainties from internal and external sources, determined by the numerical convolution procedure are labelled $\sigma_\text{syst}^\text{internal}$ and $\sigma_\text{syst}^\text{external}$.}
    \label{tab:finalresult}
\bgroup
\def\arraystretch{1.5} % this stretches the rows a bit so they don't look as cramped
\input{tabs/final_table}
\egroup
\end{table}

%% file: tabs/resonances.tex
\begin{tabular}{c|ccr|cc|rr|cc}
    \toprule
    Resonance & $J^P$ & $m_0$ & $\Gamma_0$\hphantom{.5} & $\Delta m_0$ & $\Delta\Gamma_0$ & $\sigma_{m_0}$ & $\sigma_{\Gamma_0}$ & $l$ & $L$ \\
    \midrule
    $\Lz(1405)$ & $\nicefrac{1}{2}^-$ & 1405  &  50.5 & $\pm1.3$    & $\pm2$  & 1.3 & 2 & 0 & 0, 1 \\
    $\Lz(1520)$ & $\nicefrac{3}{2}^-$ & 1519  &  16\hphantom{.5} & 1518 -- 1520    & \hphantom{1}15 -- \hphantom{1}17  & 1\hphantom{.3} & 1 & 2& 0, 1, 2 \\
    $\Lz(1600)$ & $\nicefrac{1}{2}^+$ & 1600  & 200\hphantom{.5} & 1570 -- 1630 & 150 -- 250  & 30\hphantom{.3} & 50 & 1 & 0, 1 \\
    $\Lz(1670)$ & $\nicefrac{1}{2}^-$ & 1674  &  30\hphantom{.5} & 1670 -- 1678 & \hphantom{1}25 -- \hphantom{1}35   & 4\hphantom{.3} & 5 & 0 & 0, 1 \\
    $\Lz(1690)$ & $\nicefrac{3}{2}^-$ & 1690  &  70\hphantom{.5} & 1685 -- 1695 & \hphantom{1}50 -- \hphantom{1}70   & 5\hphantom{.3} & 10 & 2& 0, 1, 2 \\
    $\Lz(1800)$ & $\nicefrac{1}{2}^-$ & 1800  & 200\hphantom{.5} & 1750 -- 1850 & 150 -- 250 & 50\hphantom{.3} & 50 & 0 & 0, 1 \\
    $\Lz(1810)$ & $\nicefrac{1}{2}^+$ & 1790  & 110\hphantom{.5} & 1740 -- 1840 & \hphantom{1}50 -- 170  & 50\hphantom{.3} & 60 & 1 & 0, 1 \\
    $\Lz(1820)$ & $\nicefrac{5}{2}^+$ & 1820  &  80\hphantom{.5} & 1815 -- 1825 & \hphantom{1}70 -- \hphantom{1}90   & 5\hphantom{.3} & 10 & 3 & 1, 2, 3 \\
    $\Lz(1830)$ & $\nicefrac{5}{2}^-$ & 1825  &  90\hphantom{.5} & 1820 -- 1830 & \hphantom{1}60 -- 120  & 5\hphantom{.3} & 30 & 2 & 1, 2, 3 \\
    $\Lz(1890)$ & $\nicefrac{3}{2}^+$ & 1890  & 120\hphantom{.5} & 1870 -- 1910 & \hphantom{1}80 -- 160  & 20\hphantom{.3} & 40 & 1 & 0, 1, 2 \\
    $\Lz(2100)$ & $\nicefrac{7}{2}^-$ & 2100  & 200\hphantom{.5} & 2090 -- 2110 & 100 -- 250 & 10\hphantom{.3} & 100 & 4 & 2, 3, 4 \\
    $\Lz(2110)$ & $\nicefrac{5}{2}^+$ & 2090  & 250\hphantom{.5} & 2050 -- 2130 & 200 -- 300 & 40\hphantom{.3} & 50 & 3 & 1, 2, 3 \\
    $\Lz(2350)$ & $\nicefrac{9}{2}^+$ & 2350  & 150\hphantom{.5} & 2340 -- 2370 & 100 -- 250 & 20\hphantom{.3} & 100 & 5 & 3, 4, 5 \\
    \bottomrule
\end{tabular}

%% file: tabs/systematics.tex
\begin{tabular}{c|rrrr|rrr|rrr}
\toprule
    & \multicolumn{4}{c|}{Amplitude model}
    & \multicolumn{3}{c|}{Acceptance model}
    & \multicolumn{3}{c}{Mass fit model} \\
Observable & $\sigma^\Lz_\text{BW}$  & $\sigma^\Lz_\text{radius}$  & $\sigma_\text{amp.}$  & $\sigma_\text{res.}$  & $\sigma_\text{finite}$  & $\sigma_\text{acc.}$  & $\sigma_\text{kin.}$ & $\sigma_{pK}$  & $\sigma_{p\gamma}$  & $\sigma_\text{comb.}$ \\ 
\midrule
$\Lz(1405)$& $_{-0.7}^{+1.2}$& $_{-0.0}^{+0.0}$& $_{+0.2}^{+0.9}$& $_{-0.4}^{+0.0}$& $_{-0.2}^{+0.2}$& $_{-0.2}^{+0.2}$& $_{-0.0}^{+0.0}$& $_{-0.1}^{+0.0}$& $_{-0.0}^{+0.1}$& $_{-0.0}^{+0.0}$ \\ 
$\Lz(1520)$& $_{-1.3}^{+1.0}$& $_{-1.1}^{+1.1}$& $_{+0.0}^{+0.3}$& $_{-0.1}^{+0.0}$& $_{-0.2}^{+0.2}$& $_{-0.2}^{+0.2}$& $_{-0.1}^{+0.1}$& $_{-0.0}^{+0.3}$& $_{-0.0}^{+0.1}$& $_{-0.1}^{+0.0}$ \\ 
$\Lz(1600)$& $_{-4.5}^{+3.6}$& $_{-1.8}^{+1.8}$& $_{+0.0}^{+0.5}$& $_{-0.2}^{+0.3}$& $_{-0.3}^{+0.3}$& $_{-0.2}^{+0.2}$& $_{-0.1}^{+0.1}$& $_{-0.1}^{+0.0}$& $_{-0.0}^{+0.1}$& $_{-0.0}^{+0.0}$ \\ 
$\Lz(1670)$& $_{-0.3}^{+1.1}$& $_{-0.2}^{+0.2}$& $_{-0.2}^{+0.2}$& $_{-0.2}^{+0.2}$& $_{-0.1}^{+0.1}$& $_{-0.0}^{+0.0}$& $_{-0.0}^{+0.0}$& $_{-0.0}^{+0.0}$& $_{-0.0}^{+0.0}$& $_{-0.0}^{+0.0}$ \\ 
$\Lz(1690)$& $_{-0.3}^{+4.1}$& $_{-2.0}^{+2.0}$& $_{+0.2}^{+1.5}$& $_{-0.5}^{+0.6}$& $_{-0.2}^{+0.2}$& $_{-0.1}^{+0.1}$& $_{-0.0}^{+0.0}$& $_{-0.0}^{+0.1}$& $_{-0.1}^{+0.0}$& $_{-0.0}^{+0.0}$ \\ 
$\Lz(1800)$& $_{-5.9}^{+3.0}$& $_{-1.1}^{+1.1}$& $_{-0.8}^{+0.1}$& $_{-1.5}^{+0.8}$& $_{-0.3}^{+0.3}$& $_{-0.1}^{+0.1}$& $_{-0.1}^{+0.1}$& $_{-0.0}^{+0.0}$& $_{-0.0}^{+0.6}$& $_{-0.0}^{+0.4}$ \\ 
$\Lz(1810)$& $_{-0.7}^{+3.7}$& $_{-1.1}^{+1.1}$& $_{+0.1}^{+1.5}$& $_{-1.4}^{+0.5}$& $_{-0.2}^{+0.2}$& $_{-0.1}^{+0.1}$& $_{-0.0}^{+0.0}$& $_{-0.0}^{+0.1}$& $_{-0.0}^{+0.2}$& $_{-0.0}^{+0.0}$ \\ 
$\Lz(1820)$& $_{-4.9}^{+1.8}$& $_{-0.2}^{+0.2}$& $_{-0.9}^{-0.0}$& $_{-0.4}^{+0.3}$& $_{-0.3}^{+0.3}$& $_{-0.1}^{+0.1}$& $_{-0.0}^{+0.0}$& $_{-0.3}^{+0.0}$& $_{-0.0}^{+0.1}$& $_{-0.1}^{+0.0}$ \\ 
$\Lz(1830)$& $_{-0.9}^{+1.3}$& $_{-0.6}^{+0.6}$& $_{-0.4}^{+0.3}$& $_{-0.5}^{+0.3}$& $_{-0.1}^{+0.1}$& $_{-0.1}^{+0.1}$& $_{-0.0}^{+0.0}$& $_{-0.0}^{+0.2}$& $_{-0.0}^{+0.1}$& $_{-0.0}^{+0.0}$ \\ 
$\Lz(1890)$& $_{-5.1}^{+4.2}$& $_{-0.8}^{+0.8}$& $_{-0.4}^{+0.4}$& $_{-0.4}^{+0.1}$& $_{-0.2}^{+0.2}$& $_{-0.1}^{+0.1}$& $_{-0.0}^{+0.0}$& $_{-0.0}^{+0.1}$& $_{-0.0}^{+0.1}$& $_{-0.0}^{+0.0}$ \\ 
$\Lz(2100)$& $_{-2.6}^{+1.0}$& $_{-0.8}^{+0.8}$& $_{-0.7}^{+0.9}$& $_{-0.2}^{+0.2}$& $_{-0.1}^{+0.1}$& $_{-0.0}^{+0.0}$& $_{-0.0}^{+0.0}$& $_{-0.0}^{+0.0}$& $_{-0.0}^{+0.1}$& $_{-0.0}^{+0.1}$ \\ 
$\Lz(2110)$& $_{-0.6}^{+5.0}$& $_{-1.5}^{+1.5}$& $_{-0.1}^{+1.5}$& $_{-0.2}^{+0.3}$& $_{-0.1}^{+0.1}$& $_{-0.1}^{+0.1}$& $_{-0.0}^{+0.0}$& $_{-0.2}^{+0.0}$& $_{-0.0}^{+0.0}$& $_{-0.0}^{+0.2}$ \\ 
$\Lz(2350)$& $_{-0.1}^{+0.0}$& $_{-0.0}^{+0.0}$& $_{-0.2}^{+0.6}$& $_{-0.0}^{+0.0}$& $_{-0.0}^{+0.0}$& $_{-0.0}^{+0.0}$& $_{-0.0}^{+0.0}$& $_{-0.0}^{+0.1}$& $_{-0.0}^{+0.1}$& $_{-0.0}^{+0.1}$ \\ 
NR$(\tfrac{3}{2}^-)$& $_{+0.3}^{+2.9}$& $_{-0.4}^{+0.4}$& $_{-2.4}^{+1.0}$& $_{-0.6}^{+0.0}$& $_{-0.1}^{+0.1}$& $_{-0.1}^{+0.1}$& $_{-0.0}^{+0.0}$& $_{-0.1}^{+0.0}$& $_{-0.3}^{+0.0}$& $_{-0.0}^{+0.0}$ \\ 
\midrule
$\Lz(1405),\Lz(1670)$& $_{-0.7}^{+0.4}$& $_{-0.3}^{+0.3}$& $_{-0.0}^{+0.2}$& $_{-0.1}^{+0.1}$& $_{-0.1}^{+0.1}$& $_{-0.0}^{+0.0}$& $_{-0.0}^{+0.0}$& $_{-0.0}^{+0.0}$& $_{-0.0}^{+0.0}$& $_{-0.1}^{+0.0}$ \\ 
$\Lz(1405),\Lz(1800)$& $_{-3.6}^{+0.5}$& $_{-0.3}^{+0.3}$& $_{-1.9}^{+0.1}$& $_{-0.4}^{+1.7}$& $_{-0.2}^{+0.2}$& $_{-0.2}^{+0.2}$& $_{-0.0}^{+0.0}$& $_{-0.0}^{+0.0}$& $_{-0.3}^{+0.0}$& $_{-0.0}^{+0.1}$ \\ 
$\Lz(1520),\Lz(1690)$& $_{-2.3}^{+0.3}$& $_{-0.9}^{+0.9}$& $_{-0.7}^{-0.1}$& $_{-0.4}^{+0.5}$& $_{-0.1}^{+0.1}$& $_{-0.0}^{+0.0}$& $_{-0.0}^{+0.0}$& $_{-0.1}^{+0.0}$& $_{-0.0}^{+0.0}$& $_{-0.0}^{+0.0}$ \\ 
$\Lz(1520),$ NR$(\tfrac{3}{2}^-)$& $_{-2.4}^{+1.2}$& $_{-1.5}^{+1.5}$& $_{-0.5}^{+0.5}$& $_{-0.4}^{+0.8}$& $_{-0.1}^{+0.1}$& $_{-0.1}^{+0.1}$& $_{-0.0}^{+0.0}$& $_{-0.0}^{+0.0}$& $_{-0.1}^{+0.0}$& $_{-0.0}^{+0.0}$ \\ 
$\Lz(1600),\Lz(1810)$& $_{-2.8}^{+4.1}$& $_{-0.6}^{+0.6}$& $_{-0.7}^{+1.5}$& $_{-0.4}^{+0.9}$& $_{-0.3}^{+0.3}$& $_{-0.2}^{+0.2}$& $_{-0.0}^{+0.0}$& $_{-0.0}^{+0.0}$& $_{-0.4}^{+0.0}$& $_{-0.4}^{+0.0}$ \\ 
$\Lz(1670),\Lz(1800)$& $_{-1.9}^{+1.5}$& $_{-0.4}^{+0.4}$& $_{-0.2}^{+0.3}$& $_{-0.4}^{+0.4}$& $_{-0.1}^{+0.1}$& $_{-0.1}^{+0.1}$& $_{-0.0}^{+0.0}$& $_{-0.0}^{+0.0}$& $_{-0.0}^{+0.0}$& $_{-0.1}^{+0.0}$ \\ 
$\Lz(1690),$ NR$(\tfrac{3}{2}^-)$& $_{-2.2}^{+0.9}$& $_{-1.1}^{+1.1}$& $_{-2.7}^{+0.2}$& $_{-0.5}^{+0.2}$& $_{-0.1}^{+0.1}$& $_{-0.1}^{+0.1}$& $_{-0.0}^{+0.0}$& $_{-0.0}^{+0.0}$& $_{-0.1}^{+0.0}$& $_{-0.0}^{+0.0}$ \\ 
$\Lz(1820),\Lz(2110)$& $_{-3.1}^{+2.4}$& $_{-1.6}^{+1.6}$& $_{-1.6}^{+0.5}$& $_{-0.5}^{+0.3}$& $_{-0.2}^{+0.2}$& $_{-0.1}^{+0.1}$& $_{-0.0}^{+0.0}$& $_{-0.0}^{+0.2}$& $_{-0.3}^{+0.0}$& $_{-0.2}^{+0.0}$ \\ 
\bottomrule
\end{tabular}

%% file: tabs/final_table.tex
\begin{tabular}{c|r>{\large}r|>{\large}r>{\large}r>{\large}r}
\toprule
Observable & Value & {\normalsize$\sigma_\text{stat}$} & {\normalsize$\sigma^\text{internal}_\text{syst}$} & {\normalsize$\sigma^\text{external}_\text{syst}$} & {\normalsize$\sigma_\text{syst}$} \\ 
\midrule
$\Lz(1405)$ & 3.5 & $_{-0.4}^{+0.3}$& $_{-0.0}^{+0.9}$& $_{-0.6}^{+1.3}$& $_{-0.3}^{+1.9}$\\
$\Lz(1520)$ & 10.4 & $_{-0.2}^{+0.4}$& $_{-0.0}^{+0.7}$& $_{-1.6}^{+1.7}$& $_{-1.2}^{+2.2}$\\
$\Lz(1600)$ & 15.6 & $_{-0.9}^{+0.6}$& $_{-0.2}^{+0.8}$& $_{-5.0}^{+3.9}$& $_{-4.6}^{+4.3}$\\
$\Lz(1670)$ & 1.3 & $_{-0.2}^{+0.2}$& $_{-0.2}^{+0.3}$& $_{-0.3}^{+1.2}$& $_{-0.2}^{+1.3}$\\
$\Lz(1690)$ & 7.7 & $_{-0.8}^{+0.4}$& $_{-0.1}^{+1.8}$& $_{-1.0}^{+5.1}$& $_{-0.2}^{+6.2}$\\
$\Lz(1800)$ & 18.3 & $_{-1.6}^{+1.3}$& $_{-1.1}^{+1.4}$& $_{-6.0}^{+3.2}$& $_{-6.2}^{+3.2}$\\
$\Lz(1810)$ & 0.1 & $_{-0.4}^{+0.9}$& $_{-0.4}^{+1.7}$& $_{-0.7}^{+4.0}$& $_{-0.7}^{+4.8}$\\
$\Lz(1820)$ & 8.3 & $_{-0.7}^{+0.4}$& $_{-1.4}^{-0.2}$& $_{-4.8}^{+1.9}$& $_{-5.7}^{+1.0}$\\
$\Lz(1830)$ & 0.3 & $_{-0.4}^{+0.4}$& $_{-0.5}^{+0.6}$& $_{-0.9}^{+1.5}$& $_{-0.9}^{+1.6}$\\
$\Lz(1890)$ & 11.2 & $_{-0.6}^{+0.7}$& $_{-0.6}^{+0.5}$& $_{-5.1}^{+4.3}$& $_{-4.9}^{+4.6}$\\
$\Lz(2100)$ & 7.3 & $_{-0.5}^{+0.5}$& $_{-0.6}^{+1.1}$& $_{-2.8}^{+1.1}$& $_{-2.9}^{+1.4}$\\
$\Lz(2110)$ & 6.5 & $_{-0.7}^{+0.6}$& $_{-0.0}^{+1.7}$& $_{-0.9}^{+5.4}$& $_{-0.2}^{+6.3}$\\
$\Lz(2350)$ & 1.0 & $_{-0.1}^{+0.2}$& $_{-0.0}^{+0.8}$& $_{-0.2}^{+0.0}$& $_{-0.1}^{+0.8}$\\
NR$(\nicefrac{3}{2}^-)$ & 2.8 & $_{-0.4}^{+0.5}$& $_{-1.9}^{+0.2}$& $_{+0.3}^{+3.0}$& $_{-1.3}^{+2.4}$\\
\midrule
$\Lz(1405),\Lz(1670)$ & $-0.7$ & $_{-0.2}^{+0.1}$& $_{-0.2}^{+0.2}$& $_{-0.8}^{+0.5}$& $_{-0.9}^{+0.5}$\\
$\Lz(1405),\Lz(1800)$ & 7.6 & $_{-0.8}^{+0.7}$& $_{-2.0}^{+1.2}$& $_{-3.5}^{+0.6}$& $_{-4.6}^{+0.9}$\\
$\Lz(1520),\Lz(1690)$ & 0.5 & $_{-0.3}^{+0.5}$& $_{-0.9}^{+0.3}$& $_{-2.6}^{+0.6}$& $_{-3.0}^{+0.5}$\\
$\Lz(1520)$, NR$(\nicefrac{3}{2}^-)$ & $-0.6$ & $_{-0.4}^{+0.4}$& $_{-0.6}^{+1.0}$& $_{-3.2}^{+1.6}$& $_{-3.0}^{+2.1}$\\
$\Lz(1600),\Lz(1810)$ & $-1.9$ & $_{-1.0}^{+1.5}$& $_{-1.5}^{+1.3}$& $_{-2.9}^{+4.1}$& $_{-3.6}^{+3.9}$\\
$\Lz(1670),\Lz(1800)$ & $-4.8$ & $_{-0.4}^{+0.5}$& $_{-0.6}^{+0.4}$& $_{-2.0}^{+1.5}$& $_{-2.1}^{+1.5}$\\
$\Lz(1690)$, NR$(\nicefrac{3}{2}^-)$ & 3.9 & $_{-0.4}^{+0.4}$& $_{-3.0}^{+0.1}$& $_{-2.7}^{+1.2}$& $_{-4.7}^{+0.3}$\\
$\Lz(1820),\Lz(2110)$ & 1.1 & $_{-0.5}^{+0.7}$& $_{-2.1}^{+0.2}$& $_{-3.9}^{+2.5}$& $_{-4.8}^{+1.9}$\\
\bottomrule
\end{tabular}

%% file: acknowledgements.tex
\section*{Acknowledgements}
\noindent We express our gratitude to our colleagues in the CERN
accelerator departments for the excellent performance of the LHC. We
thank the technical and administrative staff at the LHCb
institutes.
We acknowledge support from CERN and from the national agencies:
CAPES, CNPq, FAPERJ and FINEP (Brazil); 
MOST and NSFC (China); 
CNRS/IN2P3 (France); 
BMBF, DFG and MPG (Germany); 
INFN (Italy); 
NWO (Netherlands); 
MNiSW and NCN (Poland); 
MCID/IFA (Romania);
MICINN (Spain); 
SNSF and SER (Switzerland); 
NASU (Ukraine); 
STFC (United Kingdom); 
DOE NP and NSF (USA).
We acknowledge the computing resources that are provided by CERN, IN2P3
(France), KIT and DESY (Germany), INFN (Italy), SURF (Netherlands),
PIC (Spain), GridPP (United Kingdom),
CSCS (Switzerland), IFIN-HH (Romania), CBPF (Brazil),
and Polish WLCG (Poland).
We are indebted to the communities behind the multiple open-source
software packages on which we depend.
Individual groups or members have received support from
ARC and ARDC (Australia);
Key Research Program of Frontier Sciences of CAS, CAS PIFI, CAS CCEPP, 
Fundamental Research Funds for the Central Universities, 
and Sci. \& Tech. Program of Guangzhou (China);
Minciencias (Colombia);
EPLANET, Marie Sk\l{}odowska-Curie Actions, ERC and NextGenerationEU (European Union);
A*MIDEX, ANR, IPhU and Labex P2IO, and R\'{e}gion Auvergne-Rh\^{o}ne-Alpes (France);
AvH Foundation (Germany);
ICSC (Italy); 
GVA, XuntaGal, GENCAT, Inditex, InTalent and Prog.~Atracci\'on Talento, CM (Spain);
SRC (Sweden);
the Leverhulme Trust, the Royal Society
 and UKRI (United Kingdom).
A portion of this work was awarded the Martin Schmei{\ss}er Medal.

%% file: appendix.tex
% ===============================================================================
% Purpose: appendix to the standard template: standard symbol alises from Ulrik
% Author: Tomasz Skwarnicki
% Created on: 2009-09-24
% ===============================================================================
\clearpage
\section*{Appendices}

\appendix

\section[\boldmath Logarithmic scale plots of \texorpdfstring{$m_{\Lb}(pK^-)$}{mpK}]{\boldmath Logarithmic scale plots of $m_{\Lb}(pK^-)$}\label{app:logscale:best}
\begin{figure}[htb]
    \centering
    \includegraphics[width=.45\textwidth]{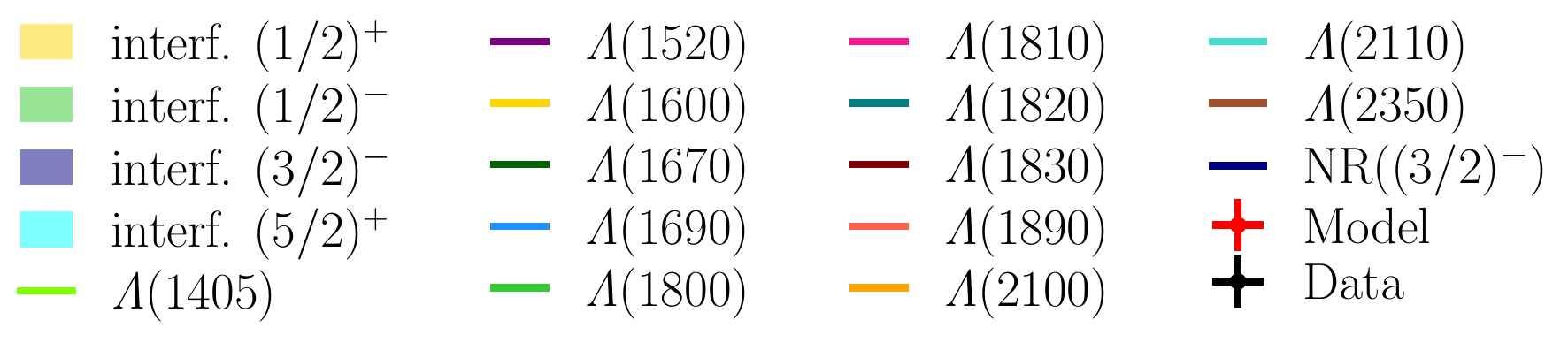} \\
    \includegraphics[width=.45\textwidth]{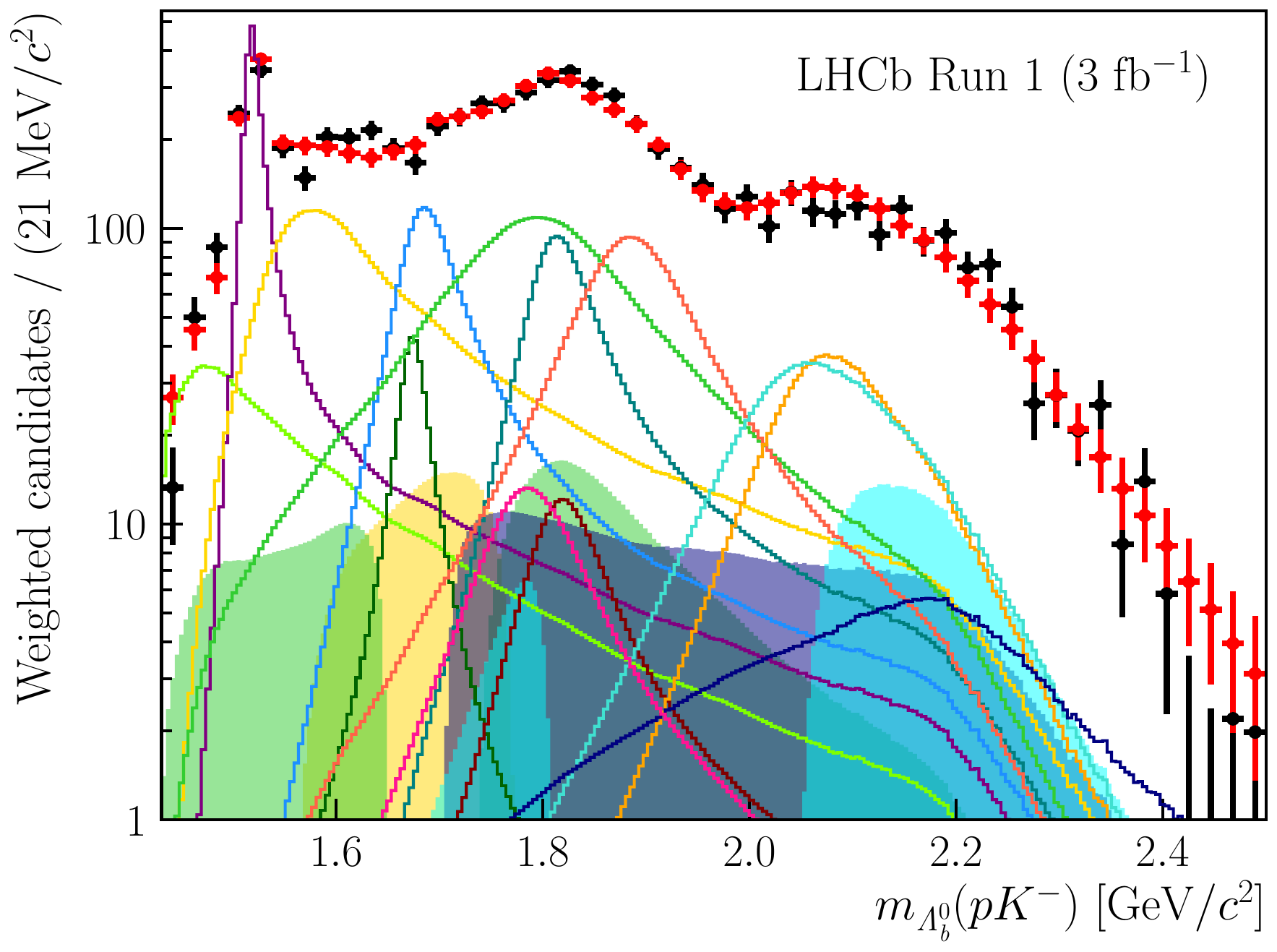}\hspace{.05\textwidth}%
    \includegraphics[width=.45\textwidth]{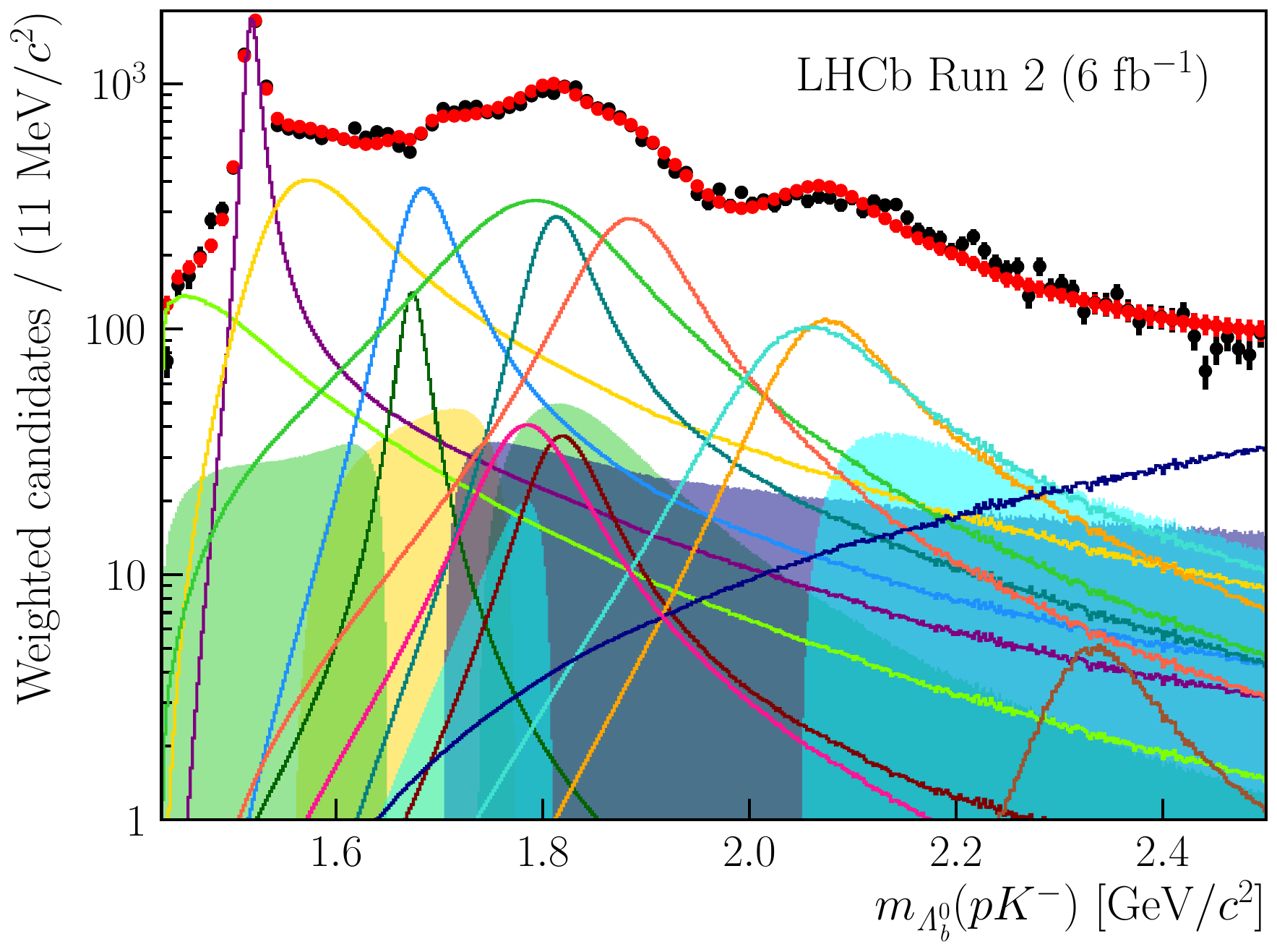}
    \caption{%
        Background-subtracted distribution of the proton-kaon invariant-mass (black dots) for the (left) Run 1 and (right) Run 2 data samples on a logarithmic scale.
        Also shown is a sample generated according to the result of a simultaneous fit of the default model to the data (red dots) and its components (lines) as well as the contributions due to interference between states with the same quantum numbers $J^P$ (shaded areas).
    }
    \label{fig:best-fits:log}
\end{figure}

The plots in Fig.~\ref{fig:best-fits:log} are  equivalent to the top plots in Fig.~\ref{fig:best-fits} with a logarithmic vertical axis in order to make all components visible.
This means the plots contain the background corrected data distributions in the proton-kaon invariant mass.
The plots also contain the full fit model and the individual components.
Note that there are regions where the interference terms become negative but this cannot be displayed on a logarithmic scale.

\section{Projections for the reduced and second best models}\label{app:plots:altmodels}
The \textit{reduced model} consists of the $\Lz$ resonances in Table~\ref{tab:lambdas}.
The \textit{best} and second best model are based on the \textit{reduced model}.
Contrary to the \textit{best model}, the second best model has no nonresonant component but instead the mass and width of the $\Lz(2100)$ and $\Lz(2110)$ states are floated in the fit.

\begin{figure}
    \centering
	\includegraphics[width=.43\textwidth]{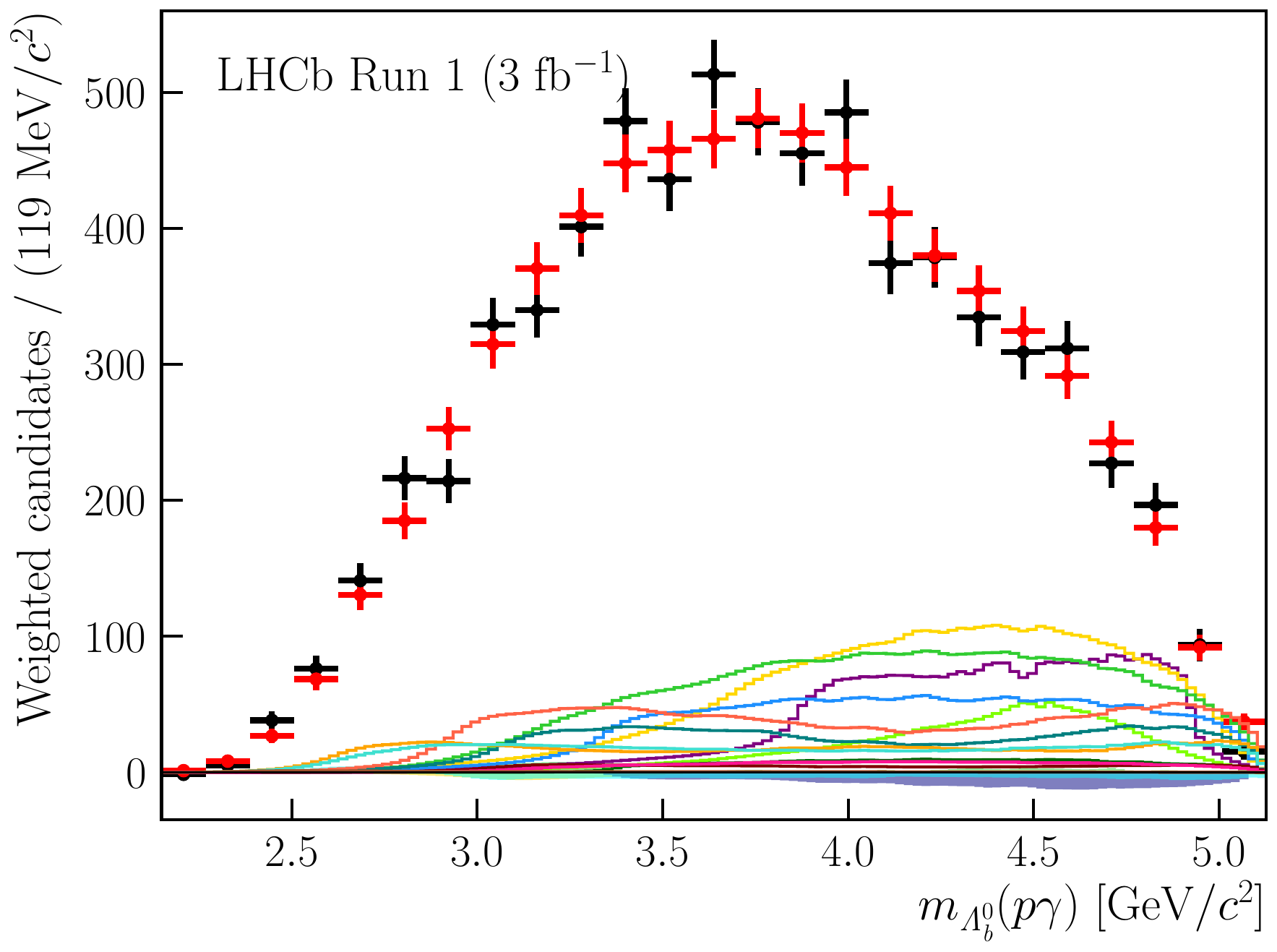}\hspace{.05\textwidth}%
	\includegraphics[width=.43\textwidth]{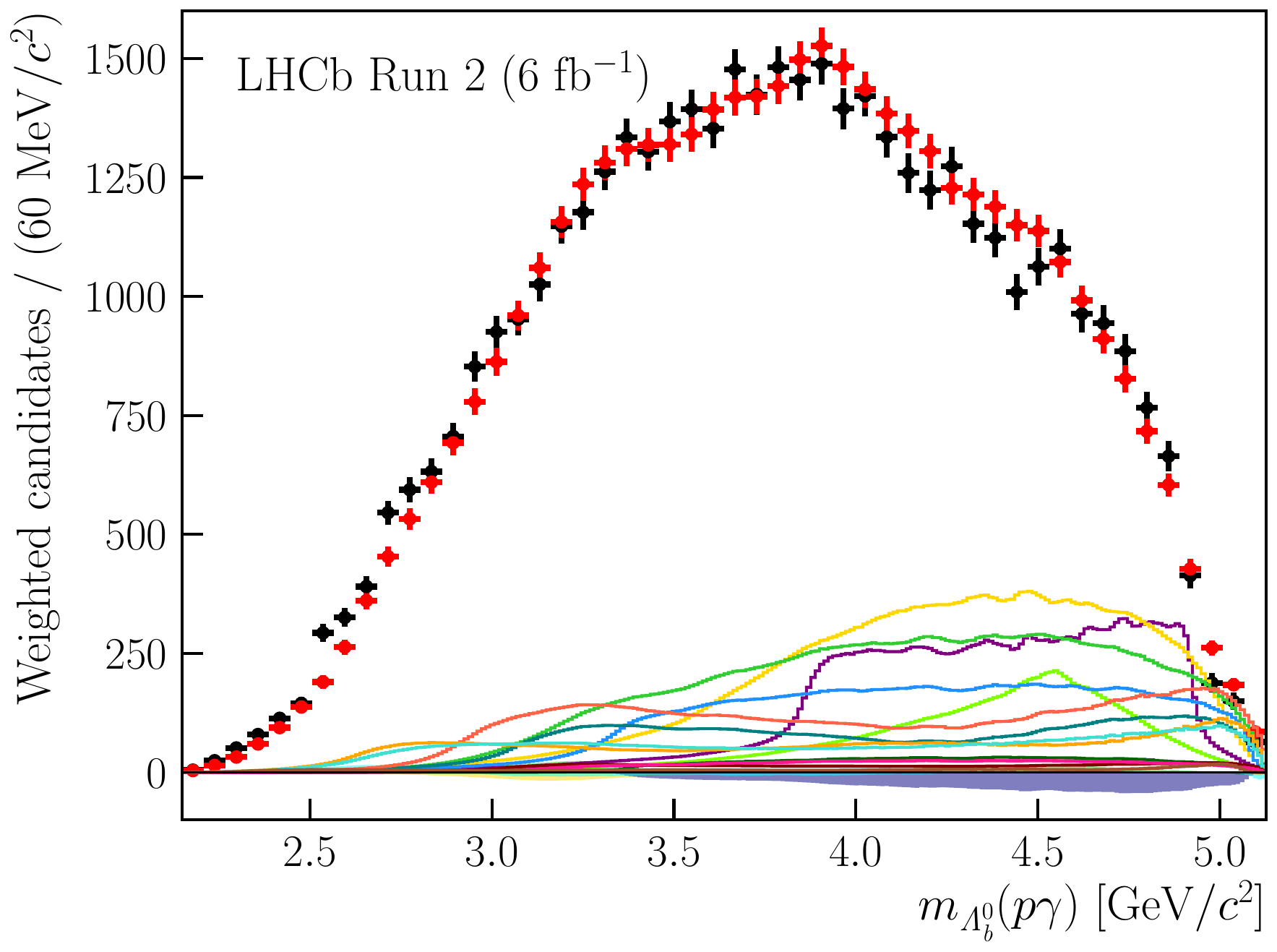}
	\includegraphics[width=.43\textwidth]{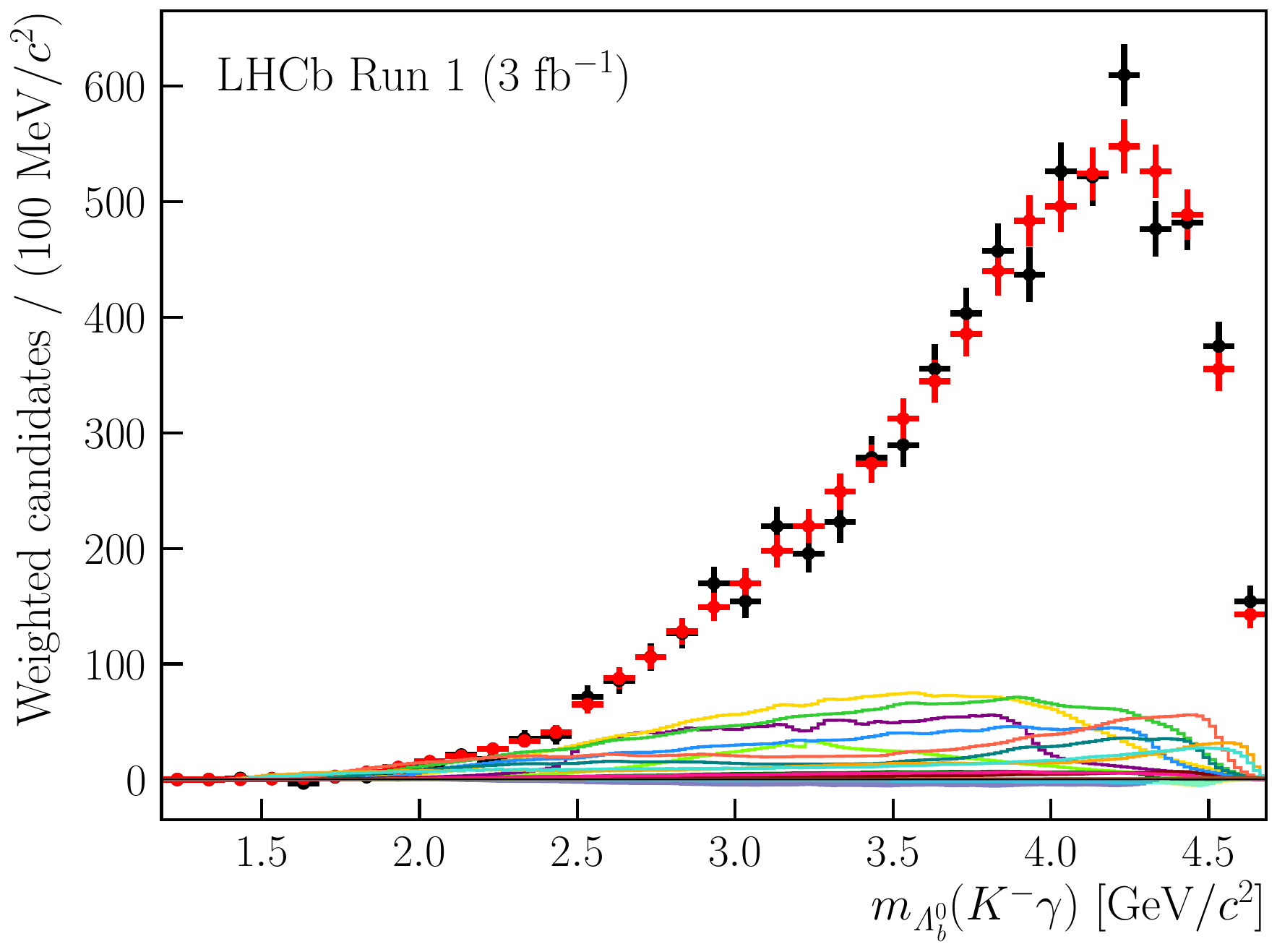}\hspace{.05\textwidth}%
	\includegraphics[width=.43\textwidth]{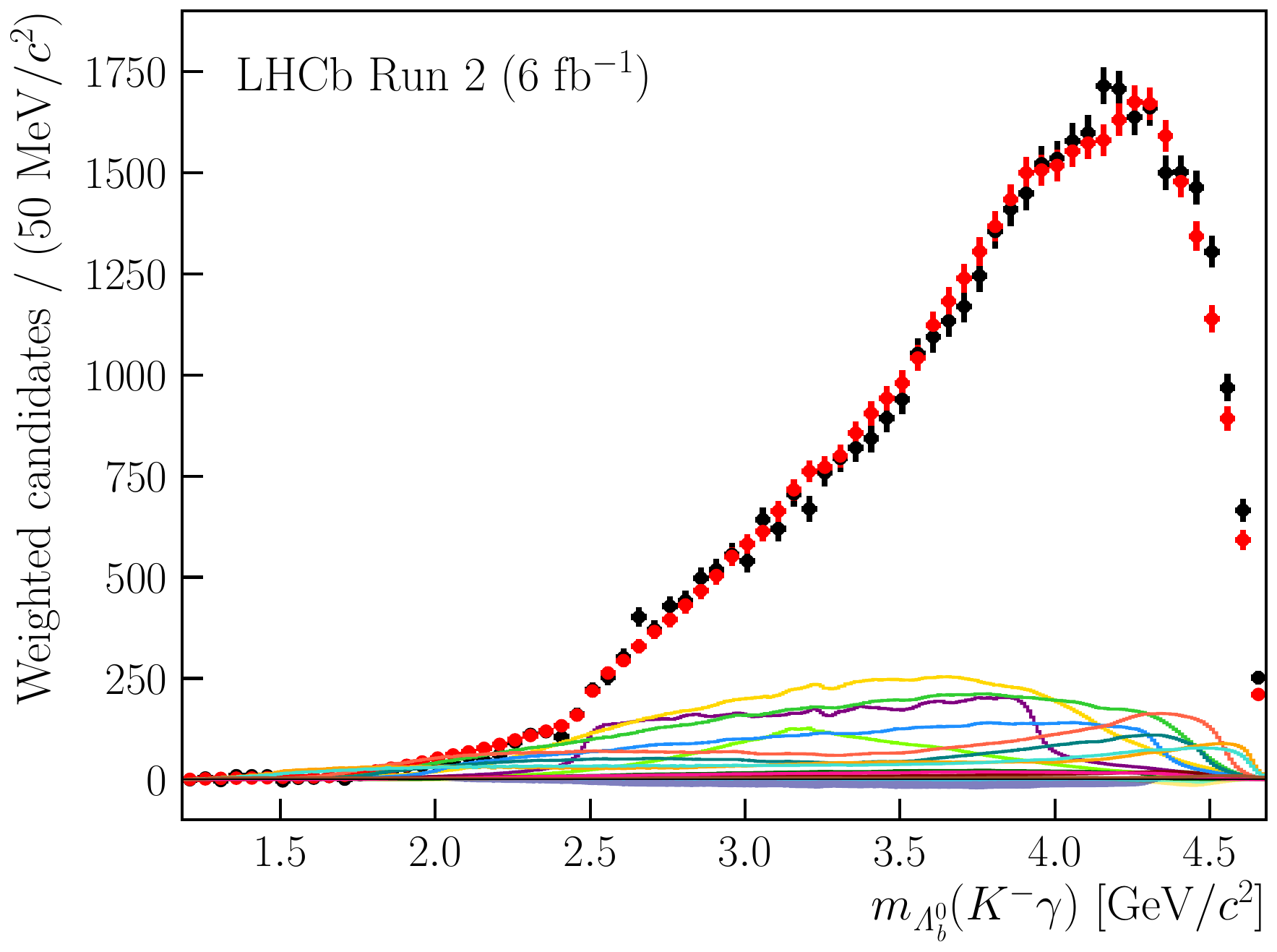}
	\includegraphics[width=.43\textwidth]{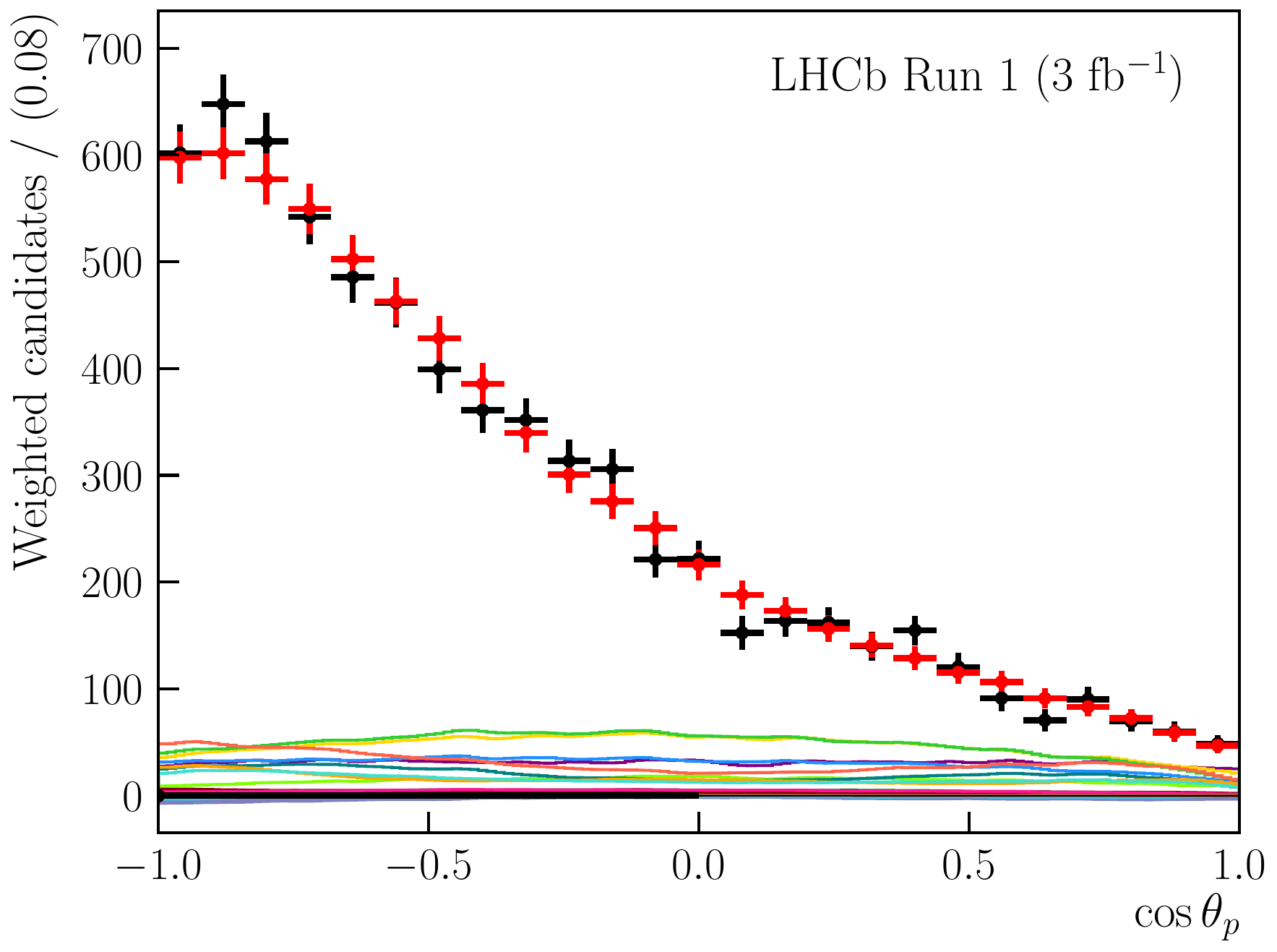}\hspace{.05\textwidth}%
	\includegraphics[width=.43\textwidth]{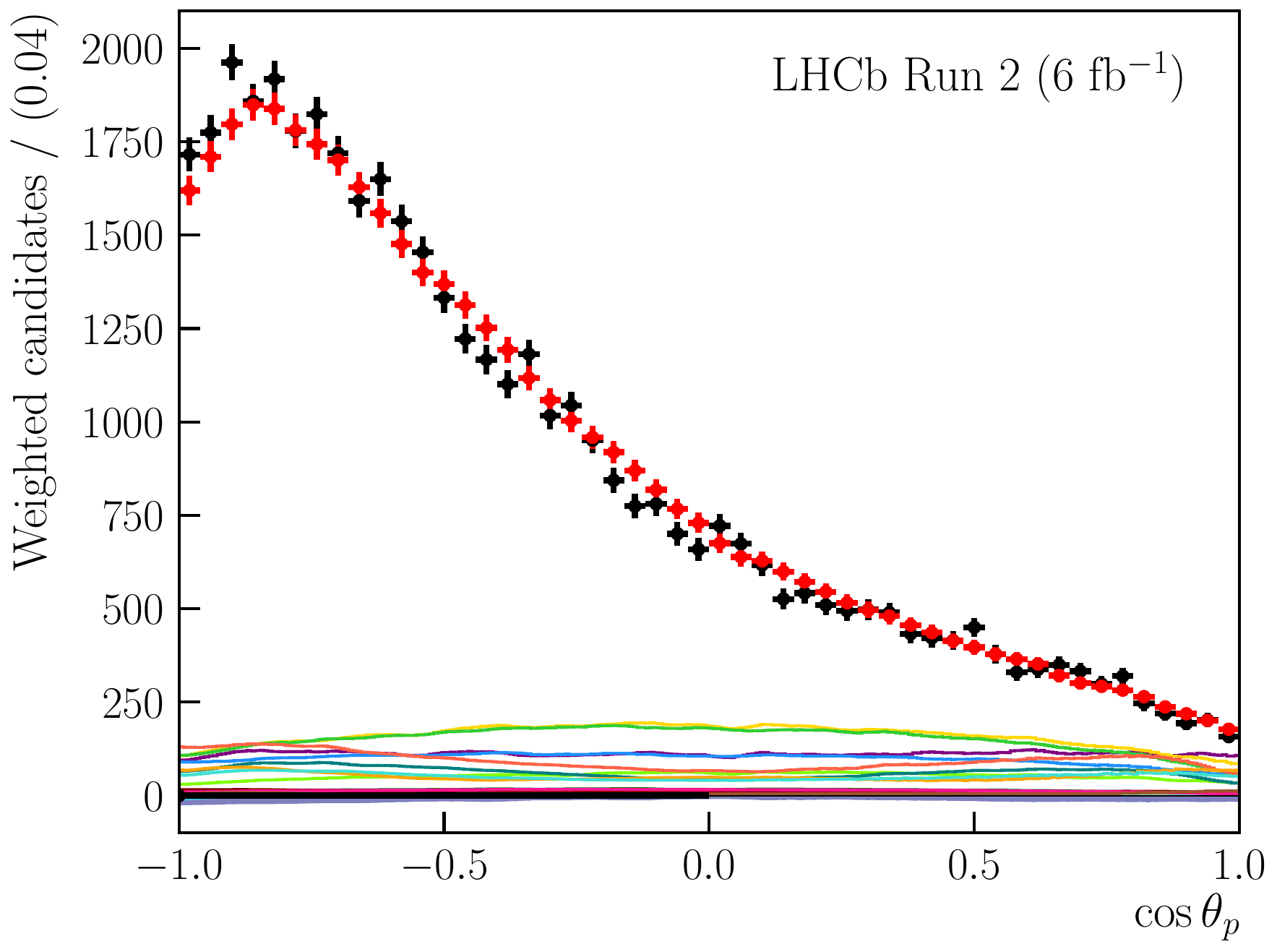}
       \caption{%
        Background-subtracted distribution of (top) the proton-photon invariant-mass, (middle) the kaon-photon invariant-mass, and (bottom) the proton helicity angle (black dots) for the (left) Run 1 and (right) Run 2 data samples.
        Also shown is a sample generated according to the result of a simultaneous fit of the \textit{reduced model} to the data (red dots) and its components (lines) as well as the contributions due to interference between states with the same quantum numbers $J^P$ (shaded areas).
        See Fig.~\ref{fig:reduced:mpk} for the legend.}
    \label{fig:reduced}
\end{figure}

\begin{figure}
    \centering
	\includegraphics[width=.43\textwidth]{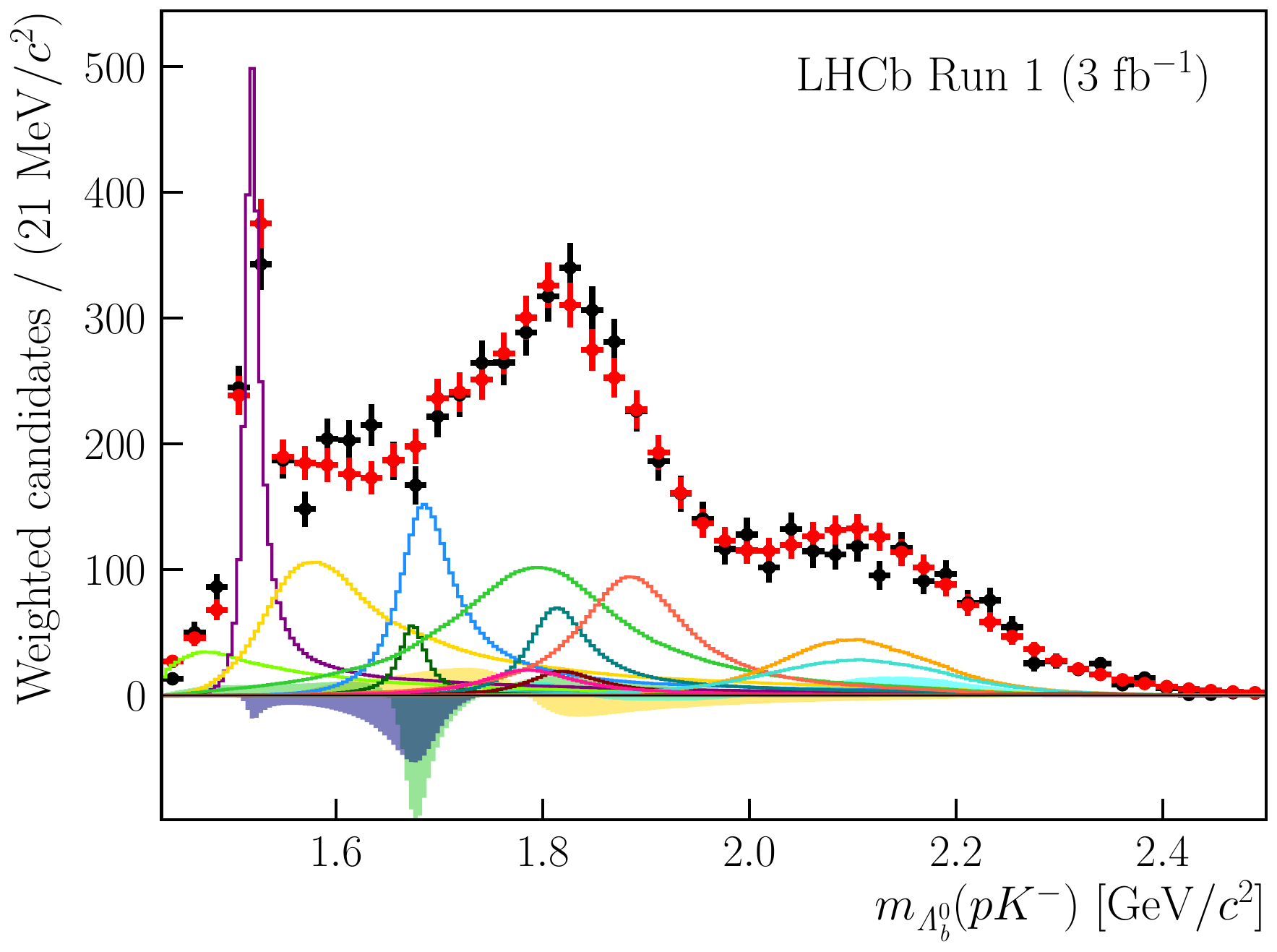}\hspace{.05\textwidth}%
	\includegraphics[width=.43\textwidth]{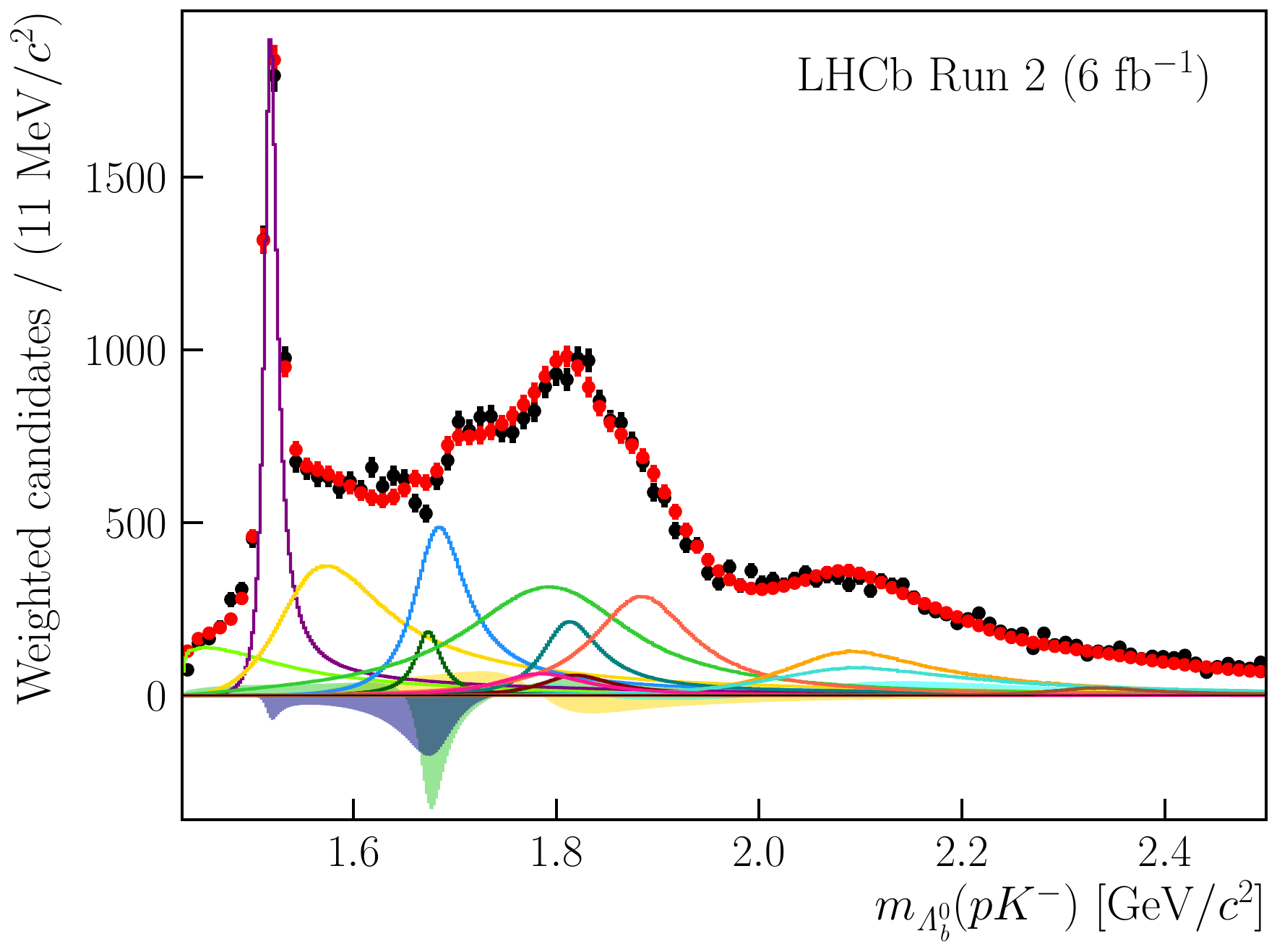}
	\includegraphics[width=.43\textwidth]{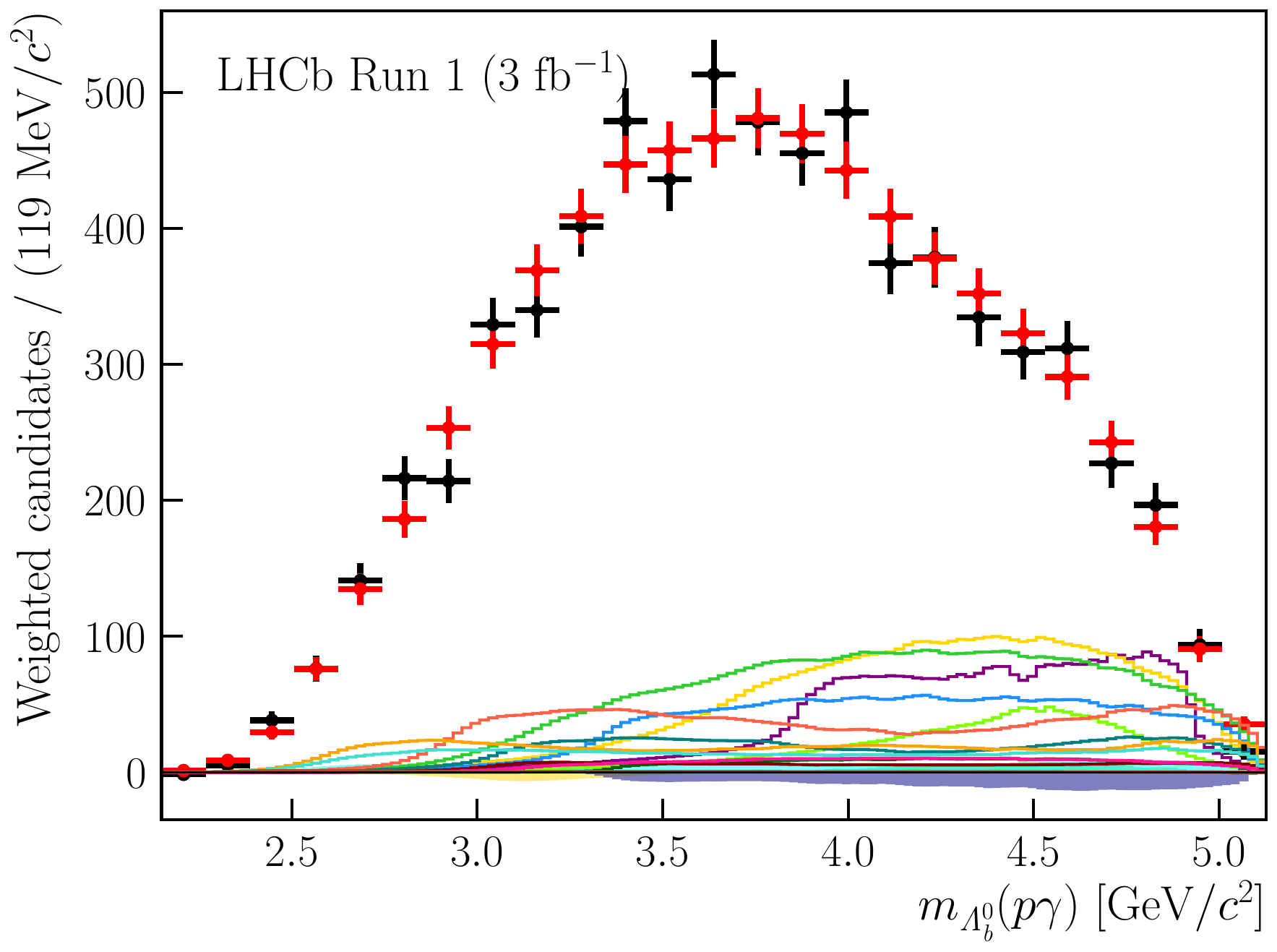}\hspace{.05\textwidth}%
	\includegraphics[width=.43\textwidth]{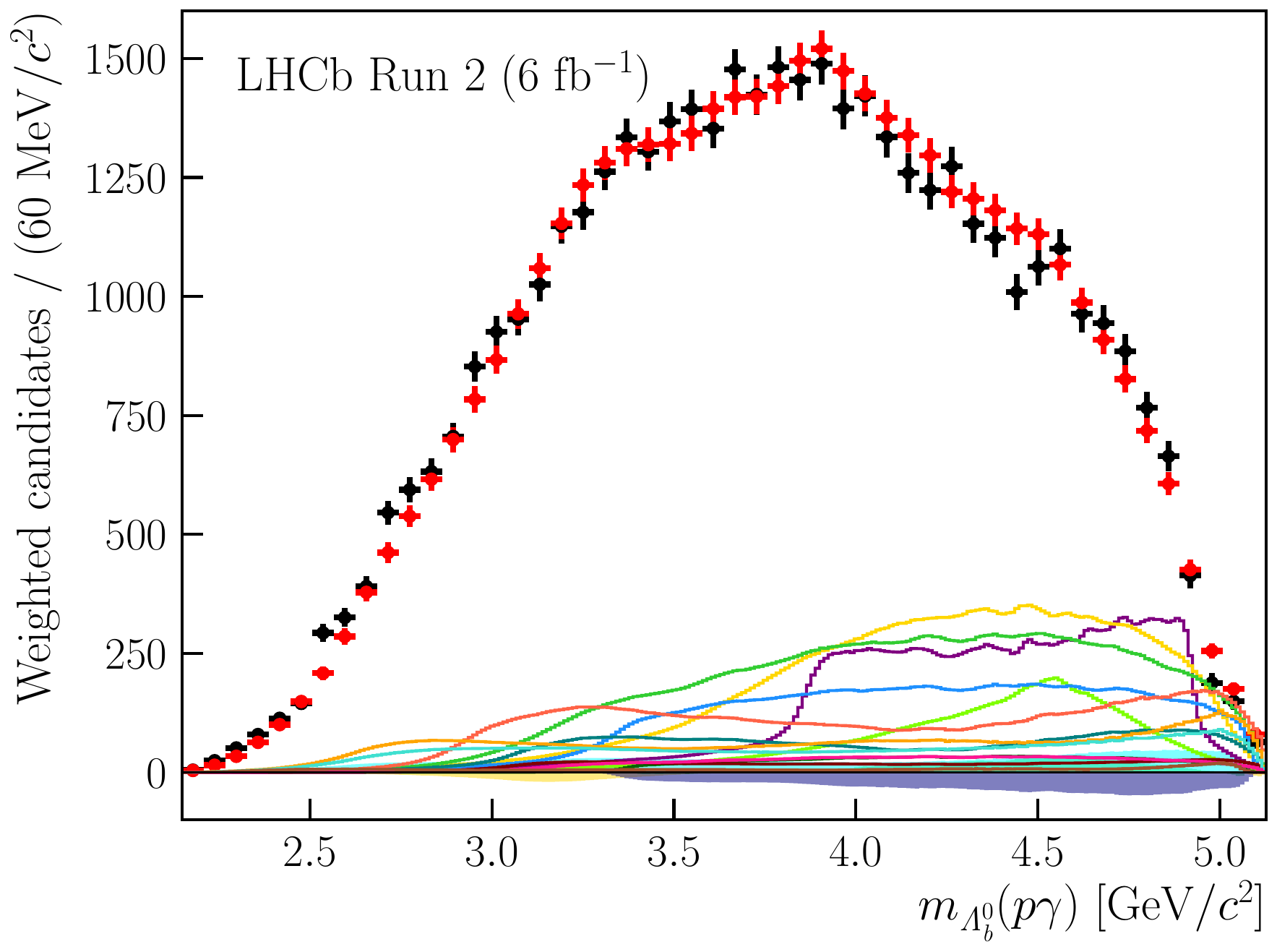}
	\includegraphics[width=.43\textwidth]{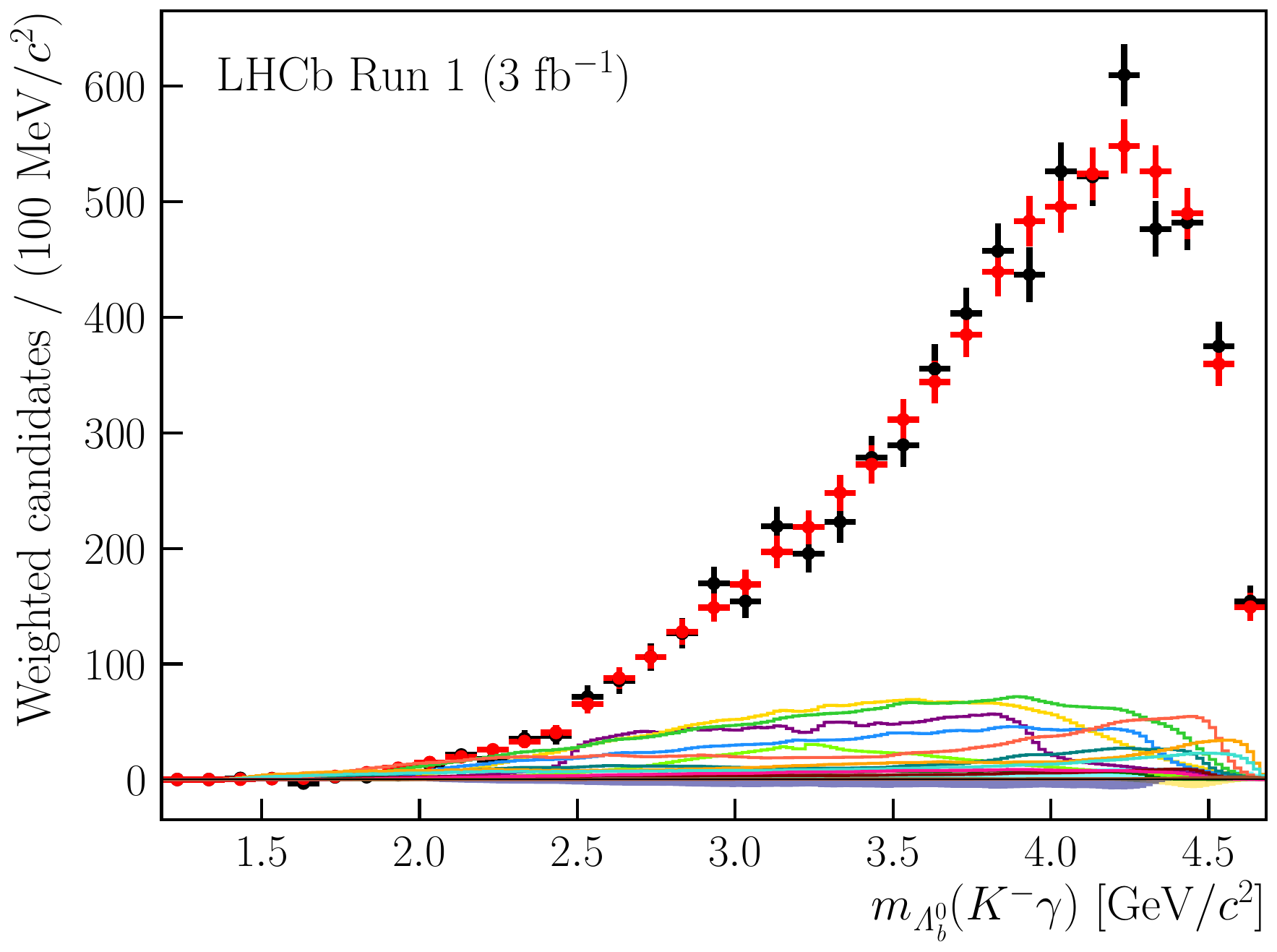}\hspace{.05\textwidth}%
	\includegraphics[width=.43\textwidth]{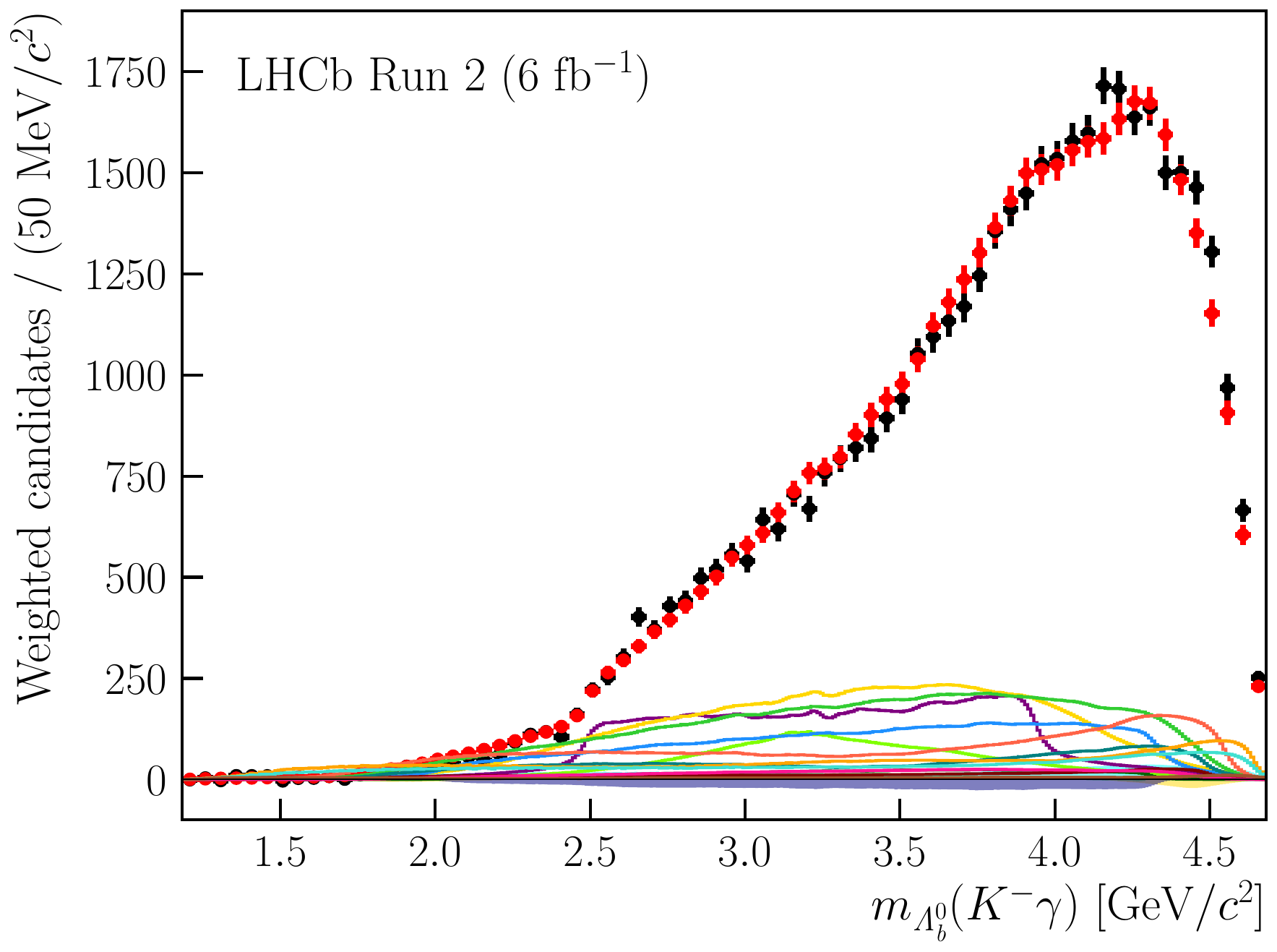}
	\includegraphics[width=.43\textwidth]{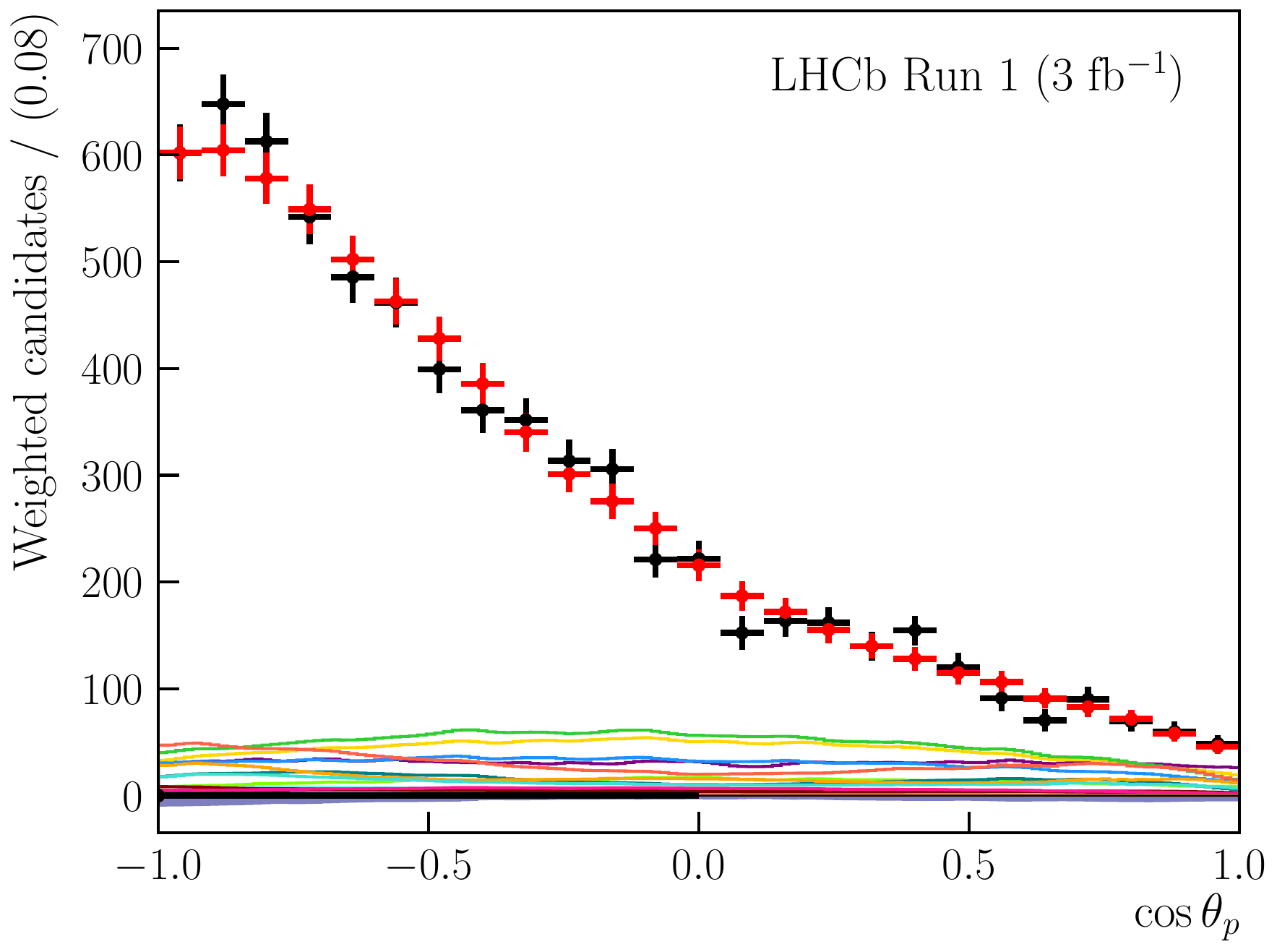}\hspace{.05\textwidth}%
	\includegraphics[width=.43\textwidth]{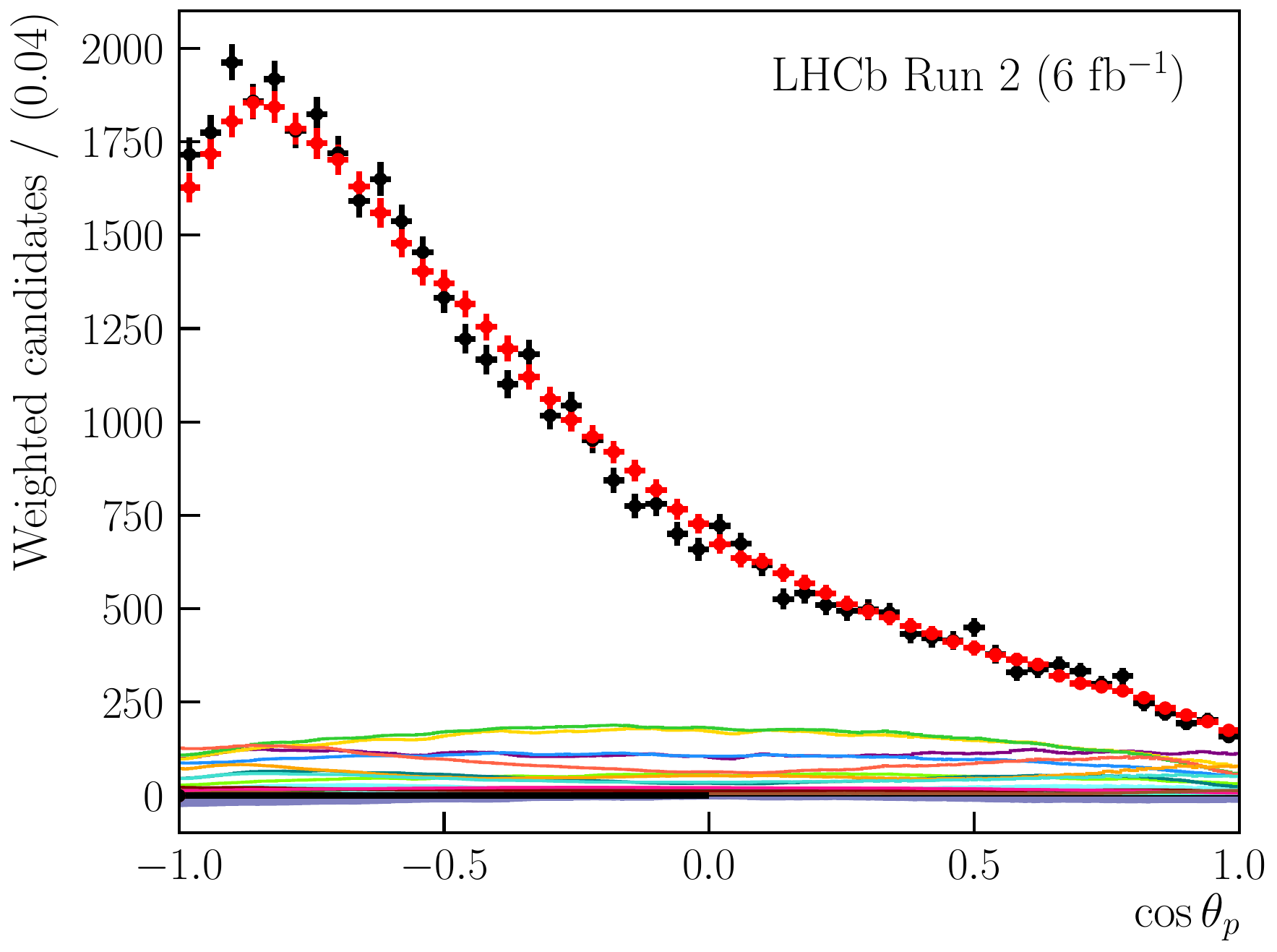}
       \caption{%
        Background-subtracted distribution of (top three rows) the two-body invariant-masses and (bottom row) the proton helicity angle (black dots) for the (left) Run 1 and (right) Run 2 data samples.
        Also shown is a sample generated according to the result of a simultaneous fit of the second best model to the data (red dots) and its components (lines) as well as the contributions due to interference between states with the same quantum numbers $J^P$ (shaded areas).
        See Fig.~\ref{fig:reduced:mpk} for the legend.}
    \label{fig:second-best-fits}
\end{figure}

\begin{figure}
    \centering
    \includegraphics[width=.45\textwidth]{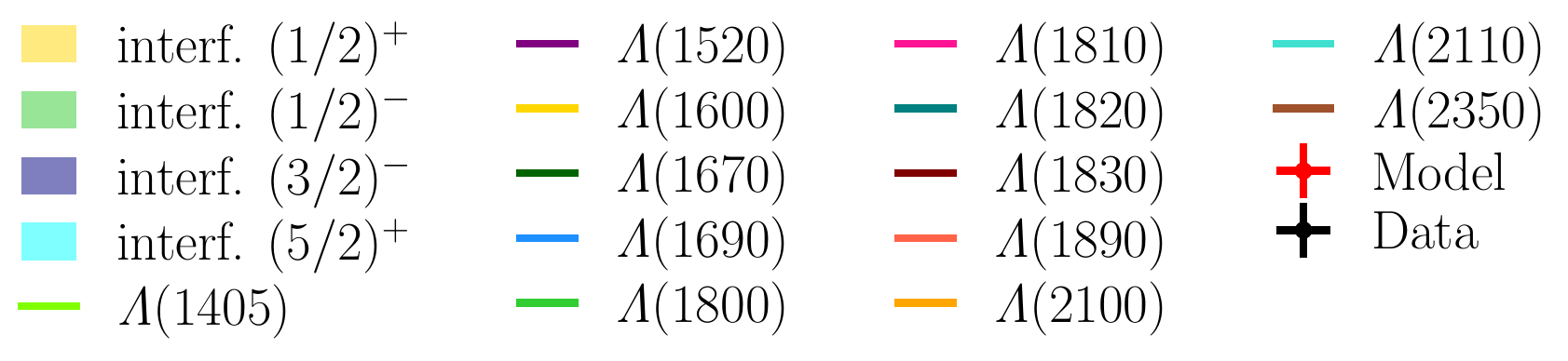} \\
    \includegraphics[width=.45\textwidth]{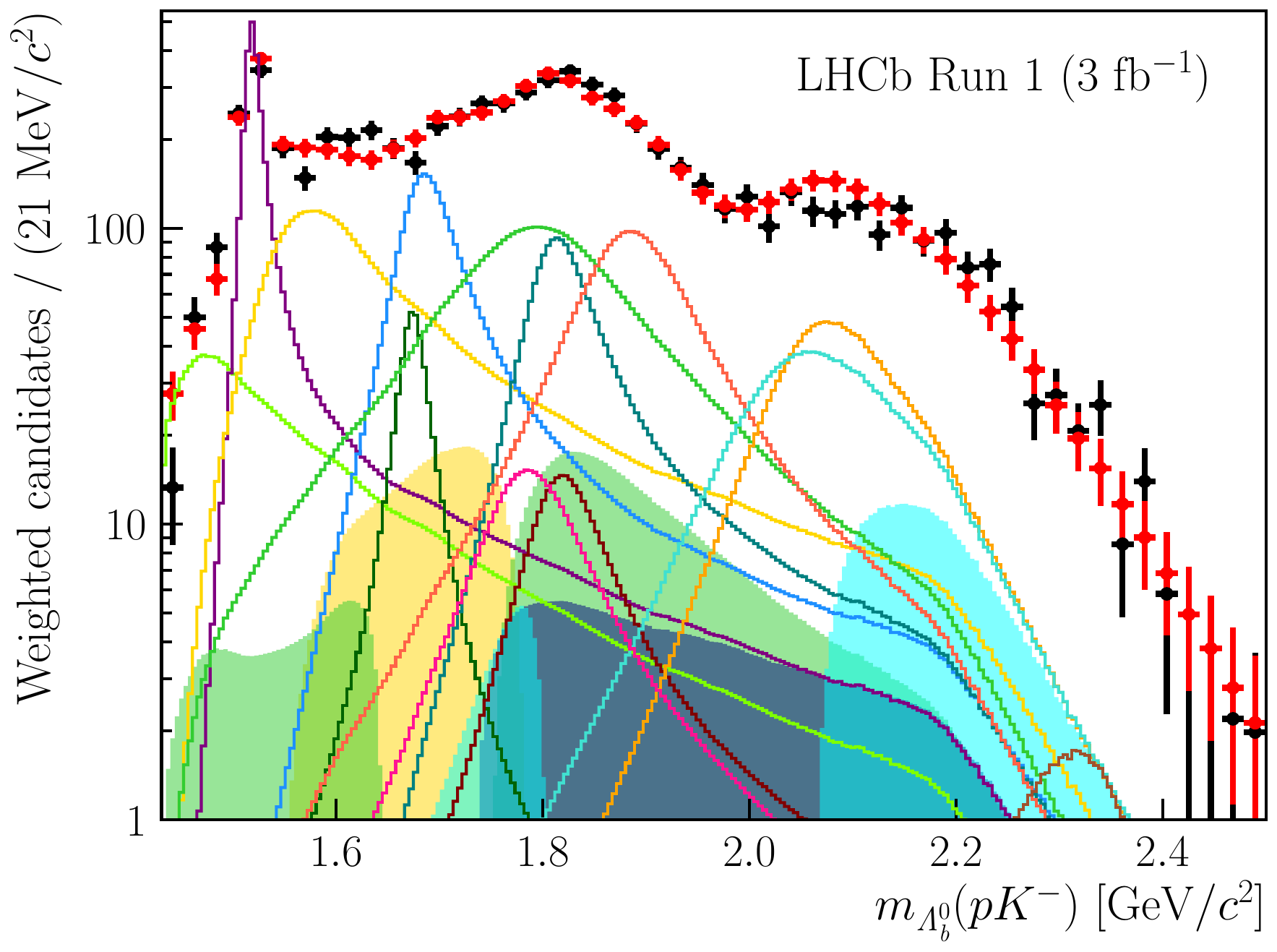}\hspace{.05\textwidth}%
    \includegraphics[width=.45\textwidth]{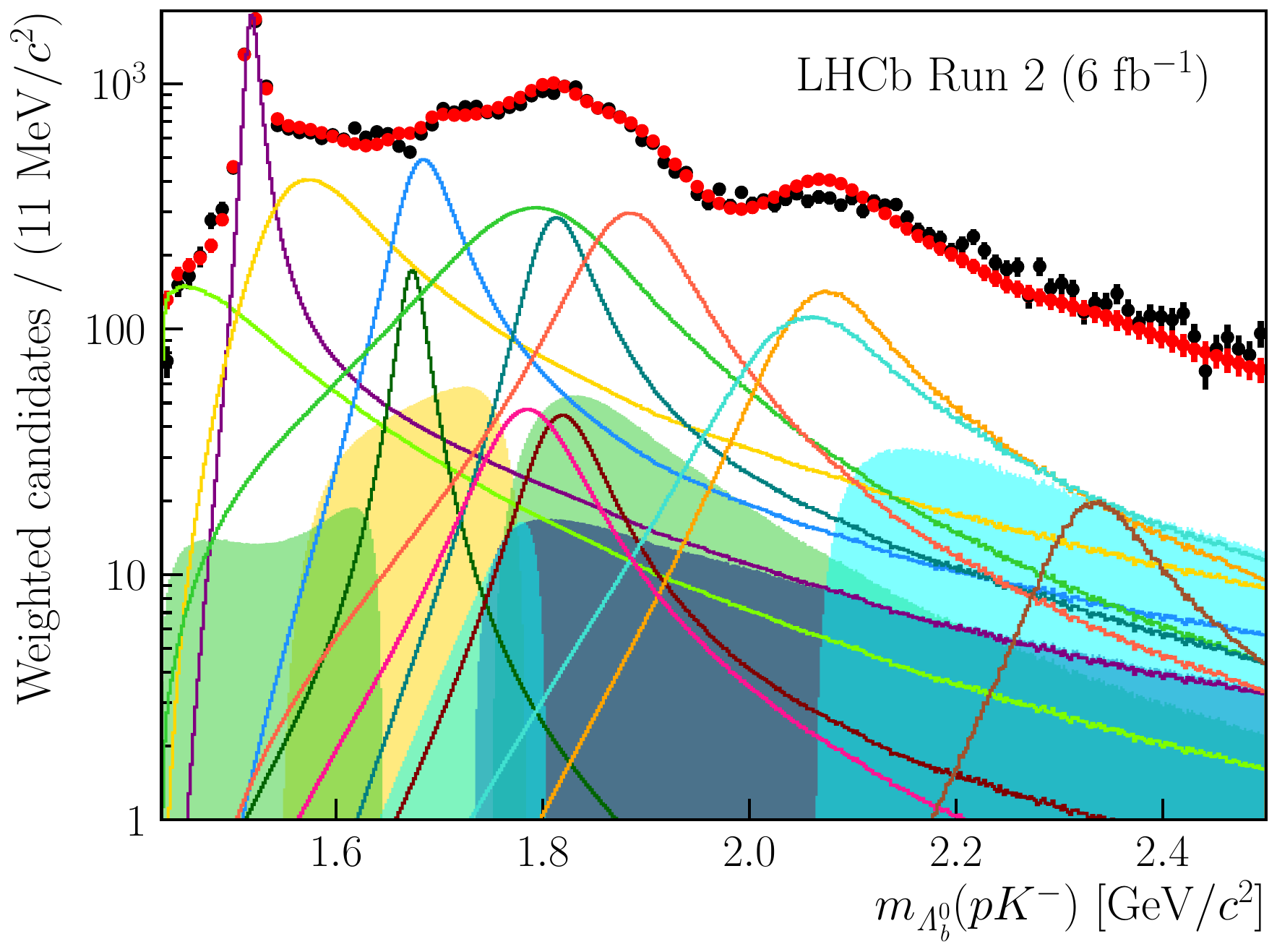}
       \caption{%
        Background-subtracted distribution of the proton-kaon invariant-mass (black dots) for the (left) Run 1 and (right) Run 2 data samples on a logarithmic scale.
        Also shown is a sample generated according to the result of a simultaneous fit of the \textit{reduced model} to the data (red dots) and its components (lines) as well as the contributions due to interference between states with the same quantum numbers $J^P$ (shaded areas).}
    \label{fig:reduced:log}
\end{figure}

\begin{figure}
    \centering
    \includegraphics[width=.45\textwidth]{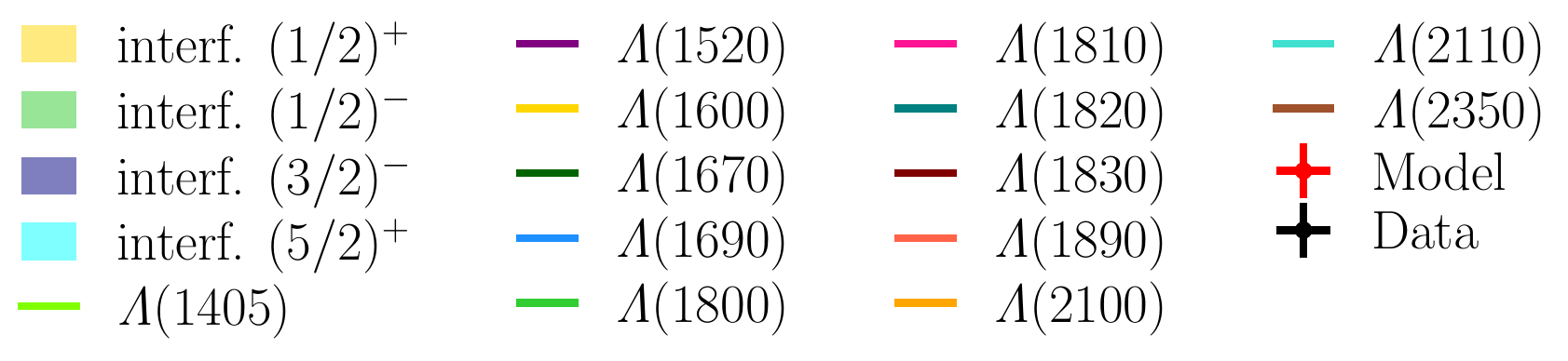} \\
    \includegraphics[width=.45\textwidth]{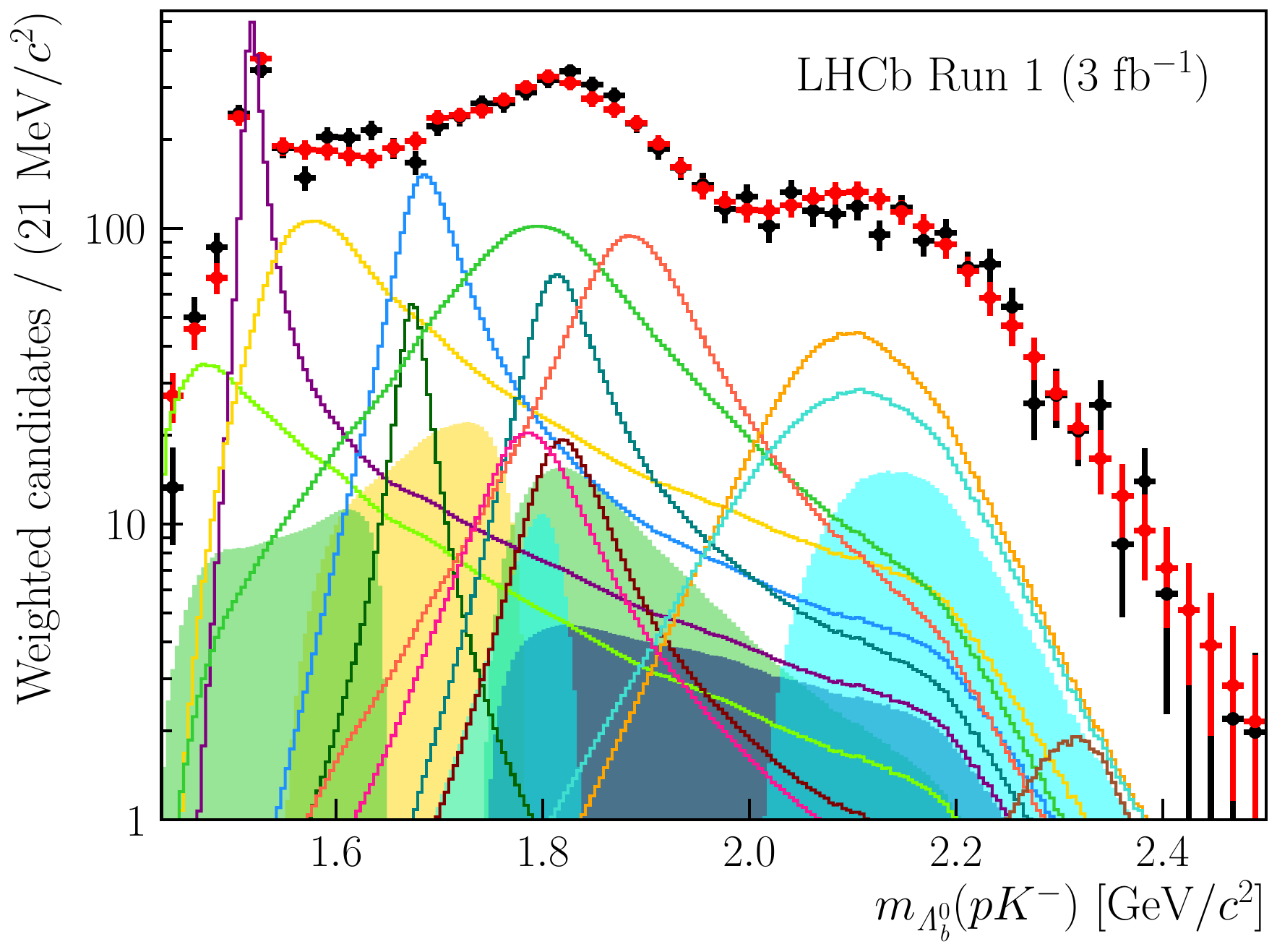}\hspace{.05\textwidth}%
    \includegraphics[width=.45\textwidth]{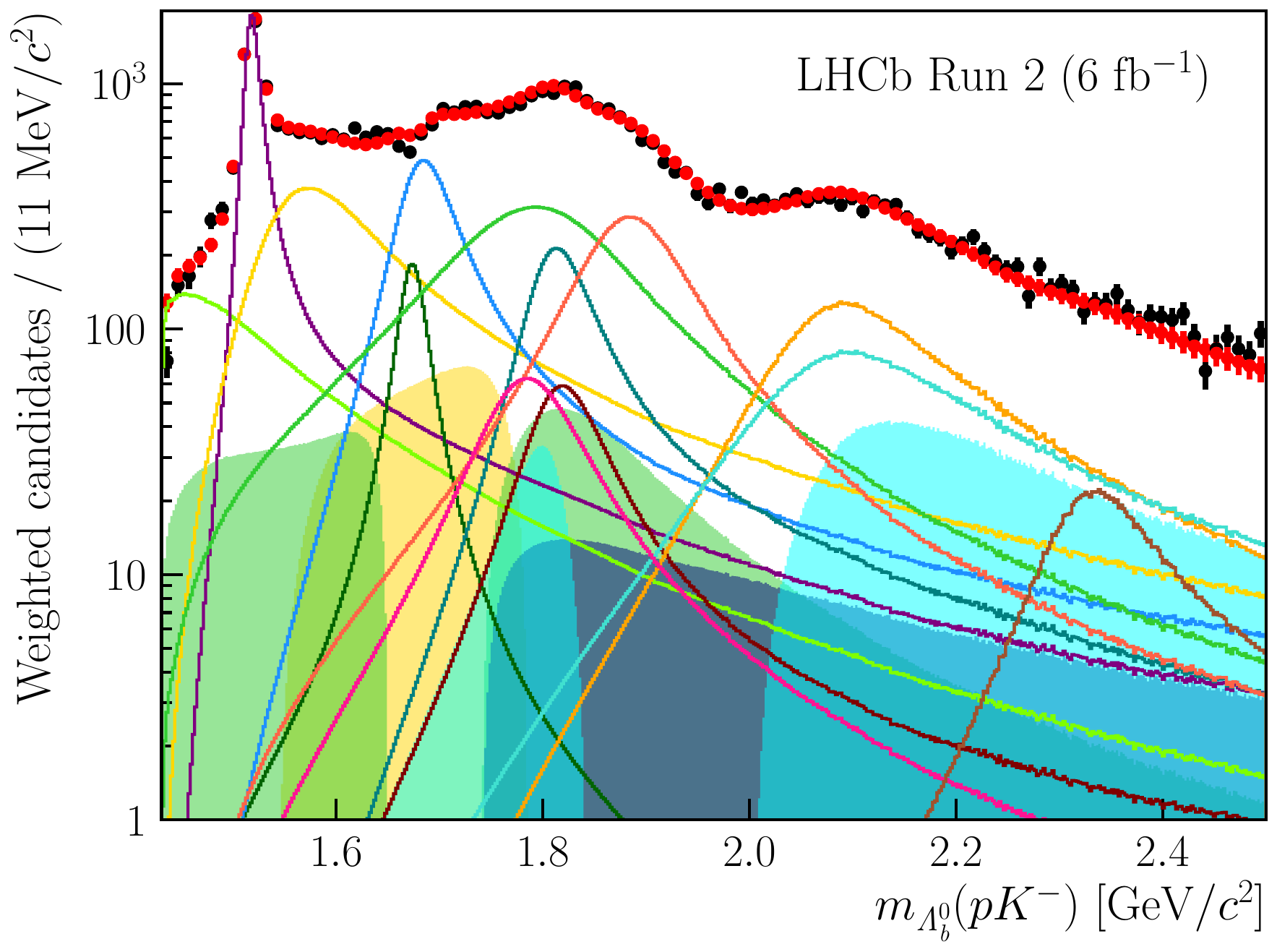}
       \caption{%
        Background-subtracted distribution of the proton-kaon invariant-mass (black dots) for the (left) Run 1 and (right) Run 2 data samples on a logarithmic scale.
        Also shown is a sample generated according to the result of a simultaneous fit of the second best model to the data (red dots) and its components (lines) as well as the contributions due to interference between states with the same quantum numbers $J^P$ (shaded areas).}
    \label{fig:second-best-fits:log}
\end{figure}

Figure~\ref{fig:reduced} shows the fit projections on the proton-photon and kaon-photon invariant-mass, as well as the proton helicity angle for the \textit{reduced model}.
Figure~\ref{fig:second-best-fits} shows the fit projections on the two-body invariant masses and the proton helicity angle for this fit.
Figures~\ref{fig:reduced:log} and \ref{fig:second-best-fits:log} show the projections on the proton-kaon invariant mass for the \textit{reduced model} and the second best model using a logarithmic scale.

\clearpage

\section{Couplings at the best fit point}\label{app:couplings}
Tables~\ref{tab:couplings:1} and \ref{tab:couplings:2} give the value of the couplings, $A_{LS}$, obtained from the fit of the default model to data.
These values serve primarily to construct the model and cannot be interpreted as measurements.
Uncertainties for the couplings are not calculated as they generally are unstable such that minor changes (as are done when estimating systematic uncertainties) can result in very different couplings.
The rightmost column indicates which couplings are dependent on the others via Eq.~\ref{eq:photon_constraint}.

\begin{table}
    \centering
    \caption{Magnitude, $|A_{LS}|$, and phase, $\arg(A_{LS})$, of the couplings at the best fit point of the default model for resonances $\Lz(1405)$, $\Lz(1520)$, $\Lz(1600)$, $\Lz(1670)$, $\Lz(1690)$, $\Lz(1800)$, $\Lz(1810)$, and $\Lz(1820)$.}
    \label{tab:couplings:1}
    \scalebox{0.95}{%
    \input{tabs/couplings}}
\end{table}

\begin{table}
    \centering
    \caption{Magnitude, $|A_{LS}|$, and phase, $\arg(A_{LS})$, of the couplings at the best fit point of the default model for resonances $\Lz(1830)$, $\Lz(1890)$, $\Lz(2100)$, $\Lz(2110)$, $\Lz(2350)$, and the nonresonant component.}
    \label{tab:couplings:2}
    \scalebox{0.95}{%
    \input{tabs/couplings2}}
\end{table}

\section{Statistical correlations}\label{app:correlations}
Table~\ref{tab:corr} provides the statistical correlations between the observables.
\begin{landscape}
\begin{table}[htb]
    \centering
    \caption{Statistical correlations between the observables in percent obtained from bootstrapping the data.
    In the interest of space, the $J^P$ specification of the nonresonant component of the best model, NR($\tfrac{3}{2}^-$), is dropped and the resonances are only referred to by their mass.
    A single mass refers to the fit fraction of a state, two masses refer to the interference fit fraction of the given combination.
    }
    \label{tab:corr}
    \scalebox{0.63}{%
    \input{tabs/correlations.tex}}
\end{table}
\end{landscape}

%% file: tabs/couplings.tex
\begin{tabular}{l|cc|crc}
    \toprule
    Resonance & $2L$ & $2S$ & $|A_{LS}|$ & $\arg(A_{LS})$ & additional comment \\
    \midrule
    \multirow{4}{*}{$\Lz(1405)$} & 0 & 1 & 2.890 & $-0.672$ &  \\
    & 2 & 1 & 2.137 & $-2.633$ &  \\
    & 2 & 3 & 1.511 & $0.509$ & dependent via Eq.~\ref{eq:photon_constraint} \\
    & 4 & 3 & 2.044 & $2.470$ & dependent via Eq.~\ref{eq:photon_constraint} \\
    \midrule
    \multirow{6}{*}{$\Lz(1520)$} & 0 & 1 & 0.400 & $-0.016$ &  \\
    & 2 & 1 & 0.542 & $0.147$ &  \\
    & 2 & 3 & 2.063 & $1.649$ &  \\
    & 4 & 3 & 1.142 & $2.083$ &  \\
    & 4 & 5 & 0.590 & $-0.607$ & dependent via Eq.~\ref{eq:photon_constraint} \\
    & 6 & 5 & 0.773 & $-0.944$ & dependent via Eq.~\ref{eq:photon_constraint} \\
    \midrule
    \multirow{4}{*}{$\Lz(1600)$} & 0 & 1 & 7.000 & $0.970$ &  \\
    & 2 & 1 & 4.127 & $3.057$ &  \\
    & 2 & 3 & 2.918 & $-0.085$ & dependent via Eq.~\ref{eq:photon_constraint} \\
    & 4 & 3 & 4.950 & $-2.171$ & dependent via Eq.~\ref{eq:photon_constraint} \\
    \midrule
    \multirow{4}{*}{$\Lz(1670)$} & 0 & 1 & 0.182 & $2.694$ &  \\
    & 2 & 1 & 0.394 & $0.549$ &  \\
    & 2 & 3 & 0.279 & $-2.592$ & dependent via Eq.~\ref{eq:photon_constraint} \\
    & 4 & 3 & 0.129 & $-0.447$ & dependent via Eq.~\ref{eq:photon_constraint} \\
    \midrule
    \multirow{6}{*}{$\Lz(1690)$} & 0 & 1 & 0.371 & $-2.977$ &  \\
    & 2 & 1 & 2.426 & $0.694$ &  \\
    & 2 & 3 & 1.328 & $0.263$ &  \\
    & 4 & 3 & 2.918 & $0.648$ &  \\
    & 4 & 5 & 1.225 & $-2.599$ & dependent via Eq.~\ref{eq:photon_constraint} \\
    & 6 & 5 & 1.418 & $0.824$ & dependent via Eq.~\ref{eq:photon_constraint} \\
    \midrule
    \multirow{4}{*}{$\Lz(1800)$} & 0 & 1 & 1.000 & $0.000$ & fixed in the fit \\
    & 2 & 1 & 4.418 & $-1.498$ &  \\
    & 2 & 3 & 3.124 & $1.643$ & dependent via Eq.~\ref{eq:photon_constraint} \\
    & 4 & 3 & 0.707 & $3.142$ & dependent via Eq.~\ref{eq:photon_constraint} \\
    \midrule
    \multirow{4}{*}{$\Lz(1810)$} & 0 & 1 & 1.453 & $-2.702$ &  \\
    & 2 & 1 & 0.374 & $-1.129$ &  \\
    & 2 & 3 & 0.264 & $2.012$ & dependent via Eq.~\ref{eq:photon_constraint} \\
    & 4 & 3 & 1.027 & $0.440$ & dependent via Eq.~\ref{eq:photon_constraint} \\
    \midrule
    \multirow{6}{*}{$\Lz(1820)$} & 2 & 3 & 0.692 & $-2.965$ &  \\
    & 4 & 3 & 3.166 & $-1.937$ &  \\
    & 4 & 5 & 2.258 & $-1.733$ &  \\
    & 6 & 5 & 3.023 & $-1.299$ &  \\
    & 6 & 7 & 0.931 & $2.510$ & dependent via Eq.~\ref{eq:photon_constraint} \\
    & 8 & 7 & 2.157 & $-1.984$ & dependent via Eq.~\ref{eq:photon_constraint} \\
    \bottomrule
\end{tabular}

%% file: tabs/couplings2.tex
\begin{tabular}{l|cc|crc}
    \toprule
    Resonance & $2L$ & $2S$ & $|A_{LS}|$ & $\arg(A_{LS})$ & additional comment \\
    \midrule
    \multirow{6}{*}{$\Lz(1830)$} & 2 & 3 & 0.646 & $2.337$ &  \\
    & 4 & 3 & 0.881 & $-2.763$ &  \\
    & 4 & 5 & 0.894 & $0.643$ &  \\
    & 6 & 5 & 0.764 & $-2.560$ &  \\
    & 6 & 7 & 0.536 & $2.018$ & dependent via Eq.~\ref{eq:photon_constraint} \\
    & 8 & 7 & 0.931 & $-2.707$ & dependent via Eq.~\ref{eq:photon_constraint} \\
    \midrule
    \multirow{6}{*}{$\Lz(1890)$} & 0 & 1 & 2.070 & $-1.103$ &  \\
    & 2 & 1 & 1.312 & $0.175$ &  \\
    & 2 & 3 & 1.449 & $3.118$ &  \\
    & 4 & 3 & 3.918 & $1.441$ &  \\
    & 4 & 5 & 2.724 & $-1.376$ & dependent via Eq.~\ref{eq:photon_constraint} \\
    & 6 & 5 & 1.455 & $0.109$ & dependent via Eq.~\ref{eq:photon_constraint} \\
    \midrule
    \multirow{6}{*}{$\Lz(2100)$} & 4 & 5 & 2.378 & $-0.537$ &  \\
    & 6 & 5 & 5.087 & $1.139$ &  \\
    & 6 & 7 & 3.192 & $1.384$ &  \\
    & 8 & 7 & 6.932 & $1.924$ &  \\
    & 8 & 9 & 3.088 & $-0.778$ & dependent via Eq.~\ref{eq:photon_constraint} \\
    & 10 & 9 & 3.946 & $1.106$ & dependent via Eq.~\ref{eq:photon_constraint} \\
    \midrule
    \multirow{6}{*}{$\Lz(2110)$} & 2 & 3 & 2.093 & $2.382$ &  \\
    & 4 & 3 & 6.217 & $-2.358$ &  \\
    & 4 & 5 & 2.348 & $-1.664$ &  \\
    & 6 & 5 & 6.899 & $-1.546$ &  \\
    & 6 & 7 & 3.044 & $2.015$ & dependent via Eq.~\ref{eq:photon_constraint} \\
    & 8 & 7 & 4.810 & $-2.427$ & dependent via Eq.~\ref{eq:photon_constraint} \\
    \midrule
    \multirow{6}{*}{$\Lz(2350)$} & 6 & 7 & 0.509 & $-1.829$ &  \\
    & 8 & 7 & 0.670 & $-0.946$ &  \\
    & 8 & 9 & 1.751 & $-1.344$ &  \\
    & 10 & 9 & 0.848 & $-0.286$ &  \\
    & 10 & 11 & 0.471 & $-2.076$ & dependent via Eq.~\ref{eq:photon_constraint} \\
    & 12 & 11 & 0.396 & $-0.711$ & dependent via Eq.~\ref{eq:photon_constraint} \\
    \midrule
    \multirow{6}{*}{NR$(\nicefrac{3}{2}^-)$} & 0 & 1 & 1.368 & $0.598$ &  \\
    & 2 & 1 & 8.811 & $3.131$ &  \\
    & 2 & 3 & 7.525 & $-0.006$ &  \\
    & 4 & 3 & 5.694 & $-2.808$ &  \\
    & 4 & 5 & 2.895 & $0.426$ & dependent via Eq.~\ref{eq:photon_constraint} \\
    & 6 & 5 & 9.075 & $3.132$ & dependent via Eq.~\ref{eq:photon_constraint} \\
    \bottomrule
\end{tabular}

%% file: tabs/correlations.tex
\begin{tabular}{c|rrrrrrrrrrrrrr|rrrrrrrr}
\toprule
& 1405 & 1520 & 1600 & 1670 & 1690 & 1800 & 1810 & 1820 & 1830 & 1890 & 2100 & 2110 & 2350 & NR & 1405, 1670 & 1520, 1690 & 1405, 1800 & 1670, 1800 & 1600, 1810 & 1820, 2110 & 1520, NR & 1690, NR \\ 
 \hline
1405 & 100 &  &  &  &  &  &  &  &  &  &  &  &  &  &  &  &  &  &  &  &  & \\ 
1520 & 8 & 100 &  &  &  &  &  &  &  &  &  &  &  &  &  &  &  &  &  &  &  & \\ 
1600 & 11 & -14 & 100 &  &  &  &  &  &  &  &  &  &  &  &  &  &  &  &  &  &  & \\ 
1670 & -37 & -9 & -33 & 100 &  &  &  &  &  &  &  &  &  &  &  &  &  &  &  &  &  & \\ 
1690 & -7 & 25 & -33 & 30 & 100 &  &  &  &  &  &  &  &  &  &  &  &  &  &  &  &  & \\ 
1800 & -24 & -13 & -8 & 34 & 14 & 100 &  &  &  &  &  &  &  &  &  &  &  &  &  &  &  & \\ 
1810 & 21 & -3 & 13 & -20 & -31 & -41 & 100 &  &  &  &  &  &  &  &  &  &  &  &  &  &  & \\ 
1820 & 2 & 3 & 2 & 3 & 6 & -11 & 3 & 100 &  &  &  &  &  &  &  &  &  &  &  &  &  & \\ 
1830 & 16 & 20 & -8 & -20 & -27 & -36 & 33 & -11 & 100 &  &  &  &  &  &  &  &  &  &  &  &  & \\ 
1890 & -3 & 4 & 11 & -3 & -25 & -33 & 50 & 1 & 14 & 100 &  &  &  &  &  &  &  &  &  &  &  & \\ 
2100 & -21 & -14 & -16 & 21 & -24 & -0 & 8 & 9 & 28 & 3 & 100 &  &  &  &  &  &  &  &  &  &  & \\ 
2110 & -10 & 20 & -5 & -9 & -2 & 27 & 15 & -3 & -9 & 23 & -26 & 100 &  &  &  &  &  &  &  &  &  & \\ 
2350 & -10 & -14 & -15 & 14 & 4 & 36 & 14 & -14 & -12 & -4 & -3 & 19 & 100 &  &  &  &  &  &  &  &  & \\ 
NR & 5 & 7 & -9 & -6 & -36 & -2 & 7 & -28 & 28 & 3 & 4 & 5 & 9 & 100 &  &  &  &  &  &  &  & \\ 
1405, 1670 & 15 & 5 & 47 & -52 & -7 & -19 & -5 & -6 & -5 & -8 & -31 & -20 & -21 & -23 & 100 &  &  &  &  &  &  & \\ 
1520, 1690 & -2 & -11 & 34 & -11 & -53 & -26 & 5 & 2 & 17 & 2 & 22 & -28 & -18 & 2 & 36 & 100 &  &  &  &  &  & \\ 
1405, 1800 & -40 & -13 & -45 & 18 & 10 & -11 & -15 & -11 & 6 & -9 & 26 & -8 & 12 & 16 & -34 & 2 & 100 &  &  &  &  & \\ 
1670, 1800 & 45 & 14 & 26 & -93 & -27 & -58 & 36 & 2 & 26 & 11 & -19 & 2 & -17 & 2 & 44 & 12 & -20 & 100 &  &  &  & \\ 
1600, 1810 & -3 & -3 & -16 & 5 & 21 & -0 & -66 & -18 & -39 & -35 & -30 & -22 & -32 & -13 & 20 & -13 & -9 & -4 & 100 &  &  & \\ 
1820, 2110 & 22 & -18 & 2 & -3 & 5 & -37 & -12 & -5 & 11 & -34 & 2 & -73 & -31 & -16 & 19 & 20 & 1 & 12 & 31 & 100 &  & \\ 
1520, NR & 8 & -6 & -6 & 1 & 2 & -39 & 45 & -5 & 9 & 20 & -21 & -6 & 6 & -22 & -8 & -7 & -4 & 16 & -12 & 19 & 100 & \\ 
1690, NR & -17 & -28 & -12 & 17 & 9 & -5 & -31 & 2 & -22 & -33 & 2 & -54 & -3 & -8 & 14 & 12 & 13 & -14 & 40 & 35 & -15 & 100 \\
\bottomrule
\end{tabular}

%% file: Authorship_LHCb-PAPER-2023-036.tex
% LHCb collaboration author list
% Data extracted on November 23rd, 2023 at 10:39am for paper reference LHCb-PAPER-2023-036
\centerline
{\large\bf LHCb collaboration}
\begin
{flushleft}
\small
R.~Aaij$^{35}$\lhcborcid{0000-0003-0533-1952},
A.S.W.~Abdelmotteleb$^{54}$\lhcborcid{0000-0001-7905-0542},
C.~Abellan~Beteta$^{48}$,
F.~Abudin{\'e}n$^{54}$\lhcborcid{0000-0002-6737-3528},
T.~Ackernley$^{58}$\lhcborcid{0000-0002-5951-3498},
B.~Adeva$^{44}$\lhcborcid{0000-0001-9756-3712},
M.~Adinolfi$^{52}$\lhcborcid{0000-0002-1326-1264},
P.~Adlarson$^{78}$\lhcborcid{0000-0001-6280-3851},
C.~Agapopoulou$^{46}$\lhcborcid{0000-0002-2368-0147},
C.A.~Aidala$^{79}$\lhcborcid{0000-0001-9540-4988}
% ...

{\footnotesize \it

% ...
$^{35}$Nikhef National Institute for Subatomic Physics, Amsterdam, Netherlands\\
% ...
$^{44}$Instituto Galego de F{\'\i}sica de Altas Enerx{\'\i}as (IGFAE), Universidade de Santiago de Compostela, Santiago de Compostela, Spain\\
% ...
$^{46}$European Organization for Nuclear Research (CERN), Geneva, Switzerland\\
% ...
$^{48}$Physik-Institut, Universit{\"a}t Z{\"u}rich, Z{\"u}rich, Switzerland\\
% ...
$^{52}$H.H. Wills Physics Laboratory, University of Bristol, Bristol, United Kingdom\\
% ...
$^{54}$Department of Physics, University of Warwick, Coventry, United Kingdom\\
% ...
$^{57}$School of Physics and Astronomy, University of Glasgow, Glasgow, United Kingdom\\
$^{58}$Oliver Lodge Laboratory, University of Liverpool, Liverpool, United Kingdom\\
% ...
$^{66}$Syracuse University, Syracuse, NY, United States\\
% ...
$^{78}$Department of Physics and Astronomy, Uppsala University, Uppsala, Sweden, associated to $^{57}$\\
$^{79}$University of Michigan, Ann Arbor, MI, United States, associated to $^{66}$\\
% ...
}
\end{flushleft}